\documentclass[10pt,journal,onecolumn]{IEEEtran}
\usepackage[noadjust]{cite}
\usepackage{mathrsfs}
\usepackage{amsfonts}
\usepackage{amsmath,amssymb,amsthm}
\usepackage{graphicx}
\newtheorem{theorem}{Theorem}

\newtheorem{lemma}[theorem]{Lemma}
\newtheorem{corollary}[theorem]{Corollary}
\usepackage{enumitem}
\begin{document}
\title{On the Dependence of Linear Coding Rates on the Characteristic of the Finite Field}
\author{Niladri Das and Brijesh Kumar Rai}
%
%
\maketitle
\begin{abstract}
It is known that for any finite/co-finite set of primes there exists a network which has a rate $1$ solution if and only if the characteristic of the finite field belongs to the given set. We generalize this result to show that for any positive rational number $k/n$, and for any given finite/co-finite set of primes, there exists a network which has a rate $k/n$ fractional linear
network coding solution if and only if the characteristic of the finite field belongs to the given set. For this purpose we construct two networks: $\mathcal{N}_1$ and $\mathcal{N}_2$; the network $\mathcal{N}_1$ has a $k/n$ fractional linear network coding solution if and only if the characteristic of the finite field belongs to the given finite set of primes, and the network $\mathcal{N}_2$ has a $k/n$ fractional linear network coding solution if and only if the characteristic of the finite field belongs to the given co-finite set of primes.   

Recently, a method has been introduced where characteristic-dependent linear rank inequalities are produced from networks whose linear coding capacity depends on the characteristic of the finite field. By employing this method on the networks $\mathcal{N}_1$ and $\mathcal{N}_2$, we construct two classes of characteristic-dependent linear rank inequalities. For any given set of primes, the first class contains an inequality which holds if the characteristic of the finite field does not belong to the given set of primes but may not hold otherwise; the second class contains an inequality which holds if the characteristic of the finite field belongs to the given set of primes but may not hold otherwise. We then use these inequalities to obtain an upper-bound on the linear coding capacity of $\mathcal{N}_1$ and $\mathcal{N}_2$.
\end{abstract}

%

\section{Introduction}
%
In the year 2000, Ahlswede \textit{et al.} \cite{ahlswede} showed that the min-cut bound on the capacity of multicast networks can be achieved by allowing the nodes of the network to compute functions of the incoming symbols. It has been later shown that a restricted version of network coding, called the linear network coding is sufficient to achieve the capacity of multicast networks \cite{li}. In linear network coding, the source alphabet is a ring or a finite field, and all symbols outgoing from a node is a linear function of the symbols the node receive. Li \textit{et al.} \cite{li} showed that such functions always exist (for multicast networks) if the underlying finite field is sufficiently large. Moreover, there are efficient algorithms to design these linear functions \cite{jaggi,ho}. Though, recently it has been shown that a multicast network being linearly solvable over a sufficiently large finite field does not necessarily guarantee solvability over every larger field \cite{sun1}.

For non-multicast networks though, linear network coding may not always achieve the capacity of the network \cite{insuff}. A network was presented in \cite{insuff} where linear coding capacity is strictly less than the coding capacity. It has been shown that the linear coding capacity of a network cannot be improved even if the source alphabet is a ring instead of a field \cite{ringc}. Reference \cite{ringc} also shows that, over finite fields, linear coding capacity of a network depends only on the characteristic of the finite field. 

In a network code a block of symbols (say $k$ symbols) is considered at every source, and each edge forwards a block of symbols (say $n$ symbols) to its outgoing edges where these symbols are functions of incoming symbols to the nodes. The ratio $k/n$ is called the rate of the network code. In this paper we consider two specific issues related to linear coding rates. The first problem is related to the dependency of the linear coding rate on the characteristic of the finite field in a linear network coding problem. The second problem deals with producing linear rank inequalities that bound the linear coding rates of a network. In the rest of this section we discuss prior works related to these issues and present our contributions. We end this section detailing the organization of the rest of the paper.
\begin{itemize}[noitemsep,topsep=0pt,leftmargin=*]
\vspace*{2pt}
\item \textit{Dependency on the characteristic of the finite field}\\
In case of multicast networks, the characteristic of the finite field does not play an important role in the sense that there does not exist a multicast network which has a scalar/vector linear solution if and only if the characteristic of the finite field belongs to a certain set of values. However, this is not true for non-multicast networks. In \cite{insuff}, Dougherty \textit{et al.} showed a network known as the Fano network which has a rate $1$ linear solution over any finite field of even characteristic, but over  finite fields of odd characteristics, no rate more than $4/5$ is achievable using linear network coding. References \cite{matroid} and \cite{insuff} show another network known as the non-Fano network which has a rate $1$ linear solution over finite fields of odd characteristics, but over even characteristics no rate more than $5/6$ is achievable using linear network coding ($5/6$ upper-bound has been shown in \cite{rateregion}). Furthermore, it has been shown in \cite{poly} that given any system of polynomial equations over integers, there exists a network which has a scalar linear network coding solution over a finite field if and only if the system of polynomial equations has a root in the same finite field. This showed that for any finite/co-finite set of primes, there exists a network which has a scalar linear solution if and only if the characteristic of the finite field belongs to the given set of primes. Afterwards, Rai \textit{et al.} showed in \cite{rai} that given finite/co-finite set of primes there exists a network which has a vector linear solution if and only if the characteristic of the finite field belong to the given set of primes.

%

\vspace*{4pt}
\item \textit{Characteristic-dependent linear rank inequalities}\\
Determining the coding capacity (or even linear coding capacity) of a general network is considered to be a very difficult problem.  
Although the capacity/linear capacity computation of various small networks have been presented in the literature. However, in such computations ad-hoc methods have been used. 
Harvey \textit{et al.} \cite{capacity} presented a method to obtain an upper-bound on the coding capacity by combining Shannon inequalities with topological properties of the network (informational dominance and independence of source symbols). In some cases, the bound obtained from this method may be improved by additionally incorporating non-Shannon information inequalities. A network named as the V$\acute{\text{a}}$mos network is a good example to see how these inequalities come together. V$\acute{\text{a}}$mos network was first considered in \cite{matroid}, where by applying non-Shannon inequalities it has been shown that its coding capacity is upper-bounded by $10/11$. This bound has been further improved to $19/21$ by applying other non-Shannon inequalities in \cite{four}.

\vspace*{4pt}
To determine an upper-bound on the linear coding capacity, in addition to the Shannon and non-Shannon information inequalities, linear rank inequalities may also be applied. 
%
Linear rank inequalities are inequalities that are obeyed by ranks (dimensions) of any collection of vector subspaces of a finite dimensional vector space. For example, if $A$ and $B$ are vector subspaces of $V$, then, $dim(A) + dim(B) \geq dim(A+B) + dim(A\cap B)$ is a linear rank inequality. On the contrary, information inequalities are the Shannon and the non-Shannon inequalities which are obeyed by random variables. When applying an information inequality to a network, the messages are taken as random variables distributed over the source alphabet. When applying a linear rank inequality, the messages are taken as vector subspaces of a finite dimensional vector space over a finite field. 
For any collection of vector subspaces of a finite dimensional vector space, in p. 452 of \cite{hammer} a way is shown to construct a corresponding set of random variables such that the dimension of any collection of the vector subspaces is equal to the joint entropy of the corresponding random variables (upto a scale factor). As a result, all subspaces of a vector space also obey the information inequalities (assuming the underlying conversion from vector subspaces to random variables). This implies that all information inequalities are also linear rank inequalities. However, the opposite is not true, \textit{i.e.} not all linear rank inequalities are information inequalities (Theorem~4 of \cite{hammer}). This implies that the best upper-bound obtained using information inequalities may not be a tight upper-bound on the linear coding capacity.


\vspace*{4pt}

Hammer \textit{et al.} showed that for upto three variables, there exists no linear rank inequality which is not an information inequality (Theorem~3 of \cite{hammer}). They also showed that for four variables, the only linear rank inequality that is not an information inequality is the Ingleton inequality upto permutations of the variables (Theorem~5 of \cite{hammer}). A list of twenty four new linear rank inequalities on five variables which are not information inequalities has been shown in \cite{five}. 
Reference \cite{six} shows that even an incomplete list of six variable linear rank inequalities crosses one billion. For seven or more variables, it has been shown in \cite{anna,rateregion} and \cite{field3} that there exist linear rank inequalities that hold if the characteristic of the field is among a certain set of values, but may not hold otherwise (this is expected as linear coding capacity has been shown to be dependent on the characteristic of the finite field). Such an inequality is called as a characteristic-dependent linear rank inequality.
\vspace*{4pt}

First, Blasiak \textit{et al.} showed two such seven variable inequalities: one holds over finite fields of even characteristic, and the other holds over finite fields of odd characteristic. Thereafter, Dougherty \textit{et al.}  showed two more seven variable characteristic-dependent linear rank inequalities in \cite{rateregion}. Subsequently, two new eight variable inequalities has been presented in \cite{field3}. Application (finding upper-bounds on the linear coding capacity of networks) of the inequalities shown in \cite{rateregion} and \cite{field3}  has been also shown in the respective papers.
\vspace*{4pt}

For producing these inequalities, the authors of \cite{rateregion} and \cite{field3} developed a novel method where these inequalities were yielded from the very networks they intended to find the linear coding capacity of. Hereafter, we will refer this method as the DFZ method. In reference \cite{rateregion} two linear rank inequalities have been obtained: one holds over all finite fields of odd characteristic but may not hold otherwise (produced from the Fano network); and another holds over all finite fields of even characteristic but may not hold otherwise (produced from the non-Fano network). In reference \cite{field3}, first an inequality that holds over all finite fields of characteristic not equal to $3$ but may not hold otherwise was produced from the T8 network; and then another inequality that holds over all finite fields of characteristic equal to $3$ but may not hold otherwise was produced from the non-T8 network.

%
\end{itemize}
\subsection{Contributions of this paper}
\begin{itemize}
\vspace*{2pt}
\item \textit{First contribution of the paper}\\
In the works of \cite{poly} and \cite{rai}, the dependency on the characteristic of the field is shown only for either scalar linear network coding or for vector linear network coding. In this paper we show that for any positive rational number $\frac{k}{n}$ and for any given finite/co-finite set of prime numbers, there exists a network which has a rate $\frac{k}{n}$ fractional linear network code solution if and only of the characteristic of the finite belongs to the given finite/co-finite set of primes.



\vspace{2pt}
\item \textit{Second contribution of the paper}\\
In the second result of this paper, we construct two classes of characteristic-dependent linear rank inequalities. Given a set of primes, the first class contains an inequality that holds if the characteristic of the finite field does not belong to the given set; and the second class contains an inequality that holds if the characteristic belongs to the given set. We also show that the inequalities in the first class may not hold if the characteristic belongs to the given set of primes; and the inequalities in the second class may not hold if the characteristic does not belong to the given set of primes. This contribution can be seen as a generalization of the works in \cite{rateregion} and \cite{field3}.



\end{itemize}

\subsection{Organization of the paper}
In Section~\ref{sec2} we reproduce the standard definitions of fractional linear network coding, vector linear network and scalar linear network coding. In Section~\ref{subsec1}, for any positive rational number $\frac{k}{n}$, and for any finite set of primes, we present a network $\mathcal{N}_1$ which has a rate $\frac{k}{n}$ fractional linear network coding solution if and only if the characteristic of the finite field belongs to the given set of primes. For the ease of readability, a part of this proof is shifted to Appendix~\ref{appA}. In Fig.~\ref{genfano1/n} we show a network $\mathcal{N}_1^\prime$ which we use to construct $\mathcal{N}_1$. 
In Section~\ref{subsec2}, for any positive rational number $\frac{k}{n}$, and for any finite set of primes, we present a network $\mathcal{N}_2$ which has a rate $\frac{k}{n}$ fractional linear network coding solution if and only if the characteristic of the finite field does not belong to the given set of primes. As earlier, a part of this proof is deferred until Appendix~\ref{appB}. We construct the network $\mathcal{N}_2$ by using another network $\mathcal{N}_2^\prime$ shown in Fig.~\ref{nofano1/n}.

In Theorem~\ref{Thm1}, Section~\ref{sec5}, using the network $\mathcal{N}_1^\prime$, we also construct a characteristic-dependent linear rank inequality that holds if the characteristic of the finite field does not belong to the given set of primes but may not hold otherwise. The proof of this theorem is presented in Appendix~\ref{ineq1}. Then, using $\mathcal{N}_2^\prime$, we construct a characteristic-dependent linear rank inequality that holds if the characteristic of the finite field belongs to the given set of primes but may not hold otherwise. This inequality is presented in Theorem~\ref{Thm2} of Section~\ref{sec5}, and proved in Appendix~\ref{appD}. Usage of these inequalities in computing upper-bounds on the linear coding capacity of $\mathcal{N}_1$ and $\mathcal{N}_2$ are also shown in Section~\ref{sec5}.

%
%
%
%
%

\section{Preliminaries}\label{sec2}
A network is represented by a graph $G(V,E)$. The set $V$ is partitioned into three disjoint sets: the set of sources $S$, the set of terminals $T$, and the set of intermediate nodes $V^\prime$. Without loss of generality, the sources are assumed to have no incoming edge and the terminals are assumed to have no outgoing edge. Each source generates an i.i.d random process uniformly distributed over an alphabet $\mathcal{A}$. The source process at any source is independent of all source processes generated at other sources. Each terminal demands the information generated by a subset of the sources. An edge $e$ originating from node $u$ and ending at node $v$ is denoted by $(u,v)$; where $u$ is denoted by $tail(e)$, and $v$ is denoted by $head(e)$. For a node $v\in V$, the set of edges $e$ for which $head(e) = v$ is denoted by $In(v)$. The information carried by an edge $e$ is denoted by $Y_e$. Without loss of generality it is assumed that all the edges in the network are unit capacity edges (meaning, in one usage of an edge it carries one symbol from $\mathcal{A}$).

In a $(k,n)$ fractional linear network code when the alphabet $\mathcal{A}$ is a finite field $\mathbb{F}_q$ is defined as follows. Each source $s_i\in S$ generates a symbol $X_i$ from the finite field $\mathbb{F}_q^k$. For any edge $e$, if $tail(e) = s_i$ for any $s_i \in S$, then $Y_e = A_{\{s_i,e\}}X_i$ where $Y_e \in \mathbb{F}_q^n$, $A_{\{s_i,e\}} \in \mathbb{F}_q^{n\times k}$ and $X_i\in \mathbb{F}_q^k$. If $tail(e) = v$ where $v\in V^{\prime}$, then $Y_e = \sum_{\forall e^\prime \in In(v)} A_{\{e^\prime,e\}}Y_{e^\prime}$ where $Y_e,Y_{e^\prime} \in \mathbb{F}_q^n$, and $A_{\{e^\prime,e\}} \in \mathbb{F}_q^{n\times n}$. For any terminal $t\in T$, if $t$ computes symbol $X_t$, then $X_t = \sum_{\forall e^\prime \in In(t)} A_{\{e^\prime,t\}}Y_{e^\prime}$ where $X_t\in \mathbb{F}_q^k, A_{\{e^\prime,t\}} \in \mathbb{F}_q^{k\times n}$ and $Y_{e^\prime}\in \mathbb{F}_q^n$. The matrices $A_{\{s_i,e\}}, A_{\{e^\prime,e\}}$ and  $A_{\{e^\prime,t\}}$ are called as the local coding matrices.

Using a $(k,n)$ fractional linear network code, if all terminals can compute the symbols it demand, then the network is said to have a $(k,n)$ fractional linear network coding solution. The ratio $\frac{k}{n}$ is called the rate. A network is said to have a rate $\frac{k}{n}$ fractional linear network coding solution if it has a $(dk,dn)$ fractional linear network coding solution for any non-zero positive integer $d$. A $(k,k)$ fractional linear network code is called as a $k$ dimensional vector linear network code; and $k$ is called as the vector dimension or as the message dimension. If a network has a $(k,k)$ fractional linear network coding solution then it is said that the network has a vector linear solution for $k$ vector dimension. If a network has a $(1,1)$ vector linear network coding solution then the network is said to be scalar linearly solvable.

\section{A network having a rate $\frac{k}{n}$ fractional linear network coding solution iff the characteristic belongs to a given finite/co-finite set of primes}\label{sec3}

\subsection{Network having $\frac{k}{n}$ solution iff the characteristic of the finite field belongs to a given finite set of primes.}\label{subsec1}
\begin{figure*}
\centering
\includegraphics[width=.96\textwidth]{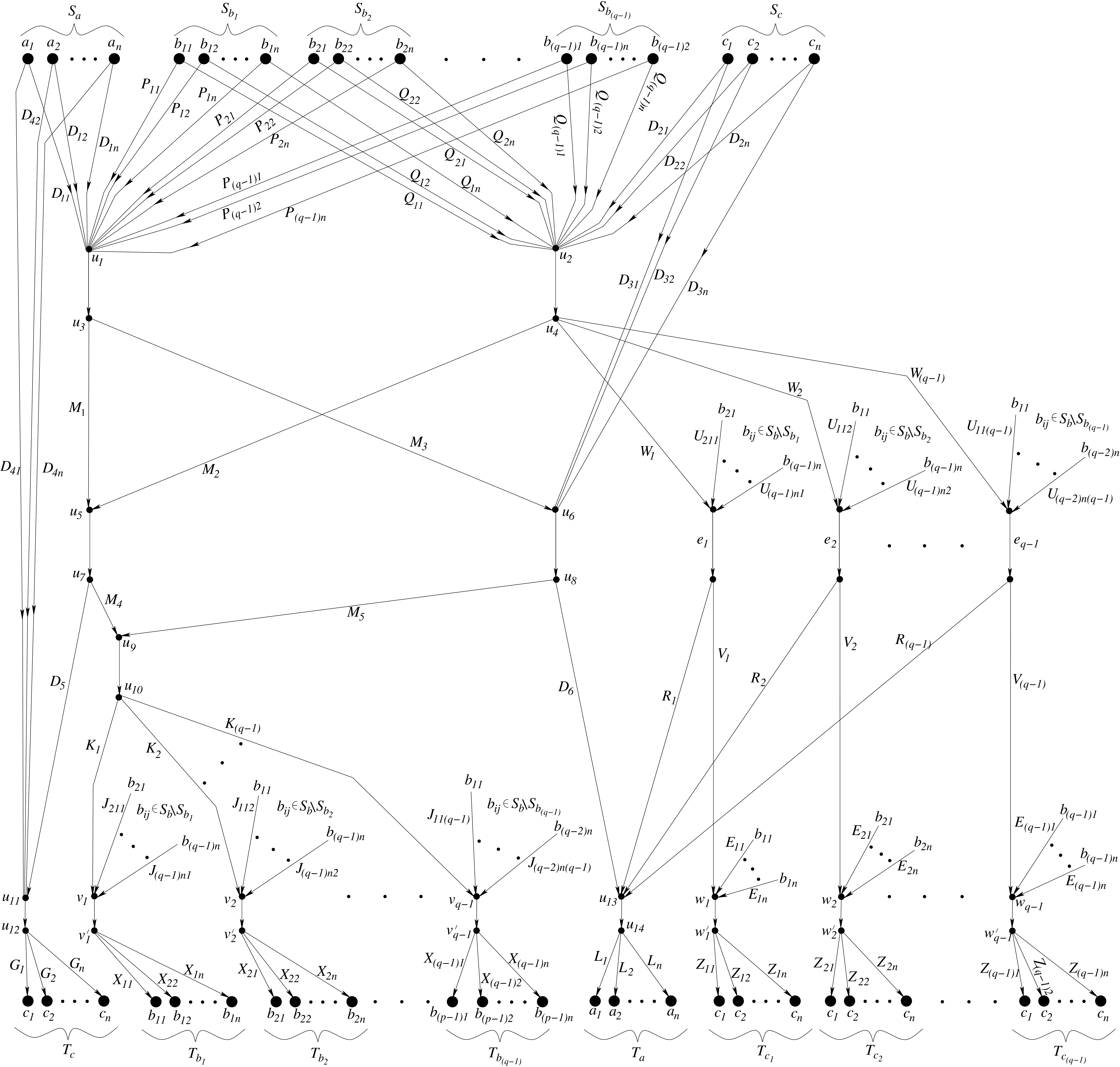}
\caption{Network $\mathcal{N}_1^\prime$ which has a rate $\frac{1}{n}$ fractional linear network coding solution if and only if the characteristic of the finite field divides $q$}
\label{genfano1/n}
\end{figure*}
First we show that for any positive non-zero rational number $\frac{k}{n}$, and for any given finite set of primes, there exists a network which has a rate $\frac{k}{n}$ fractional linear network coding solution if and only if the characteristic of the finite field belongs to the given set. Our proof is constructive. Consider the network $\mathcal{N}_1^\prime$ presented in Fig.~\ref{genfano1/n}. 
The network has $(q+1)$ sets of sources: $S_a = \{a_1,a_2,\ldots,a_n\}$, $S_{b_i} = \{b_{i1},b_{i2},\ldots,b_{in}\}$ for $1\leq i\leq (q-1)$, and $S_c = \{c_1,c_2,\ldots,c_n\}$. The source $s_i\in S_a $ generates the message $a_i$. For $1\leq i\leq (q-1)$ and $1\leq j\leq n$ the source $s_j \in S_{b_i}$ generates the message $b_{ij}$. And the source $s_i\in S_c$ generates the message $c_i$. In the figure, the source nodes are indicated by the massage it generates. 
There are $2q$ sets of terminals: $T_c, T_a$, $T_{b_i}$ for $1\leq i\leq (q-1)$, and $T_{c_i}$ for $1\leq i\leq (q-1)$. Each individual terminal is indicated by the source message it demands.

List of edges emanating from a source node:
\begin{enumerate}
\item $(s,u_1)$ for $\forall s \in \{S_a,S_{b_1},S_{b_2},\ldots,S_{b_{q-1}} \} $.
\item $(s,u_2)$ for $\forall s \in \{S_{b_1},S_{b_2},\ldots,S_{b_{q-1}},S_c\} $.
\item $(a_i,u_{11})$ for $1\leq i\leq n$.
\item $(c_i,u_6)$ for $1\leq i\leq n$.
\item $(b_{ij},tail(e_k))$ for $1\leq i,k \leq (q-1)$, $i\neq k$, $1\leq j\leq n$.
\item $(b_{ij},v_k)$ for $1\leq i,k \leq (q-1)$, $i\neq k$, $1\leq j\leq n$.
\item $(b_{ij},w_i)$ for $1\leq i \leq (q-1)$, $1\leq j\leq n$.
\end{enumerate} 
List the edges which originates at an intermediate node and ends at a intermediate node:
\begin{enumerate}
\item $(u_i,u_{i+2})$ for $1\leq i\leq 7, i\neq 4$.
\item $(u_i,u_{i+1})$ for $i=4,8,9,11,13$.
\item $(u_3,u_6)$, $(u_7,u_{11})$, and $(u_8,u_{13})$
\item $e_i$ for $1\leq i\leq (q-1)$
\item $(u_4,tail(e_i))$ for $1\leq i\leq (q-1)$
\item $(head(e_i),u_{13})$ and $(head(e_i),w_i)$ for $1\leq i\leq (q-1)$
\item $(u_{10},v_i)$ and $(v_i,v^\prime_i)$ for $1\leq i\leq (q-1)$
\item $(w_i,w^\prime_i)$ for $1\leq i\leq (q-1)$
\end{enumerate}
For any terminal $t_i\in T_c$ there exists an edge $(u_{12},t_i)$ and $t_i$ demands the message $c_i$. For any terminal $t_j\in T_{b_i}$ for $1\leq i\leq (q-1), 1\leq j\leq n$, there exits an edge $(v^\prime_i,t_j)$ where the terminal $t_j$ demands the message $b_{ij}$. For any terminal $t_i\in T_a$ there exits an edge $(u_{14},t_i)$ and $t_i$ demands the message $a_i$. For $1\leq i\leq (q-1)$, a terminal $t_j\in T_{c_i}$ for $1\leq j\leq n$ is connected from the node $w^\prime_i$ by the edge $(w^\prime_i,t_j)$ and $t_j$ demands the message $c_j$. The local coding matrices are shown alongside the edges. 
\begin{lemma}\label{lem3}
The network in Fig.~\ref{genfano1/n} has a rate $\frac{1}{n}$ fractional linear network coding solution if and only if the characteristic of the finite field divides $q$.
\end{lemma}

The proof of this lemma is shown in Appendix~\ref{appA}.

\begin{theorem}\label{thm1}
For any non-zero positive rational number $\frac{k}{n}$ and for any finite set of prime numbers $\{p_1,p_2,\ldots,p_l\}$, there exists a network which has a rate $\frac{k}{n}$ fractional linear network coding solution if and only if the characteristic of the finite field belongs to the given set of primes.
\end{theorem}
\begin{IEEEproof}
Let us consider the union of $k$ copies of the network $\mathcal{N}_1^\prime$ shown in Fig.~\ref{genfano1/n} each for $q = p_1\times p_2\times \cdots\times p_l$. Denote the $i^{\text{th}}$ copy as $\mathcal{N}_{1i}^\prime$. Note that each source and each terminal has $k$ copies in the union. Join all copies of any source or terminal into a single source or terminal respectively. Name this new network as $\mathcal{N}_1$. We show below that $\mathcal{N}_1$ has a rate $\frac{k}{n}$ fractional linear network coding solution if and only if the characteristic of the finite field belong to the set $\{p_1,p_2,\ldots,p_l\}$. Before we proceed further, consider the following property of $\mathcal{N}_1$ and $\mathcal{N}_1^\prime$.
\begin{lemma}\label{lem2}
If $\mathcal{N}_1$ has a $(dk,dn)$ fractional linear network coding solution for any non-zero positive integer $d$, then $\mathcal{N}_1^\prime$ has a $(dk,dkn)$ fractional linear network coding solution.
\end{lemma}
\begin{IEEEproof}
This is true since the information that can be sent using the network $\mathcal{N}_1$ in $x$ times, can be sent using the network $\mathcal{N}_1^\prime$ in $kx$ times. This is because $\mathcal{N}_1^\prime$ has $k$ copies of $\mathcal{N}_1$. 
\end{IEEEproof}
First consider the only if part. Say $\mathcal{N}_1$ has a rate $\frac{k}{n}$ fractional linear network coding solution even if the characteristic does not belong to the set $\{p_1,p_2,\ldots,p_l\}$. Then from Lemma~\ref{lem2}, the network $\mathcal{N}_1^\prime$ has a rate $\frac{1}{n}$ fractional linear network coding solution even if the characteristic does not belong to the given set of primes. However, as shown in Lemma~\ref{lem3}, $\mathcal{N}_1^\prime$ has a rate $\frac{1}{n}$ fractional linear network coding solution if and only if the characteristic of the finite field divides $q$. But, as $q = p_1\times p_2\times \cdots\times p_l$, the characteristic divides $q$ if and only if the characteristic is one of the primes in the set. Hence this is a contradiction.

Now consider the if part. Since $\mathcal{N}_{1i}^\prime$ for $1\leq i\leq k$ has a $(1,n)$ fractional linear network coding solution, a $(k,n)$ fractional linear network coding solution for $\mathcal{N}_1$ can be constructed by keeping the same local coding matrices in all of the copies and sending the $i^{\text{th}}$ component of each source through $\mathcal{N}_{1i}^\prime$.
\end{IEEEproof}

\subsection{Network having $\frac{k}{n}$ solution iff the characteristic of the finite field belongs to a given co-finite set of primes.}\label{subsec2}
\begin{figure*}
\centering
\includegraphics[width=\textwidth]{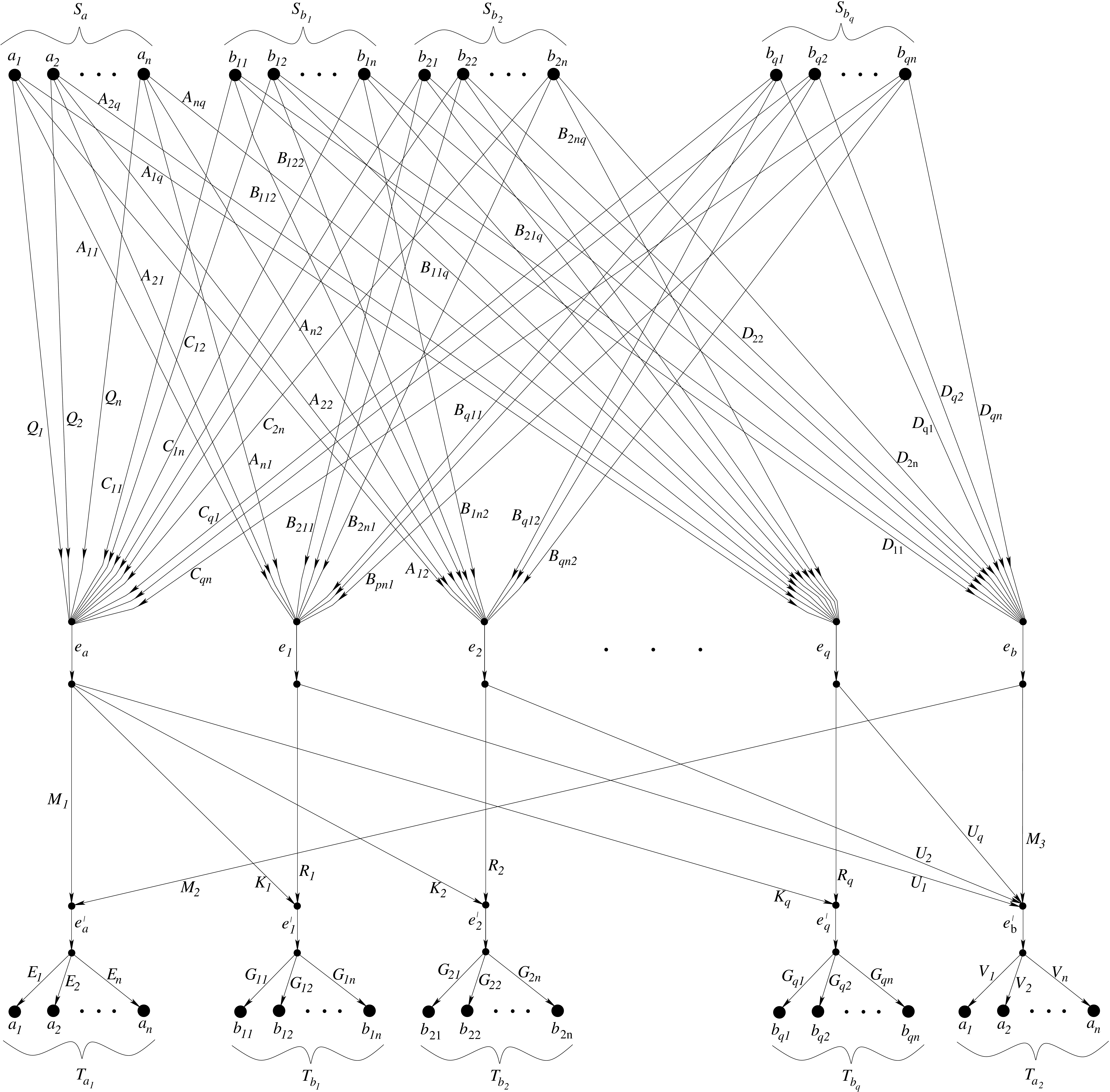}
\caption{A network $\mathcal{N}_2^\prime$ which has a rate $1/n$ fractional linear network coding solution if and only if the characteristic of the finite field does not divide $q$.}
\label{nofano1/n}
\end{figure*}
The outline of the contents in this sub-section is similar to that of the last sub-section. Consider the network $\mathcal{N}_2^\prime$ shown in Fig.~\ref{nofano1/n}. The sources are partitioned into $(q+1)$ sets: $S_a = \{a_1,a_2,\ldots,a_n\}$ and $S_{b_i} = \{b_{i1},b_{i2},\ldots,b_{in}\}$ for $1\leq i\leq q$. A source node and the message generated by the node is indicated by the same notation. The set of terminals are partitioned into $(q+2)$ disjoint sets: $T_{a_1},T_{a_2}$ and $T_{b_i}$ for $1\leq i\leq q$; where each set has $n$ terminals. Each individual terminal is indicated by the source message it demands. We have the following edges in the network.
\begin{enumerate}
\item $e_a,e_b,e_a^\prime$ and $e_b^\prime$
\item $e_i$ and $e_i^\prime$ for $1\leq i\leq q$
\item $(s,tail(e_a))$ for $\forall s\in  S_a \cup \{\cup_{i=1}^q S_{b_i}\}$
\item $(s,tail(e_i))$ for $1\leq i\leq q$ and $\forall s\in S_a \cup \{ \cup_{j=1,j\neq i}^q S_{b_i}\}$
\item $(s,tail(e_b))$ for $\forall s\in \cup_{i=1}^q S_{b_i}$
\item $(head(e_a),tail(e_a^\prime))$ and $(head(e_b),tail(e_a^\prime))$
\item $(head(e_b),tail(e_b^\prime))$
\item $(head(e_i),tail(e_i^\prime))$ for $1\leq i\leq q$
\item $(head(e_a),tail(e_i^\prime))$ for $1\leq i\leq q$
\item $(head(e_i),tail(e_b^\prime))$ for $1\leq i\leq q$
\end{enumerate}
From each of the nodes $head(e_a^\prime)$, $head(e_i^\prime)$ for $1\leq i\leq q$, and $head(e_b^\prime)
$, $n$ outgoing edges emanate, and the $head$ node of all such edges is a terminal. The set of $n$ terminals which have a path from node $head(e_a^\prime)$ are denoted by $T_{a_1}$. Similarly, the set of $n$ terminals which have a path from node $head(e_b^\prime)$ are denoted by $T_{a_2}$. And the $n$ terminals in the set $T_{b_i}$ for $1\leq i\leq q$ are connected from the node $head(e_i^\prime)$ by an edge.

\begin{lemma}\label{lem4}
The network shown in Fig.~\ref{nofano1/n} has a rate $\frac{1}{n}$ fractional linear network coding solution if and only if the characteristic of the finite field does not divide $q$.
\end{lemma}

The proof of this lemma is shown in Appendix~\ref{appB}.

\begin{theorem}\label{thm2}
For any non-zero positive rational number $\frac{k}{n}$ and for any finite set of prime numbers $\{p_1,p_2,\ldots,p_l\}$, there exists a network which has a rate $\frac{k}{n}$ fractional linear network coding solution if and only if the characteristic of the finite field does not belong to the given set of primes.
\end{theorem}
\begin{IEEEproof}
Let $q$ be equal to $p_1.p_2.\ldots .p_l$ in $\mathcal{N}_2^\prime$. Let us construct $\mathcal{N}_2$ by joining $k$ copies of $\mathcal{N}_2^\prime$ at the corresponding sources and the terminals, in a similar way $\mathcal{N}_1$ was constructed from $\mathcal{N}_1^\prime$. It can be also seen that Lemma~\ref{lem2} holds true when $\mathcal{N}_1$ and $\mathcal{N}_1^\prime$ are replaced by $\mathcal{N}_2$ and $\mathcal{N}_2^\prime$ respectively. So if $\mathcal{N}_2$ has a rate $\frac{k}{n}$ fractional linear network coding solution then $\mathcal{N}_2^\prime$ has a rate $\frac{1}{n}$ fractional linear network coding solution. 

Now say $\mathcal{N}_2$ has a rate $\frac{k}{n}$ fractional linear network coding solution even if the characteristic of the finite belongs to the set $\{p_1,p_2,\ldots,p_l\}$. Then, as $q = p_1.p_2.\ldots .p_l$, the characteristic of the finite field divides $q$. Then, $\mathcal{N}_2^\prime$ has a rate $\frac{k}{kn} = \frac{1}{n}$ fractional linear network coding solution over a finite field even if the characteristic divides $q$. However, this is in contradiction to Lemma~\ref{lem4}.

If however, the characteristic does not belong to the given set of primes, then, since there are $k$ copies of $\mathcal{N}_2^\prime$ in $\mathcal{N}_2$, and each copy has a $(1,n)$ fractional linear network coding solution, a $(k,n)$ fractional linear network coding solution can easily be constructed for $\mathcal{N}_2$.
\end{IEEEproof}

\section{A multiple-unicast network having a rate $\frac{k}{n}$ fractional linear network coding solution iff the characteristic belongs to a given finite/co-finite set of primes}\label{sec4}
\begin{figure}
\centering
\includegraphics[width=0.48\textwidth]{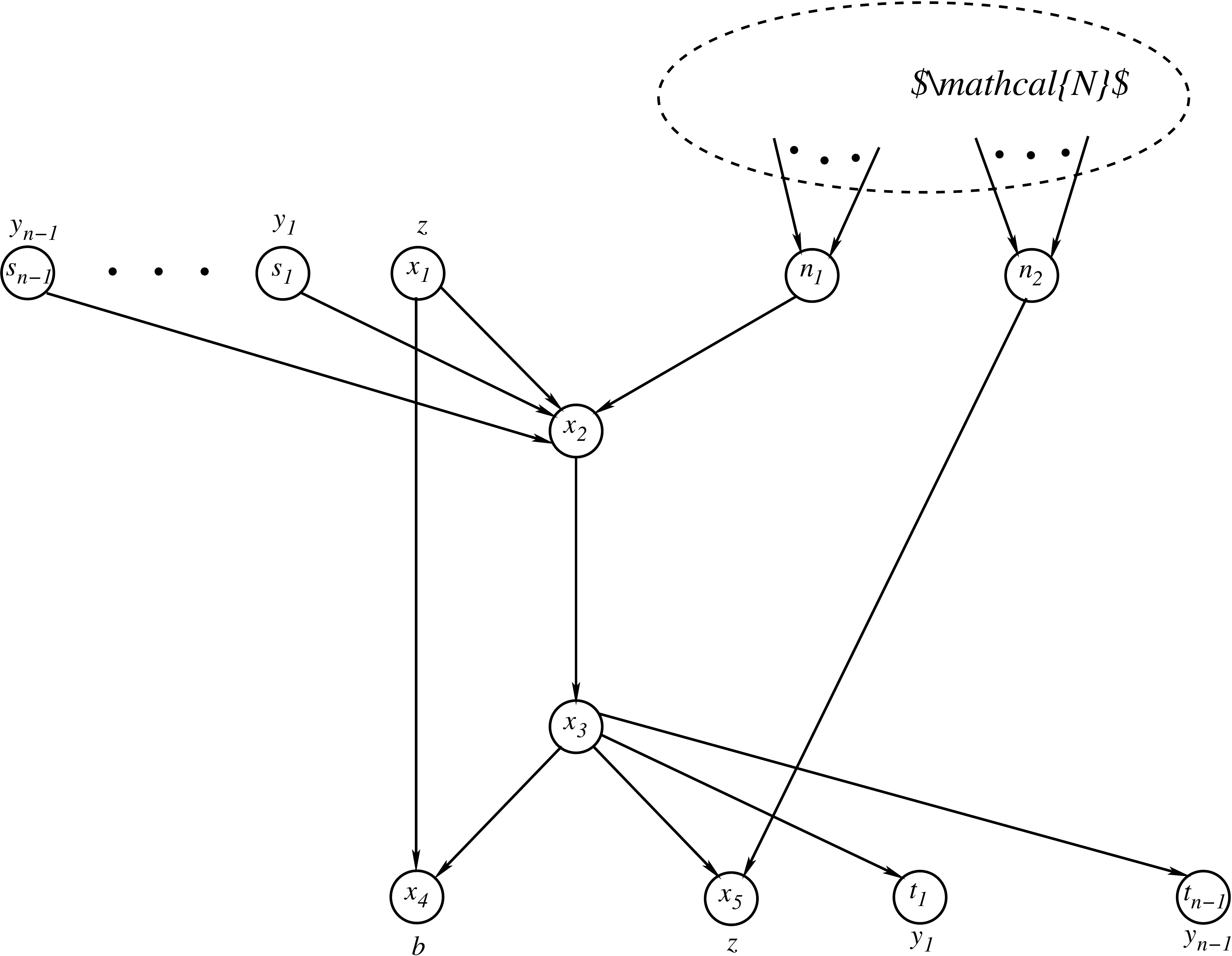}
\caption{Gadget which attaches to the two terminals (denoted by $n_1$ and $n_2$) demanding the same message (denoted by $b$) of any arbitrary network (indicated by the dotted lines). Nodes $x_1,s_1,\ldots ,s_{n-1}$ are source nodes and source $s_i$ generates the messages $y_i$ for $1\leq i\leq (n-1)$; $x_1$ generates the message $z$. The nodes $t_1, \ldots ,t_{n-1},x_4,x_5$ are terminals and $t_i$ demands $y_{i}$ for $1\leq i\leq (n-1)$; $x_4$ demands $b$, and $x_5$ demands $z$. Nodes $x_2$ and $x_3$ are the intermediate nodes.
}
\label{nw}
\end{figure}
In this section we show that for any non-zero positive rational number $\frac{k}{n}$ and for any finite/co-finite set of primes, there exists a multiple-unicast network which has a rate $\frac{k}{n}$ fractional linear network coding solution if and only if the characteristic of the finite field belongs to the given set. To prove this result, we first show that for each of the networks $\mathcal{N}_1^\prime$ and $\mathcal{N}_2^\prime$ presented in Section~\ref{sec3}, there exists a multiple-unicast network which has a $(1,n)$ fractional linear network coding solution if and only if the the corresponding network $\mathcal{N}_1^\prime$ or $\mathcal{N}_2^\prime$ has a $(1,n)$ fractional linear network coding solution.

In a multiple-unicast network, by definition, each source process is generated at only one source node and is demanded by only one terminal. Additionally, each source node generates only one source process, and each terminal demands only one source process. In both the networks $\mathcal{N}_1^\prime$ and $\mathcal{N}_2^\prime$ there exists no source processes which is generated by more than one source node, and no source node generates more than one source process. Moreover, there does not exist any terminal which demands more than one source process. However, there exists more than one terminal which demands the same source process. This is fixed in the following way. 

In \cite{reversible} it has been shown that for any network there exists a solvably equivalent multiple-unicast network. To resolve the case of more than one terminals demanding the same source message, the authors considered two such terminals at a time and added a gadget to the two terminals. The same procedure is followed here, only the gadget has been modified. This modified gadget is shown in Fig.~\ref{nw}. It is assumed that the nodes $n_1$ and $n_2$ both demanded the same message $b$ in the original network (network before attaching the gadget). After adding the gadget, the modified network has $n$ more source nodes $x_1,s_1,\ldots ,s_{n-1}$, and $n+1$ new terminal nodes $x_4,x_5,t_1,\ldots ,t_{n-1}$. Nodes $n_1$ and $n_2$ are intermediate nodes in the modified construction. This process has to be repeated iteratively for every two terminals in the original network that demand the same source process. In the same way as shown in \textit{Theorem II.1} of \cite{reversible}, it can be shown that after the completion of this process, the resulting network has a $(1,n)$ fractional linear network coding solution if and only if the original network has a $(1,n)$ fractional linear network coding solution.

Hence, as shown above, corresponding to each of the networks $\mathcal{N}_1^\prime$ and $\mathcal{N}_2^\prime$, there exist multiple-unicast networks $\mathcal{N}_1^{\prime m}$ and $\mathcal{N}_2^{\prime m}$ which have a $(1,n)$ fractional linear network coding solution if and only if $\mathcal{N}_1^\prime$ and $\mathcal{N}_2^\prime$ have a $(1,n)$ fractional linear network coding solution respectively. Now by connecting $k$ copies of $\mathcal{N}_1^{\prime m}$ and $\mathcal{N}_2^{\prime m}$ in the same way as $\mathcal{N}_1$ and $\mathcal{N}_2$ was constructed from $\mathcal{N}_1^\prime$ and $\mathcal{N}_2^\prime$ respectively, the following theorem can be proved in a similar way to Theorem~\ref{thm1} and Theorem~\ref{thm2}.
\begin{theorem}
For any non-zero positive rational number $\frac{k}{n}$ and for any finite/co-finite set of prime numbers there exists a multiple-unicast network which has a rate $\frac{k}{n}$ fractional linear network coding solution if and only if the characteristic of the finite field belongs to the given set of primes.
\end{theorem}

\section{Characteristic-dependent linear rank inequality}\label{sec5}
In this section, for any finite or co-finite set of primes, we present a characteristic-dependent linear rank inequality that holds if the characteristic of the finite field belongs to the given set, but may not hold otherwise. First we introduce some notations. To denote the dimension of a finite dimensional vector space $V$ the notations $dim(V)$ and $H(V)$ are used interchangeably. $H(U,V)$ denotes $dim(U+V)$. $H(U|V)$ denotes $dim(U+V)-dim(V)$.

\begin{theorem}\label{Thm1}
For any given set of primes $\{p_1,p_2,\ldots,p_l\}$, let $A,B_1,B_2,\ldots,B_{q-1},C,U,W,X,Y,$ $Z,V_1,V_2,\ldots,V_{q-1}$ for $q = p_1 \times p_2 \times \cdots \times p_l$, be vector subspaces of a finite dimensional vector space $V$. Then the following linear rank inequality holds if $V$ is a vector space over a finite field whose characteristic does not belong to $\{p_1,p_2,\ldots,p_l\}$, but may not hold otherwise:
\begin{IEEEeqnarray*}{l}
(2q-1)H(A) + (2q-2)H(C) + \sum_{i=1}^{q-1} 2H(B_i) \leq (q-1)(H(U) + H(Y) + H(W) + 2H(X)) + \sum_{i=1}^{q-1} H(V_i)  \\+\> (7q-6)H(U|A,B_1,\ldots,B_{q-1}) + (6q-5)H(Y|B_1,\ldots,B_{q-1},C) + \sum_{i=1}^{q-1} (2q)H(V_i|Y,B_1,\ldots, B_{i-1},B_{i+1},\ldots, B_{q-1}) \\+\> (3q-3)H(W|U,Y) + (4q-3)H(X|U,C) + (2q-2)H(Z|W,X) + (2q-1)H(A|X,V_1,\ldots ,V_{q-1}) \\+\> (q-1)H(C|A,W) + \sum_{i=1}^{q-1} 2H(B_i|Z,B_1,\ldots,B_{i-1},B_{i+1},\ldots, B_{q-1}) + \sum_{i=1}^{q-1} H(C|V_i,B_i) \\+\> (5q-4)(H(A) - H(A,B_1,B_2,\ldots ,B_{q-1},C)) + (6q-5)(\sum_{i=1}^{q-1} H(B_{i}) + H(C)) - (q-1)H(B_1,\ldots ,B_{q-1},C)\IEEEeqnarraynumspace\IEEEyesnumber\label{eq0}
\end{IEEEeqnarray*}
\end{theorem}
The proof of this inequality can be found in Appendix~\ref{ineq1}. Here we show that this inequality may not hold if $q = 0$ over the finite field (note $q=0$ when the characteristic belong to the given set of primes). Let $V$ be the vector space $V(q+1,\mathbb{F}_{p^\alpha})$ where $p \in \{p_1,p_2,\ldots,p_l\}$ and $\alpha$ is some positive integer. Let $u_i$ be the $1$ dimensional vector space spanned by the $q+1$-length vector whose $i^{\text{th}}$ element is 1 and all other elements are zero. Now, consider the following vector subspaces of $V$. 
(We construct these subspaces from the fact that the network $\mathcal{N}_1^\prime$ has a rate $1$ linear solution when $q = 0$ over the finite field and $n=1$.)
\begin{IEEEeqnarray*}{l}
A = u_1 \qquad \text{for } 1\leq i\leq q-1:\; B_i = u_{i+1} \qquad C = u_{q+1} \qquad
U = \sum_{i=1}^q u_i \qquad Y = \sum_{i=2}^{q+1} u_i\\ W = u_1 - u_{q+1} \qquad X = \sum_{i=1}^{q} u_i - u_{q+1} \qquad \text{for } 1\leq i\leq q-1:\; V_i = u_{i+1} + u_{q+1}\qquad
 Z = \sum_{i=2}^q u_i
\end{IEEEeqnarray*}
Now note $V_i = Y - \sum_{j=1,j\neq i}^q u_j$, $W = U - Y$, $X = U - u_{q+1}$, $Z = X - W$, $A = X - \sum_{i=1}^{q-1} V_i$, $C = A - W$, $B_i = Z - \sum_{j=1,j\neq i}^{q-1} B_j$, $C = Y_i - B_i$. Hence all the conditional terms in equation (\ref{eq0}) becomes zero; and the inequality returns $(6q-5) \leq (6q-6)$, or, $6 \leq 5$. Hence the inequality in equation (\ref{eq0}) is not valid over such a finite field. 

It can be easily seen that, when inequality \ref{eq0} is applied to $\mathcal{N}_1$, it results an upper-bound equal to $\frac{(6q-6)k}{(6q-5)n}$.


\begin{theorem}\label{Thm2}
For any given set of primes $\{p_1,p_2,\ldots,p_l\}$, let $A$, $B_1,B_2,\ldots,B_q$, $X$, $Z$, $Y_1,\ldots,Y_q$ for $q = p_1 \times p_2 \times \cdots \times p_l$, be vector subspaces of a finite dimensional vector space $V$. Then the following linear rank inequality holds if $V$ is a vector space over a finite field whose characteristic belongs to $\{p_1,p_2,\ldots,p_l\}$, but may not hold otherwise:
\begin{IEEEeqnarray*}{l}
2H(A) + (q+1)H(B_1) + \sum_{i=2}^q 2H(B_i) \leq (2q-1)H(X) + \sum_{i=1}^q H(Y_i) + H(Z) + (3q)H(X|A,B_1,\ldots ,B_q) \\
+\> (q+2)H(Y_1|A,\cup_{j=2}^q B_j)  + \sum_{i=2}^{q} 3H(Y_i|A,\cup_{j=1,j\neq i}^q B_j)  + 2H(Z|B_1,\ldots ,B_q) + H(A|Y_1,\ldots ,Y_q,Z) + H(A|X,Z) \\+\> (q+1)H(B_1|X,Y_1) + \sum_{i=2}^q 2H(B_i|X,Y_i) + (3q+1)(H(A) + \sum_{i=1}^q H(B_i) - H(A,B_1,\ldots ,B_q))\IEEEeqnarraynumspace\IEEEyesnumber\label{thmeq1}
\end{IEEEeqnarray*}
\end{theorem}
The proof of this inequality can be found in Appendix~\ref{appD}. Here we show that this inequality may not hold if $q$ has an inverse over the finite field (thereby meaning the characteristic of the finite field does not belong to $\{p_1,p_2,\ldots,p_l\}$). Let $V$ be the vector space $V(q+1,\mathbb{F}_{p^\alpha})$ where $p \notin \{p_1,p_2,\ldots,p_l\}$ and $\alpha$ is some positive integer. Let $u_i$ be the $1$ dimensional vector space spanned by the $q+1$-length vector whose $i^{\text{th}}$ element is 1 and all other elements are zero. (We construct these subspaces by using the fact that the network $\mathcal{N}_2^\prime$ has a rate $1$ linear solution when $q \neq 0$ over the finite field and $n=1$.)
\begin{IEEEeqnarray*}{l}
A = u_1 \qquad \text{for } 1\leq i\leq q:\; B_i = u_{i+1} \qquad X = \sum_{i=1}^{q+1} u_i \qquad \text{for } 1\leq i\leq q:\; Y_i = \sum_{j=1,j\neq i+1}^{q+1} u_i \qquad Z = \sum_{i=2}^{q+1} u_i
\end{IEEEeqnarray*}
Now note $X = A + \sum_{i=1}^q B_i$, $Y_i = A + \sum_{j=1,j\neq i}^q B_i$, $Z = \sum_{i=1}^q B_i$, $A = q^{-1}(\sum_{i=1}^q Y_i - (q-1)Z)$, $A = X - Z$, and $B_i = X - Y_i$. Hence all the conditional terms in equation (\ref{thmeq1}) becomes zero; and the inequality returns $(3q+1) \leq (3q)$, or, $1 \leq 0$.

Now, note inequality \ref{thmeq1} when applied to $\mathcal{N}_2$ results an upper-bound equal to $\frac{(3q)k}{(3q+1)n}$.

\section{Conclusion}\label{sec6}
We have showed that for any given finite/co-finite set of primes, and for any given positive rational number $k/n$, there exists a network which has a $k/n$ fractional linear network coding solution if and only if the characteristic of the finite field belongs to the given set. 




Next, for any given set of primes we have presented two characteristic-dependent linear rank inequalities: one holds if the characteristic of the finite field does not belong to the given set but may not hold otherwise; and the other holds if the characteristic of the finite field belongs to the given set but may not hold otherwise. 


%

\appendices
\section{}\label{appA}
\subsection{Proof of lemma~\ref{lem3}}
We prove this lemma by first forming a set of equations that the local coding matrices must satisfy for the network to be linearly solvable, and then we find an expression to show that these equations hold only if $q = 0$ over the finite field. The `if' part is shown by forming a rate $1$ linear solution when $q = 0$.

Consider a $(d,dn)$ fractional linear network coding solution of the network $\mathcal{N}_1^\prime$ where $d$ is any positive integer. The sizes of the local coding matrices are as follows. For $1\leq i\leq n$, the matrices $D_{4i}$ and $D_{1i}$ are of size $dn\times d$, and it left multiplies the information $a_i$ which is a $d$ length vector. Matrices $P_{ij}$, $Q_{ij}$, $U_{ijk}$, $J_{ijk}$ and $E_{ij}$ for $1\leq i,k\leq (q-1),i\neq k$ and $1\leq j\leq n$ are of size $dn\times d$ and it left multiplies the $d$ length vector $b_{ij}$. For $1\leq i\leq n$, the matrices $D_{2i}$ and $D_{3i}$ are of size $dn\times d$ and it left multiplies the information $c_i$. The following matrices are of size $dn\times dn$: $D_5,D_6$, $M_i$, $K_i,R_i,V_i$ and $W_i$ for $1\leq i\leq (q-1)$. And, the following are the matrices of size $d\times dn$: $G_j, X_{ij}, L_j$ and $Z_{ij}$ for $1\leq i\leq (q-1)$ and $1\leq j\leq n$. Also let $I_d$ be a $d\times d$ identity matrix. Then, from the definition of network coding we have:
\begin{IEEEeqnarray}{l}
Y_{(u_1,u_3)} = \sum_{i=1}^n D_{1i}a_i + \sum_{i=1}^{q-1} \sum_{j=1}^n P_{ij}b_{ij}\\
Y_{(u_2,u_4)} = \sum_{i=1}^{q-1} \sum_{j=1}^n Q_{ij}b_{ij} + \sum_{i=1}^n D_{2i}c_{i}\\
Y_{(u_5,u_7)} = M_1Y_{(u_1,u_3)} + M_2Y_{(u_2,u_4)} = \sum_{i=1}^n M_1D_{1i}a_i + \sum_{i=1}^{q-1} \sum_{j=1}^n (M_1P_{ij} + M_2Q_{ij}) b_{ij} + \sum_{i=1}^n M_2D_{2i}c_{i}\IEEEeqnarraynumspace\\
Y_{(u_6,u_8)} = M_3Y_{(u_1,u_3)} + \sum_{i=1}^n D_{3i}c_{i} = \sum_{i=1}^n M_3D_{1i}a_i {+} \sum_{i=1}^{q-1} \sum_{j=1}^n M_3P_{ij}b_{ij} + \sum_{i=1}^n D_{3i}c_{i}\IEEEeqnarraynumspace\\
Y_{(u_9,u_{10})} = M_4Y_{(u_5,u_7)} + M_5Y_{(u_6,u_8)} = \sum_{i=1}^n (M_4M_1D_{1i} + M_5M_3D_{1i})a_i + \sum_{i=1}^{q-1} \sum_{j=1}^n \{M_4(M_1P_{ij} + M_2Q_{ij}) + M_5M_3P_{ij}\} b_{ij} \IEEEnonumber\\ \hfill +\> \sum_{i=1}^n (M_4M_2D_{2i} + M_5D_{3i})c_{i}\IEEEeqnarraynumspace\\
Y_{(u_{11},u_{12})} = \sum_{i=1}^n D_{4i}a_i + D_5Y_{(u_5,u_7)} = \sum_{i=1}^n (D_{4i} + D_5M_1D_{1i})a_i + \sum_{i=1}^{q-1} \sum_{j=1}^n D_5(M_1P_{ij} + M_2Q_{ij})b_{ij} + \sum_{i=1}^n D_5M_2D_{2i}c_{i}\IEEEeqnarraynumspace\label{t1}\\
\text{for } 1\leq i\leq (q-1):\IEEEnonumber\\
Y_{e_i} = W_iY_{(u_2,u_4)} + \sum_{j=1,j\neq i}^{q-1} \sum_{k=1}^n U_{jki}b_{jk} = \sum_{k=1}^n W_iQ_{ik}b_{ik} + \sum_{j=1,j\neq i}^{q-1} \sum_{k=1}^n (W_iQ_{jk} + U_{jki})b_{jk} + \sum_{k=1}^n W_iD_{2k}c_{k}\IEEEeqnarraynumspace\\
\text{for } 1\leq i\leq (q-1):
Y_{(v_i,v_i^\prime)} = K_iY_{(u_9,u_{10})} + \sum_{j=1,j\neq i}^{q-1} \sum_{k=1}^n J_{jki}b_{jk} = \sum_{k=1}^n K_i(M_4M_1D_{1k} + M_5M_3D_{1k})a_k \IEEEnonumber\\+\>  \sum_{j=1}^n K_i\{ M_4(M_1P_{ij} + M_2Q_{ij}) + M_5M_3P_{ij}\}b_{ij} + \sum_{k=1,k\neq i}^{q-1} \sum_{j=1}^n \{J_{kji} + K_i(M_4(M_1P_{kj} + M_2Q_{kj}) + M_5M_3P_{kj})\}b_{kj} \IEEEnonumber\\\hfill +\> \sum_{j=1}^n K_i(M_4M_2D_{2j} + M_5D_{3j})c_{j}\IEEEeqnarraynumspace\label{t2}\\
\text{for } 1\leq i\leq (q-1):\IEEEnonumber\\
Y_{(w_i,w_i^\prime)} = V_iY_{e_i} + \sum_{j=1}^n E_{ij}b_{ij} = \sum_{k=1}^n (V_iW_iQ_{ik} + E_{ik})b_{ik} + \sum_{j=1,j\neq i}^{q-1}\sum_{k=1}^n \{V_i(W_iQ_{jk} + U_{jki})\}b_{jk} + \sum_{k=1}^n V_iW_iD_{2k}c_{k}\IEEEeqnarraynumspace\label{t4}\\
Y_{(u_{13},u_{14})} = D_6Y_{(u_6,u_8)} + \sum_{i=1}^{q-1} R_iY_{e_i} = \sum_{i=1}^n D_6M_3D_{1i}a_i + \sum_{i=1}^{q-1} \sum_{j=1}^n \{D_6M_3P_{ij} + R_iW_iQ_{ij} + \sum_{k=1,k\neq i}^{q-1} R_k(W_kQ_{ij}+U_{ijk})\}b_{ij}\IEEEnonumber\\\hfill +\> \sum_{i=1}^n\{D_6D_{3i} + (\sum_{k=1}^{q-1} R_kW_k)D_{2i}\}c_i \IEEEeqnarraynumspace\label{t3}
\end{IEEEeqnarray}

It can be seen that some of the above considered local coding matrices are rectangular matrices. As a rectangular matrix do not have a unique inverse (which would be all-important as we progress), we use the following lemma, which shows how in some cases many different rectangular matrices can be combined to form a square matrix having a unique inverse. 

\noindent Let $A = {\begin{bmatrix} A_1 & A_2 & \cdots & A_n  \end{bmatrix}}^T$ and $B = \begin{bmatrix} B_1 & B_2 &\cdots & B_n \end{bmatrix}$ where $A_i$ and $B_i$ for $1\leq i\leq n$ are matrices of size $d\times dn$ and $dn\times d$ respectively (So $A$ and $B$ are both of size $dn \times dn$).
\begin{lemma}\label{lem1}
For $1\leq i,j\leq n, i\neq j$, if $A_iB_i = I_d$ and $A_iB_j = 0$, then $AB = I_{dn}$.
\end{lemma}
\begin{IEEEproof}
\begin{IEEEeqnarray*}{rl}
AB &= \begin{bmatrix} A_1 \\ A_2 \\ \vdots \\ A_n  \end{bmatrix} \begin{bmatrix} B_1 & B_2 & \cdots & B_n  \end{bmatrix} =  \begin{bmatrix}
A_1B_1 & A_1B_2 & \cdots & A_1B_n\\
A_2B_1 & A_2B_2 & \cdots & A_2B_n\\
\vdots &\vdots &\vdots &\vdots\\
A_nB_1 & A_nB_2 & \cdots & A_nB_n
\end{bmatrix}
= \begin{bmatrix}
I_d & 0 & \cdots & 0\\
0 & I_d & \cdots & 0\\
\vdots &\vdots &\vdots &\vdots\\
0 & 0 & \cdots & I_d
\end{bmatrix} = I_{dn}
\end{IEEEeqnarray*}
\end{IEEEproof}
\begin{corollary}\label{coro1}
For $1\leq i,j\leq n$, if $A_iB_j = 0$, then $AB = 0$.
\end{corollary}

As the components of $b_{ij}$ is also zero at all $t_k\in T_c$, for $1\leq i\leq (q-1)$ and $1\leq j,k\leq n$, from equation (\ref{t1}) and Fig.~\ref{genfano1/n} we have:
\begin{equation}
G_k\{D_5(M_1P_{ij} + M_2Q_{ij})\} = 0\label{etcba}
\end{equation}
\begin{IEEEeqnarray}{ll}
\text{Let, }\; & G = {\begin{bmatrix} G_1 & G_2 & \cdots & G_n\end{bmatrix}}^T,\label{g}\\
& P_i = {\begin{bmatrix} P_{i1} & P_{i2} & \cdots & P_{in}\end{bmatrix}},\label{p}\\
\text{and }& Q_i = {\begin{bmatrix} Q_{i1} & Q_{i2} & \cdots & Q_{in}\end{bmatrix}}. \label{q}\\
\text{Hence, } & D_5(M_1P_{i} + M_2Q_{i}) = \begin{bmatrix}D_5(M_1P_{i1} + M_2Q_{i1}) & D_5(M_1P_{i2} + M_2Q_{i2}) & \cdots & D_5(M_1P_{in} + M_2Q_{in})  \end{bmatrix}\label{d5}
\end{IEEEeqnarray}
Then, applying corollary~\ref{coro1} on equations (\ref{etcba}), (\ref{g}) and (\ref{d5}) we have: 
\begin{equation}
\text{for } 1\leq i\leq (q-1):\; G\{D_5(M_1P_{i} + M_2Q_{i})\} = 0 \label{etcb2}
\end{equation}
Now, since the terminal $t_i\in T_c$ retrieves $c_i$ for $1\leq i,j\leq n$, $j\neq i$ from equation (\ref{t1}) and Fig.~\ref{genfano1/n} we have:
\begin{IEEEeqnarray}{l}
G_i(D_5M_2D_{2i}) = I_d\label{etc3a}\\
G_i(D_5M_2D_{2j}) = 0\label{etc3b}
\end{IEEEeqnarray}
\begin{IEEEeqnarray}{ll}
\text{Let, }& D_2 = \begin{bmatrix} D_{21} & D_{22} & \cdots & D_{2n} \end{bmatrix}.\\
\text{Then, }& D_5M_2D_2 = \begin{bmatrix} D_5M_2D_{21} & D_5M_2D_{22} & \cdots & D_5M_2D_{2n} \end{bmatrix}\IEEEeqnarraynumspace\label{q1}
\end{IEEEeqnarray}
Then, applying lemma~\ref{lem1} on equations (\ref{etc3a}) and (\ref{etc3b})  we have:
\begin{equation}
G(D_5M_2D_2) = I_{dn} \label{etc3c}
\end{equation}
Now as equation~(\ref{etc3c}) implies both $G$ and $D_5$ are invertible, from equation (\ref{etcb2}) we have:
\begin{equation}
\text{for } 1\leq i\leq (q-1):\; M_1P_{i} + M_2Q_{i} = 0 \label{etcb3}
\end{equation}
Now consider the $n$ terminals in the set $T_{b_i}$ for $1\leq i\leq (q-1)$. Since the component of $a_k$ for $1\leq k\leq n$ at $t_j\in T_{b_i}$ for $1\leq j\leq n$ is zero, for $1\leq i\leq (q-1)$ and $1\leq j,k\leq n$, using equation (\ref{t2}) and Fig.~\ref{genfano1/n} we have:
\begin{equation}
X_{ij}K_i(M_4M_1D_{1k} + M_5M_3D_{1k}) = 0 \label{etba1}
\end{equation}
\begin{IEEEeqnarray}{ll}
\text{Let, }\; &  X_i = \begin{bmatrix}X_{i1} & X_{i2} & \cdots & X_{in} \end{bmatrix}^T\\
\text{and, }\; &  D_1 = \begin{bmatrix}D_{11} & D_{12} & \cdots & D_{1n} \end{bmatrix}.\label{d1}\\
\text{Then, }& X_iK_i = \begin{bmatrix}X_{i1}K_i & X_{i2}K_i & \cdots & X_{in}K_i \end{bmatrix}^T\label{r}\\
\text{and }& M_4M_1D_{1} + M_5M_3D_{1} = \begin{bmatrix}M_4M_1D_{11} + M_5M_3D_{11} & M_4M_1D_{12} + M_5M_3D_{12} & \cdots & M_4M_1D_{1n} + M_5M_3D_{1n} \end{bmatrix}\IEEEeqnarraynumspace\label{x4}
\end{IEEEeqnarray}
Now, applying corollary~\ref{coro1} on equations (\ref{etba1}), (\ref{r}), and (\ref{x4}) we have:
\begin{equation}
\text{for } 1\leq i\leq (q-1):\; X_iK_i(M_4M_1D_{1} + M_5M_3D_{1}) = 0 \label{etba2}
\end{equation}
Since the terminal $t_j\in T_{b_i}$ computes the information $b_{ij}$, we have for $1\leq i\leq (q-1), 1\leq j,m\leq n$ and $m\neq j$:
\begin{IEEEeqnarray}{l}
X_{ij}K_i\{ M_4(M_1P_{ij} + M_2Q_{ij}) + M_5M_3P_{ij}\} = I_d\label{etbb1}\\
X_{ij}K_i\{ M_4(M_1P_{im} + M_2Q_{im}) + M_5M_3P_{im}\} = 0\label{etbb2}
\end{IEEEeqnarray}
\begin{IEEEeqnarray}{l}
\text{From equations (\ref{p}) and (\ref{q}) we have: }\;M_4(M_1P_{i} + M_2Q_{i}) + M_5M_3P_{i} \IEEEnonumber\\{=}\! \begin{bmatrix} M_4(M_1P_{i1} + M_2Q_{i1}) + M_5M_3P_{i1} & M_4(M_1P_{i2} + M_2Q_{i2}) + M_5M_3P_{i2} & \!{\cdots} & M_4(M_1P_{in} + M_2Q_{in}) + M_5M_3P_{in} \end{bmatrix}\IEEEeqnarraynumspace\label{m4m2m5}
\end{IEEEeqnarray}
Using lemma~\ref{lem1} and equations (\ref{etbb1}), (\ref{etbb2}), (\ref{r}) and (\ref{m4m2m5})  we have:
\begin{equation}
\text{for } 1\leq i\leq (q-1):\; X_{i}K_i\{ M_4(M_1P_{i} + M_2Q_{i}) + M_5M_3P_{i}\} = I_{dn} \label{etbb3}
\end{equation}
Substituting equation (\ref{etcb3}) in equation (\ref{etbb3}) we have:
\begin{equation}
X_{i}K_iM_5M_3P_{i} = I_{dn} \label{etbb4}
\end{equation}
Since from equation (\ref{etbb3}) both $X_i$ and $K_i$ are invertible, we have from equation (\ref{etba2}):
\begin{equation}
M_4M_1D_{1} + M_5M_3D_{1} = 0\label{neweq1}
\end{equation}
Since the component of $c_k$ for $1\leq k\leq n$ is zero at $t_j\in T_{b_i}$, using equation (\ref{t2}) we have for $1\leq i\leq (q-1)$ and $1\leq j,k\leq n$:
\begin{equation}
X_{ij}K_i(M_4M_2D_{2k} + M_5D_{3k}) = 0 \label{etbc1}
\end{equation}
\begin{IEEEeqnarray}{l}
\text{Let, } D_2 = \begin{bmatrix}D_{21} & D_{22} & \cdots & D_{2n} \end{bmatrix},\label{d2}\\
\text{and } D_3 = \begin{bmatrix}D_{31} & D_{32} & \cdots & D_{3n} \end{bmatrix}.\label{d3}\\
\text{Then, } M_4M_2D_{2} + M_5D_{3} = \begin{bmatrix} M_4M_2D_{21} + M_5D_{31} & M_4M_2D_{22} + M_5D_{32} & \cdots & M_4M_2D_{2n} + M_5D_{3n} \end{bmatrix}\IEEEeqnarraynumspace\label{m4m5}
\end{IEEEeqnarray}
Using Corollary~\ref{coro1} on equations (\ref{etbc1}), (\ref{r}) and (\ref{m4m5}) we have:
\begin{equation}
\text{for } 1\leq i\leq (q-1):\; X_{i}K_i(M_4M_2D_{2} + M_5D_{3}) = 0 \label{etbc2}
\end{equation}
Since both $X_i$ and $K_i$ are invertible (equation (\ref{etbb3})), from equation (\ref{etbc2}) we have:
\begin{equation}
M_4M_2D_{2} + M_5D_{3} = 0 \label{etbc3}
\end{equation}

\noindent Let us consider the terminals in the set $T_{a}$. Since $t_i\in T_a$ computes the message $a_i$, for $1\leq i,j\leq n$ and $j\neq i$, using equation (\ref{t3}) and Fig.~\ref{genfano1/n} we have:
\begin{IEEEeqnarray}{l}
L_iD_6M_3D_{1i} = I_d \label{etaa1}\\
L_iD_6M_3D_{1j} = 0 \label{etaa2}
\end{IEEEeqnarray}
\begin{IEEEeqnarray}{ll}
\text{Let, } L = \begin{bmatrix}L_1 & L_2 & \cdots & L_n \end{bmatrix}^T\label{l}.\\
\text{From equation (\ref{d1}) we have: } D_6M_3D_1 = \begin{bmatrix}D_6M_3D_{11} & D_6M_3D_{12} & \cdots & D_6M_3D_{1n} \end{bmatrix}\label{q2} \IEEEeqnarraynumspace
\end{IEEEeqnarray}
Then applying lemma~\ref{lem1} on equations (\ref{etaa1}), (\ref{etaa2}), (\ref{l}) and (\ref{q2}) we have:
\begin{equation}
LD_6M_3D_1 = I_{dn} \label{etaa3}
\end{equation}
Since from equation (\ref{etaa3}) $D_1$ is invertible, from equation (\ref{neweq1}):
\begin{equation}
M_4M_1 + M_5M_3 = 0 \label{etba3}
\end{equation}
Substituting (\ref{etba3}) in (\ref{etbb3}) we have:
\begin{equation}
\text{for } 1\leq i\leq (q-1):\; X_{i}K_iM_4M_2Q_{i} = I_{dn} \label{newmethod1}
\end{equation}
At any $t_{l}\in T_a$ for $1\leq i\leq (q-1)$ and $1\leq l,j\leq n$ the component of $b_{ij}$ is zero. So we have:
\begin{equation}
L_l\{D_6M_3P_{ij} + R_iW_iQ_{ij} + \sum_{k=1,k\neq i}^{q-1} R_k(W_kQ_{ij} + U_{ijk})\} = 0\label{etab1}
\end{equation}
\begin{IEEEeqnarray}{l}
\text{Let } U_{ik} = \begin{bmatrix} U_{i1k} & U_{i2k} & \cdots & U_{ink} \end{bmatrix}.\label{uik}\\
\text{Then from equations (\ref{p}), (\ref{q}), (\ref{uik}) we have: }
 D_6M_3P_{i} + R_iW_iQ_{i} + \sum_{k=1,k\neq i}^{q-1} R_k(W_kQ_{i} + U_{ik})\IEEEnonumber\\
= \begin{bmatrix} D_6M_3P_{i1} + R_iW_iQ_{i1} + \sum_{k=1,k{\neq} i}^{q-1} R_k(W_kQ_{i1} + U_{i1k}) & \cdots & D_6M_3P_{in} + R_iW_iQ_{in} + \sum_{k=1,k{\neq} i}^{q-1} R_k(W_kQ_{in} + U_{ink}) \end{bmatrix}\IEEEnonumber
\end{IEEEeqnarray}
So using Corollary~\ref{coro1} on equation (\ref{etab1}) for $1\leq i\leq (q-1)$ we get: 
\begin{equation}
\text{for } 1\leq i\leq (q-1):\; L(D_6M_3P_{i} + R_iW_iQ_{i} + \sum_{k=1,k\neq i}^{q-1} R_k(W_kQ_{i} + U_{ik})) = 0\label{howi1}
\end{equation}
Since from equation (\ref{etaa3}) $L$ is invertible, we have:
\begin{equation}
\text{for } 1\leq i\leq (q-1):\;D_6M_3P_{i} + R_iW_iQ_{i} + \sum_{k=1,k\neq i}^{q-1} R_k(W_kQ_{i} + U_{ik}) = 0 \label{etab2}
\end{equation}
At a terminal $t_j\in T_a$, since the component of $c_i$ is zero, for $1\leq i,j\leq n$ we have:
\begin{equation}
L_j\{ D_6D_{3i} + \sum_{k=1}^{q-1} R_kW_kD_{2i}\} = 0 \label{etac1}
\end{equation}
\begin{IEEEeqnarray}{l}
\text{Using equations (\ref{d2}) and (\ref{d3}): }\;  D_6D_{3} + \sum_{k=1}^{q-1} R_kW_kD_{2} \IEEEnonumber\\\hfill = \begin{bmatrix} D_6D_{31} + \sum_{k=1}^{q-1}R_kW_kD_{21} & D_6D_{32} + \sum_{k=1}^{q-1}R_kW_kD_{22} & \cdots & D_6D_{3n} + \sum_{k=1}^{q-1}R_kW_kD_{2n}\end{bmatrix}\IEEEeqnarraynumspace\label{q2d2}
\end{IEEEeqnarray}
Using corollary~\ref{coro1}, and equations (\ref{etac1}), (\ref{l}) and (\ref{q2d2}) we have:
\begin{equation}
L(D_6D_{3} + \sum_{k=1}^{q-1} R_kW_kD_{2}) = 0\label{howi2}
\end{equation}
Since $L$ is invertible from equation (\ref{etaa3}) we have:
\begin{equation}
D_6D_{3} + \sum_{k=1}^{q-1} R_kW_kD_{2} = 0 \label{etac2}
\end{equation}
Now consider the terminals in the set $T_{c_i}$ for $1\leq i\leq q-1$. Since at $t_{l}\in T_{c_i}$, for $1\leq k\leq (q-1),k\neq i$ the component of $b_{kj}$ for $1\leq l,j\leq n$ is zero, we have:
\begin{equation}
Z_{il}\{V_i(W_iQ_{kj} + U_{kji})\} = 0 \label{etci1}
\end{equation}
\begin{IEEEeqnarray}{l}
\text{Let, }\; Z_{i} = \begin{bmatrix} Z_{i1} & Z_{i2} & \cdots & Z_{in} \end{bmatrix}^T\\
\text{Then, }\; Z_{i}V_i = \begin{bmatrix} Z_{i1}V_i & Z_{i2}V_i & \cdots & Z_{in}V_i \end{bmatrix}^T\label{z}\\
\text{From equations (\ref{q}) and (\ref{uik}): }\; W_iQ_{k} + U_{ki} = \begin{bmatrix} V_i(W_iB_{k1} + U_{k1i}) & V_i(W_iB_{k2} + U_{k2i}) & \cdots & V_i(W_iB_{kn} + U_{kni}) \end{bmatrix}\IEEEeqnarraynumspace\label{u}
\end{IEEEeqnarray}
Using corollary~\ref{coro1} on equations (\ref{etci1}), (\ref{z}) and (\ref{u}) we get:
\begin{equation}
\text{for } 1\leq i,k\leq (q-1), k\neq i:\; Z_{i}V_i(W_iQ_{k} + U_{ki}) = 0 \label{etcix}
\end{equation}
Since $t_l\in T_{c_i}$ computes $c_l$, for $1\leq l,m\leq n, l\neq m$, we have:
\begin{IEEEeqnarray}{l}
Z_{il}V_iW_iD_{2l} = I_d \label{etcc1}\\
Z_{il}V_iW_iD_{2m} = 0 \label{etcc2}
\end{IEEEeqnarray}
\begin{IEEEeqnarray}{l}
\text{From equation (\ref{d2}): }\;W_iD_{2} = \begin{bmatrix} W_iD_{21} & W_iD_{22} & \cdots & W_iD_{2n} \end{bmatrix}\label{wd2}
\end{IEEEeqnarray}
Using Lemma~\ref{lem1} on equations (\ref{etcc1}), (\ref{etcc2}), (\ref{z}) and (\ref{wd2}) we have: 
\begin{equation}
\text{for } 1\leq i\leq (q-1): \; Z_{i}V_iW_iD_{2} = I_{dn} \label{etcc3}
\end{equation}
Since from equation (\ref{etcc3}) $Z_iV_i$ is invertible, from equation (\ref{etcix}) we have:
\begin{equation}
\text{for } 1\leq i,k\leq (q-1), k\neq i:\; W_iQ_{k} + U_{ki} = 0 \label{etci3}
\end{equation}
Substituting equation (\ref{etci3}) in equation (\ref{etab2}) we have:
\begin{equation}
\text{for } 1\leq i\leq (q-1):\; D_6M_3P_{i} + R_iW_iQ_{i} = 0 \label{etab3}
\end{equation}
Hence, for $1\leq i\leq (q-1)$:
\begin{IEEEeqnarray*}{ll}
D_6M_3P_{i}Q_{i}^{-1} + R_iW_i = 0 & \text{ [$Q_i$ is invertible from (\ref{newmethod1})]}\\
D_6M_3P_{i}X_{i}K_iM_4M_2 + R_iW_i = 0&\text{ [from equation (\ref{newmethod1})]}\\
D_6M_3P_{i}X_{i}K_iM_4M_2D_2 + R_iW_iD_2 = 0 \qquad\qquad\qquad& \text{ [multiplying both sides by $D_2$]}\\
-D_6M_3P_{i}X_{i}K_iM_5D_{3} + R_iW_iD_2 = 0 & \text{ [from equation (\ref{etbc3})]}\\
-D_6M_3P_{i}P_{i}^{-1}M_3^{-1}D_{3} + R_iW_iD_2 = 0 & \text{ [from equation (\ref{etbb4})]}\\
-D_6D_{3} + R_iW_iD_2 = 0 \IEEEyesnumber\label{newmethod2}&
\end{IEEEeqnarray*}
Substituting equation (\ref{newmethod2}) in equation (\ref{etac2}) we have:
\begin{equation}
qD_6D_{3} = 0
\end{equation}
From equation (\ref{etaa3}) $D_6$ is invertible. As $M_4M_2$ is invertible from equation~(\ref{newmethod1}), and as $D_2$ is invertible from equation~(\ref{etc3c}); $M_5D_3$ is invertible from equation~(\ref{etbc3}). This implies $D_3$ is also an invertible matrix. So for $qD_6D_{3} = 0$ to hold $q$ must be equal to zero. Now, in a finite field, an element is equal to zero if and only if the characteristic divides the element. This proves that the network in Fig.~\ref{genfano1/n} has a rate $\frac{1}{n}$ fractional linear network coding solution only if the characteristic of the finite field divides $q$. Next, we show that the network $\mathcal{N}_1^\prime$ has a $(1,n)$ fractional linear network coding solution if $q=0$.

For this section, let $\bar{a}_i$ denote an $n$-length column vector whose $i^{\text{th}}$ component is $a_i$ and all other components are zero (since $k=1$, $a_i$ is an unit-length vector). Let $\bar{c}_i$ to denote an $n$-length column vector whose $i^{\text{th}}$ component is $c_i$ and all other components are zero. Also let $\bar{b}_{ij}$ denote an $n$-length column cvector whose $j^{\text{th}}$ component is $b_{ij}$ and all other components are zero. 
Now, by choosing the appropriate local coding matrices, the messages shown below can be transmitted by the corresponding edges.
\begin{IEEEeqnarray*}{l}
Y_{(u_1,u_3)} = \sum_{i=1}^{n} \bar{a}_i + \sum_{i=1}^{p-1} \sum_{j=1}^{n} \bar{b}_{ij}\\
Y_{(u_2,u_4)} = \sum_{i=1}^{q-1} \sum_{j=1}^{n} \bar{b}_{ij} + \sum_{i=1}^{n} \bar{c}_i\\
Y_{(u_5,u_7)} = Y_{(u_1,u_3)} - Y_{(u_2,u_4)} = \sum_{i=1}^{n} \bar{a}_i - \sum_{i=1}^{n} \bar{c}_i\\
Y_{(u_6,u_8)} = Y_{(u_1,u_3)} - \sum_{i=1}^{n} \bar{c}_i = \sum_{i=1}^{n} \bar{a}_i + \sum_{i=1}^{q-1} \sum_{j=1}^{n} \bar{b}_{ij} - \sum_{i=1}^{n} \bar{c}_i\\
\text{for } 1\leq i\leq q-1: \quad Y_{e_i} = \sum_{j=1}^{n} \bar{b}_{ij} + \sum_{i=1}^{n} \bar{c}_i\\
Y_{(u_9,u_{10})} = Y_{(u_6,u_8)} - Y_{(u_5,u_7)} = \sum_{i=1}^{q-1} \sum_{j=1}^{n} \bar{b}_{ij}\\
Y_{(u_{13},u_{14})} = Y_{(u_6,u_8)} - \sum_{i=1}^{q-1} Y_{e_i} = \sum_{i=1}^{n} \bar{a}_i - \sum_{i=1}^{n} \bar{c}_i - \sum_{i=1}^{q-1} \sum_{i=1}^{n} \bar{c}_i = \sum_{i=1}^{n} \bar{a}_i - \sum_{i=1}^{q} \sum_{i=1}^{n} \bar{c}_i = \sum_{i=1}^{n} \bar{a}_i -  \sum_{i=1}^{n} q\bar{c}_i = \sum_{i=1}^{n} \bar{a}_i\\
Y_{(u_{11},u_{12})} = \sum_{i=1}^{n} \bar{a}_i - Y_{(u_5,u_7)} = \sum_{i=1}^{n} \bar{c}_i\\
\text{for } 1\leq i\leq q-1:\quad Y_{(v_i,v_i^\prime)} = Y_{(u_9,u_{10})} - \sum_{k=1,k\neq i}^{q-1} \sum_{j=1}^{n} \bar{b}_{ij} = \sum_{j=1}^{n} \bar{b}_{ij}\\
\text{for } 1\leq i\leq q-1: \quad Y_{(w_i,w_i^\prime)} = Y_{e_i} - \sum_{j=1}^{n} \bar{b}_{ij} = \sum_{i=1}^{n} \bar{c}_i
\end{IEEEeqnarray*}
Let $\check{u}(i)$ be a unit row vector of length $n$ which has $i^{\text{th}}$ component equal to one and all other components are zero. Then from the vector $\sum_{i=1}^{n} \bar{a}_i$, $a_i$ for any $1\leq i\leq n$ can be determined by the dot product $\check{u}(i) \cdot (\sum_{i=1}^{n} \bar{a}_i)$. Similarly for any $1\leq i\leq (q-1)$, $b_{ij} = \check{u}(j) \cdot (\sum_{j=1}^{n} \bar{b}_{ij})$. For $1\leq i\leq n$, $c_i$ can be determined similarly from $\sum_{i=1}^{n} \bar{c}_i$. \hfill $\blacksquare$

\subsection{Proof of theorem~\ref{Thm1}:}\label{ineq1}
To produce the desired characteristic-dependent linear rank inequality, we apply DFZ method to the  network shown in Fig.~\ref{genfano1/n} for $n=1$ and $q = p_1\times p_2\times \ldots \times p_l$.

\noindent Let the message carried by an edge $(u_i,u_j)$ be denoted by $Y_{i,j}$. Also let the massage carried by the edge $e_{i}$ for $1\leq i\leq q-1$ be denoted by $Y_{e_i}$. Corresponding to each of the source messages and the massages carried by the edges, consider the vector subspaces $A,$ $B_1,\ldots,B_{q-1},$ $C$, $Y_{1,3}$, $Y_{2,4}$, $Y_{5,7}, $ $Y_{6,8}$, $Y_{9,10}$, $Y_{e_1}, \ldots, Y_{e_{q-2}}$ and $Y_{e_{q-1}}$ of a finite dimensional vector space $V$.

\noindent Corresponding to the matrices in Fig.~\ref{genfano1/n} consider the following linear functions:
\begin{IEEEeqnarray*}{llllll}
f_{D_1}: Y_{1,3} \rightarrow A \qquad & f_{D_2}: Y_{2,4} \rightarrow C \qquad & f_{D_3}: Y_{6,8} \rightarrow C \qquad & f_{D_4}: C \rightarrow A \qquad & f_{D_5}: C \rightarrow Y_{5,7} \qquad &f_{D_6}: A \rightarrow Y_{6,8}\\
f_{M_1}: Y_{5,7} \rightarrow Y_{1,3} & f_{M_2}: Y_{5,7} \rightarrow Y_{2,4} & f_{M_3}: Y_{6,8} \rightarrow Y_{1,3} & f_{M_4}: Y_{9,10} \rightarrow Y_{5,7} \qquad & f_{M_5}: Y_{9,10} \rightarrow Y_{6,8}\\
\text{for } 1\leq i\leq q-1:\;& f_{P_i}: Y_{1,3} \rightarrow B_i \qquad& f_{Q_i}: Y_{2,4} \rightarrow B_i \qquad& f_{K_i}: B_i \rightarrow Y_{9,10}\\
\text{for } 1\leq i\leq q-1:\;& f_{W_i}: Y_{e_i} \rightarrow Y_{2,4} \qquad& f_{R_i}: A \rightarrow Y_{e_i} \qquad& f_{V_i}: C \rightarrow Y_{e_i} \qquad& f_{E_i}: C \rightarrow B_i\\
\IEEEeqnarraymulticol{5}{l}{\text{for } 1\leq i,j\leq q-1, j\neq i:\quad   f_{U_{ji}}: Y_{e_i} \rightarrow B_j\qquad    f_{J_{ji}}: B_i \rightarrow B_j}
\end{IEEEeqnarray*}
The idea behind the DFZ method is as follows. First note that the linear functions shown above is in accordance with the topology of the network. Now, we have seen in the last subsection that over a finite field where $q \neq 0$ the network $\mathcal{N}_1^\prime$ does not have a rate $1$ linear solution (note $n=1$ in Fig.~\ref{genfano1/n} for this current proof). This means that if the dimension of all the above considered vector subspaces are equal, then such a functional assignment won't exists when $q \neq 0$ over the finite field (because if it had existed then the realization of these vector subspaces would have formed a rate $1$ linear solution). 
The DFZ method starts with these linear functions and tries to find an equation (relating the dimension of the corresponding vector subspaces) that must hold true for such a functional assignment to exist over a finite filed where $q\neq 0$. This equation is the desired inequality.

Now to obtain this equation, the DFZ method requires to find a subspace (say $S$) that becomes a zero subspaces when $q \neq 0$. This subspace must also be expressible as an intersection of other subspaces. Then, applying lemma~\ref{Lem1} (shown below) on $S$ results the desired inequality. At present, all the steps of the DFZ method sans finding the set $S$ is algorithmic. Intuitively, when $S$ becomes the zero subspace (which happens when $q \neq 0$ in our case) the dimension of the union of the subspaces whose intersection is equal to $S$ increases; thereby meaning that more information has to be sent (more is reflected in the increment of the dimension) when $q \neq 0$. This `more' information results the rate to be less than $1$.

For this proof, to find $S$, we use the proof of lemma~\ref{lem3} shown in the above subsection. %
%
Let us define some notations and introduce some lemmas which will be required for the rest of the proof.

If $A$ is a subspace of $V$ then co-dimension of $A$ in $V$ is $codim_V(A) = dim(V) - dim(A)$. The following lemmas are reproduced from \cite{rateregion}. The proofs of these lemmas are omitted from here and can be found in \cite{rateregion}. In all of these lemmas, $V$ is a finite dimensional vector space, and $A,B,A_1,A_2,\ldots,A_m$~are~subspaces~of $V$. Let $f: A \rightarrow B$ be a linear function. If $B^\prime$ is a subspace of $B$, then $f^{-1}(B^\prime)$ denotes a vector subspace $A^\prime$ of $A$ such that $f(A^\prime) = B^\prime$.
\begin{lemma}{\cite[Lemma 2, p. 2501]{rateregion}}: \label{Lem1}
\begin{equation*}
codim_V(\cap_{i=1}^m A_i) \leq \sum_{i=1}^m codim_V(A_i)
\end{equation*}
\end{lemma}
\begin{lemma}{\cite[Lemma 3, p. 2501]{rateregion}}:\label{Lem2}
If $B^\prime$ is a subspace of $B$, then
\begin{equation*}
codim_A(f^{-1}(B^\prime)) \leq codim_B(B^\prime)
\end{equation*}
\end{lemma}
\begin{lemma}{\cite[Lemma 4, p. 2501]{rateregion}}:\label{Lem3}
There exist linear functions $f_i: A \rightarrow A_i$ for $1\leq i\leq m$ such that $f_1 + \cdots + f_m  = I$ on a subspace $A^\prime$ of $A$ with
\begin{equation*}
codim_A(A^\prime) \leq H(A|A_1,A_2,\ldots,A_m)
\end{equation*}
\end{lemma}
\begin{lemma}{\cite[Lemma 6, p. 2502]{rateregion}}:\label{Lem4}
For $1\leq i\leq m$, let $f_i: A \rightarrow A_i$ be linear functions such that $f_1 + f_2 + \cdots + f_m  = 0$ on $A$. Then $f_1 = \cdots = f_m = 0$ on a subspace $A^\prime$ of $A$ with
\begin{equation*}
codim_A(A^\prime) \leq H(A_1) + \cdots + H(A_m) - H(A_1,\ldots,A_m)
\end{equation*}
\end{lemma}

According to Lemma~\ref{Lem3} the following holds:
\begin{IEEEeqnarray*}{l}
%
f_{D_1} + \sum_{i=1}^{q-1} f_{P_i} = I \text{ over a subspace } Y_{1,3}^\prime \text{ of } Y_{1,3} \text{ where }
codim_{Y_{1,3}}(Y_{1,3}^\prime) \leq H(Y_{1,3}|A,B_1,\ldots,B_{q-1})\IEEEeqnarraynumspace\IEEEyesnumber\label{y13}\\
\sum_{i=1}^{q-1} f_{Q_i} + f_{D_2} = I \text{ over a subspace } Y_{2,4}^\prime \text{ of } Y_{2,4} \text{ where } codim_{Y_{2,4}}(Y_{2,4}^\prime) \leq H(Y_{2,4}|B_1,\ldots,B_{q-1},C)\IEEEeqnarraynumspace\IEEEyesnumber\label{y24}\\
f_{M_1} + f_{M_2} = I \text{ over a subspace } Y_{5,7}^\prime \text{ of } Y_{5,7} \text{ where } codim_{Y_{5,7}}(Y_{5,7}^\prime) \leq H(Y_{5,7}|Y_{1,3},Y_{2,4})\IEEEeqnarraynumspace\IEEEyesnumber\label{y57}\\
f_{M_3} + f_{D_3} = I \text{ over a subspace } Y_{6,8}^\prime \text{ of } Y_{6,8} \text{ where } codim_{Y_{6,8}}(Y_{6,8}^\prime) \leq H(Y_{6,8}|Y_{1,3},C)\IEEEeqnarraynumspace\IEEEyesnumber\label{y68}\\
\text{ for } 1\leq i,j\leq q-1, j\neq i:\quad f_{W_i} + \sum_{j=1,j\neq i}^{q-1} f_{U_{ji}} = I \text{ over a subspace } Y_{e_i}^\prime \text{ of } Y_{e_i} \text{ where }\\\hfill codim_{Y_{e_i}}(Y_{e_i}^\prime) \leq H(Y_{e_i}|Y_{2,4},B_1,\ldots, B_{i-1},B_{i+1},\ldots, B_{q-1})\IEEEyesnumber\IEEEeqnarraynumspace\label{yei}\\
f_{M_4} + f_{M_5} = I \text{ over a subspace } Y_{9,10}^\prime \text{ of } Y_{9,10} \text{ where } codim_{Y_{9,10}}(Y_{9,10}^\prime) \leq H(Y_{9,10}|Y_{5,7},Y_{6,8})\IEEEeqnarraynumspace\IEEEyesnumber\label{y910}\\
f_{D_4} + f_{D_5} = I \text{ over a subspace } C^\prime \text{ of } C \text{ where } codim_{C}(C^\prime) \leq H(C|A,Y_{5,7})\IEEEeqnarraynumspace\IEEEyesnumber\label{c}\\
\text{ for } 1\leq i,j\leq q-1, j\neq i:\quad f_{K_i} + \sum_{j=1,j\neq i}^{q-1} f_{J_{ji}} = I
\text{ over a subspace } B_i^\prime \text{ of } B_i \text{ where }\\\hfill codim_{B_i}(B_i^\prime) \leq H(B_i|Y_{9,10},B_1,\ldots,B_{i-1},B_{i+1},\ldots, B_{q-1})\IEEEeqnarraynumspace\IEEEyesnumber\label{bi}\\
f_{D_6} + \sum_{i=1}^{q-1} f_{R_i} = I \text{ over a subspace } A^\prime \text{ of } A \text{ where } codim_{A}(A^\prime) \leq H(A|Y_{6,8},Y_{e_1},\ldots ,Y_{e_{q-1}})\IEEEeqnarraynumspace\IEEEyesnumber\label{a}\\
f_{V_i} + f_{E_i} = I \text{ over a subspace } C^\prime_i \text{ of } C \text{ where } codim_{C}(C_i^\prime) \leq H(C|Y_{e_i},B_i)\IEEEeqnarraynumspace\IEEEyesnumber\label{ci}
\end{IEEEeqnarray*}
\begin{IEEEeqnarray*}{l}
\text{Now, let's consider the following composite functions:}\\
f_{D_4} + f_{D_1}f_{M_1}f_{D_5}: C \rightarrow A\\
\text{for } 1\leq i\leq (q-1): (f_{P_i}f_{M_1} + f_{Q_i}f_{M_2})f_{D_5}: C \rightarrow B_i\\
f_{D_2}f_{M_2}f_{D_5}: C \rightarrow C
\end{IEEEeqnarray*}
\begin{IEEEeqnarray*}{l}
\text{Using (\ref{y13}), (\ref{y24}), and (\ref{y57}) we have:}\\
f_{D_1}f_{M_1}f_{D_5} + \sum_{i=1}^{q-1} f_{P_i}f_{M_1}f_{D_5} = f_{M_1}f_{D_5} \text{ over a subspace } f_{D_5}^{-1}f_{M_1}^{-1}(Y_{1,3}^\prime) \text{ of } C\\
f_{D_2}f_{M_2}f_{D_5} + \sum_{i=1}^{q-1} f_{Q_i}f_{M_2}f_{D_5} = f_{M_2}f_{D_5}  \text{ over a subspace } f_{D_5}^{-1}f_{M_2}^{-1}(Y_{2,4}^\prime) \text{ of } C\\
f_{M_1}f_{D_5} + f_{M_2}f_{D_5} = f_{D_5} \text{ over a subspace } f_{D_5}^{-1}(Y_{5,7}^\prime) \text{ of } C\\
\end{IEEEeqnarray*}
Also note, from equation (\ref{c}), $f_{D_4} + f_{D_5} = I$ over $C^\prime$. Then,
\begin{IEEEeqnarray*}{l}
f_{D_1}f_{M_1}f_{D_5} + \sum_{i=1}^{q-1} f_{P_i}f_{M_1}f_{D_5} + f_{D_2}f_{M_2}f_{D_5} + \sum_{i=1}^{q-1} f_{Q_i}f_{M_2}f_{D_5} + f_{D_4} = I \text{ over a subspace }\\ C^{\prime\prime} = f_{D_5}^{-1}f_{M_1}^{-1}(Y_{1,3}^\prime) \cap f_{D_5}^{-1}f_{M_2}^{-1}(Y_{2,4}^\prime) \cap f_{D_5}^{-1}(Y_{5,7}^\prime) \cap C^\prime \text{ of } C\\
\end{IEEEeqnarray*}
Then using lemma~\ref{Lem1}:
\begin{IEEEeqnarray*}{l}
codim_C(C^{\prime\prime}) \leq codim_C(f_{D_5}^{-1}f_{M_1}^{-1}(Y_{1,3}^\prime)) + codim_C(f_{D_5}^{-1}f_{M_2}^{-1}(Y_{2,4}^\prime)) + codim_C(f_{D_5}^{-1}(Y_{5,7}^\prime)) + codim_C(C^\prime)\\
\text{or, }codim_C(C^{\prime\prime}) \leq codim_{Y_{1,3}}(Y_{1,3}^\prime) + codim_{Y_{2,4}}(Y_{2,4}^\prime) + codim_{Y_{5,7}}(Y_{5,7}^\prime) + codim_{C}(C^\prime)\qquad\text{[using lemma~\ref{Lem2}]}\IEEEeqnarraynumspace\IEEEyesnumber\label{codimc}
\end{IEEEeqnarray*}
Now, according to Lemma~\ref{Lem4} there exists a subspace $\bar{C}$ of $C^{\prime\prime}$ over which:
\begin{IEEEeqnarray}{l}
f_{D_4} + f_{D_1}f_{M_1}f_{D_5} = 0\\
\text{for } 1\leq i\leq (q-1): (f_{P_i}f_{M_1} + f_{Q_i}f_{M_2})f_{D_5} = 0\IEEEeqnarraynumspace\label{cob}\\
f_{D_2}f_{M_2}f_{D_5} - I = 0\label{coc}
\end{IEEEeqnarray}
such that
\begin{IEEEeqnarray*}{l}
codim_{C^{\prime\prime}}(\bar{C}) \leq H(A) + \sum_{i=1}^{q-1} H(B_{i}) + H(C) - H(A,B_1,B_2,\ldots ,B_{q-1},C)\IEEEeqnarraynumspace\IEEEyesnumber\label{codimbarc}\\
\text{Now, } codim_C(\bar{C}) = codim_C{C^{\prime\prime}} + codim_{C^{\prime\prime}}\bar{C}. \text{ So, from equations (\ref{codimc}) and (\ref{codimbarc}) we get: }\\
codim_C(\bar{C}) \leq codim_{Y_{1,3}}(Y_{1,3}^\prime) + codim_{Y_{2,4}}(Y_{2,4}^\prime) + codim_{Y_{5,7}}(Y_{5,7}^\prime) + codim_{C}(C^\prime) + H(A) \\\hfill +\> \sum_{i=1}^{q-1} H(B_{i}) + H(C) - H(A,B_1,B_2,\ldots ,B_{q-1},C)\IEEEeqnarraynumspace\IEEEyesnumber\label{codimbarc2}
\end{IEEEeqnarray*}
Notice the similarity between equations (\ref{cob}) and (\ref{etcb2}); and between equations (\ref{coc}) and (\ref{etc3c}). 
We now want to find a subspace $\bar{B}_i$ of $B_i$, for ${1\leq i\leq q-1}$, over which the following identities hold:
\begin{IEEEeqnarray}{l}
f_{D_1}(f_{M_1}f_{M_4} + f_{M_3}f_{M_5})f_{K_i} = 0 \label{bia}\\
\{(f_{P_i}f_{M_1} + f_{Q_i}f_{M_2})f_{M_4} + f_{P_i}f_{M_3}f_{M_5}\}f_{K_i} = I\label{bibi}\\
\text{for } 1\leq j\leq q-1, j\neq i:\;
\hfill (f_{P_j}f_{M_1} + f_{Q_j}f_{M_2})f_{M_4}f_{K_i} + f_{P_j}f_{M_3}f_{M_5}f_{K_i} + f_{J_{ji}} = 0\label{biji}\\
(f_{D_2}f_{M_2}f_{M_4} + f_{D_3}f_{M_5})f_{K_i} = 0\label{bic}
\end{IEEEeqnarray}
Here also notice the similarity between equations (\ref{bia}) and (\ref{etba2}); between equations (\ref{bibi}) and (\ref{etbb3}); and between equations (\ref{bic}) and (\ref{etbc2}).
From equations (\ref{y13}), (\ref{y24}), (\ref{y57}) and (\ref{y68}) we have:
\begin{IEEEeqnarray*}{l}
\text{for } 1\leq i\leq (q-1):\;
f_{D_1}f_{M_1}f_{M_4}f_{K_i} + \sum_{i=1}^{q-1} f_{P_i}f_{M_1}f_{M_4}f_{K_i} = f_{M_1}f_{M_4}f_{K_i} \text{ over a subspace } f_{K_i}^{-1}f_{M_4}^{-1}f_{M_1}^{-1}(Y_{1,3}^\prime) \text{ of } B_i\\
f_{D_1}f_{M_3}f_{M_5}f_{K_i} + \sum_{i=1}^{q-1} f_{P_i}f_{M_3}f_{M_5}f_{K_i} = f_{M_3}f_{M_5}f_{K_i} \text{ over a subspace } f_{K_i}^{-1}f_{M_5}^{-1}f_{M_3}^{-1}(Y_{1,3}^\prime) \text{ of } B_i\\
f_{D_2}f_{M_2}f_{M_4}f_{K_i} + \sum_{i=1}^{q-1} f_{Q_i}f_{M_2}f_{M_4}f_{K_i} = f_{M_2}f_{M_4}f_{K_i} \text{ over a subspace } f_{K_i}^{-1}f_{M_4}^{-1}f_{M_2}^{-1}(Y_{2,4}^\prime) \text{ of } B_i\\
f_{M_1}f_{M_4}f_{K_i} + f_{M_2}f_{M_4}f_{K_i} = f_{M_4}f_{K_i} \text{ over a subspace } f_{K_i}^{-1}f_{M_4}^{-1}(Y_{5,7}^\prime) \text{ of } B_i\\
f_{M_3}f_{M_5}f_{K_i} + f_{D_3}f_{M_5}f_{K_i} = f_{M_5}f_{K_i} \text{ over a subspace } f_{K_i}^{-1}f_{M_5}^{-1}(Y_{6,8}^\prime) \text{ of } B_i\\
f_{M_4}f_{K_i} + f_{M_5}f_{K_i} = f_{K_i} \text{ over a subspace } f_{K_i}^{-1}(Y_{9,10}^\prime) \text{ of } B_i
%
\end{IEEEeqnarray*}
Also note that from eqn. (\ref{bi}) we have: $f_{K_i} + \sum_{j=1,j\neq i}^{q-1} f_{J_{ji}} = I$ over $B_i^\prime$.
\begin{IEEEeqnarray*}{l}
\text{Hence, } f_{D_1}f_{M_1}f_{M_4}f_{K_i} + \sum_{i=1}^{q-1} f_{P_i}f_{M_1}f_{M_4}f_{K_i} + f_{D_1}f_{M_3}f_{M_5}f_{K_i} + \sum_{i=1}^{q-1} f_{P_i}f_{M_3}f_{M_5}f_{K_i} +  f_{D_2}f_{M_2}f_{M_4}f_{K_i} \\+\> \sum_{i=1}^{q-1} f_{Q_i}f_{M_2}f_{M_4}f_{K_i} + f_{D_3}f_{M_5}f_{K_i} + \sum_{j=1,j\neq i}^{q-1} f_{J_{ji}} = I \text{ on subspace } \\B_i^{\prime\prime} {=} f_{K_i}^{-1}f_{M_4}^{-1}f_{M_1}^{-1}(Y_{1,3}^\prime) \cap f_{K_i}^{-1}f_{M_5}^{-1}f_{M_3}^{-1}(Y_{1,3}^\prime) \cap f_{K_i}^{-1}f_{M_4}^{-1}f_{M_2}^{-1}(Y_{2,4}^\prime) \cap f_{K_i}^{-1}f_{M_4}^{-1}(Y_{5,7}^\prime) \cap f_{K_i}^{-1}f_{M_5}^{-1}(Y_{6,8}^\prime)\cap f_{K_i}^{-1}(Y_{9,10}^\prime) \cap B_i^\prime \text{ of } B_i\\
\text{So, applying Lemma~\ref{Lem1}:}\\
codim_{B_i}(B_i^{\prime\prime}) \leq codim_{B_i}(f_{K_i}^{-1}f_{M_4}^{-1}f_{M_1}^{-1}(Y_{1,3}^\prime) + 
codim_{B_i}(f_{K_i}^{-1}f_{M_5}^{-1}f_{M_3}^{-1}(Y_{1,3}^\prime)) + codim_{B_i}(f_{K_i}^{-1}f_{M_4}^{-1}f_{M_2}^{-1}(Y_{2,4}^\prime) \\\hfill +\> codim_{B_i}(f_{K_i}^{-1}f_{M_4}^{-1}(Y_{5,7}^\prime)) + codim_{B_i}(f_{K_i}^{-1}f_{M_5}^{-1}(Y_{6,8}^\prime)) + codim_{B_i}(f_{K_i}^{-1}(Y_{9,10}^\prime)) + codim_{B_i}(B_i^\prime)\\
\text{Using Lemma~\ref{Lem2} we get: }\\
codim_{B_i}(B_i^{\prime\prime}) \leq 2codim_{Y_{1,3}}(Y_{1,3}^\prime) + codim_{Y_{2,4}}(Y_{2,4}^\prime) + codim_{Y_{5,7}}(Y_{5,7}^\prime) + codim_{Y_{6,8}}(Y_{6,8}^\prime) \\\hfill +\> codim_{Y_{9,10}}(Y_{9,10}^\prime) + codim_{B}(B_i^\prime) \IEEEeqnarraynumspace\IEEEyesnumber\label{biprime}
\end{IEEEeqnarray*}

\noindent So from Lemma~\ref{Lem4} over a subspace $\bar{B_i}$ of $B_i^{\prime\prime}$ equations (\ref{bia}), (\ref{bibi}), (\ref{biji}) and (\ref{bic}) holds where
\begin{IEEEeqnarray*}{l}
codim_{B_i^{\prime\prime}}(\bar{B_i}) \leq H(A) + \sum_{i=1}^{q-1} H(B_{i}) + H(C) - H(A,B_1,B_2,\ldots ,B_{q-1},C)\IEEEeqnarraynumspace\IEEEyesnumber\label{bibar}\\
\text{Now, } codim_{B_i}(\bar{B_i}) = codim_{B_i}(B_i^{\prime\prime}) + codim_{B_i^{\prime\prime}}(\bar{B}_i)\\
\text{from equations (\ref{biprime}) and (\ref{bibar}) we have}:\\
codim_{B_i}(\bar{B_i}) \leq 2codim_{Y_{1,3}}(Y_{1,3}^\prime) + codim_{Y_{2,4}}(Y_{2,4}^\prime) + codim_{Y_{5,7}}(Y_{5,7}^\prime) + codim_{Y_{6,8}}(Y_{6,8}^\prime) + codim_{Y_{9,10}}(Y_{9,10}^\prime) \\\hfill
+\> codim_{B}(B_i^\prime) + H(A) + \sum_{i=1}^{q-1} H(B_{i}) + H(C) - H(A,B_1,B_2,\ldots ,B_{q-1},C)\IEEEeqnarraynumspace\IEEEyesnumber\label{barbi}
\end{IEEEeqnarray*}

Next, we want to find an upper-bound on the co-dimension of a subspace $\bar{A}$ of $A^{\prime\prime}$ over which the following relations hold:
\begin{IEEEeqnarray}{l}
f_{D_1}f_{M_3}f_{D_6} = I\label{aa}\\
\text{for } 1\leq i\leq q-1:\quad f_{P_i}f_{M_3}f_{D_6} + f_{Q_i}f_{W_i}f_{R_i} + \sum_{j=1,j\neq i}^{q-1} (f_{Q_i}f_{W_j} + f_{U_{ij}})f_{R_j} = 0\label{abi}\\
f_{D_3}f_{D_6} + \sum_{i=1}^{q-1} f_{D_2}f_{W_i}f_{R_i} = 0\label{ac}
\end{IEEEeqnarray}
Here also notice the similarity between equations (\ref{aa}) and (\ref{etaa3}); between equations (\ref{abi}) and (\ref{howi1}); and between equations (\ref{ac}) and (\ref{howi2}).
Using equations (\ref{y13}), (\ref{y24}), (\ref{y68}) and (\ref{yei}) we have:
\begin{IEEEeqnarray*}{l}
f_{D_1}f_{M_3}f_{D_6} + \sum_{i=1}^{q-1} f_{P_i}f_{M_3}f_{D_6} = f_{M_3}f_{D_6} \text{ over a subspace } f_{D_6}^{-1}f_{M_3}^{-1}(Y_{1,3}^\prime) \text{ of } A\\
\sum_{i=1}^{q-1} f_{D_2}f_{W_i}f_{R_i} + \sum_{i=1}^{q-1} (f_{Q_i}( \sum_{j=1}^{q-1} f_{W_j}f_{R_j}) ) 
= f_{D_2} \sum_{i=1}^{q-1} f_{W_i}f_{R_i} + (\sum_{i=1}^{q-1} f_{Q_i})( \sum_{j=1}^{q-1} f_{W_j}f_{R_j}) 
= \sum_{j=1}^{q-1} f_{W_j}f_{R_j} \text{ over a subspace}\\\hfill  (\sum_{j=1}^{q-1} f_{W_j}f_{R_j})^{-1}(Y_{2,4}^{\prime}) \text{ of } A\\
%
\text{Now, } 
\sum_{i=1}^{q-1} (\sum_{j=1,j\neq i}^{q-1} f_{U_{ij}}f_{R_j}) = \sum_{j=1}^{q-1} (\sum_{i=1,i\neq j}^{q-1} f_{U_{ij}}f_{R_j})
= \sum_{i=1}^{q-1} (\sum_{j=1,j\neq i}^{q-1} f_{U_{ji}}f_{R_i})\\
\text{So, } \sum_{i=1}^{q-1} (f_{W_i}f_{R_i}) + \sum_{i=1}^{q-1} (\sum_{j=1,j\neq i}^{q-1} f_{U_{ij}}f_{R_j})
= \sum_{i=1}^{q-1} (f_{W_i}f_{R_i}) + \sum_{i=1}^{q-1} (\sum_{j=1,j\neq i}^{q-1} f_{U_{ji}}f_{R_i})
= \sum_{i=1}^{q-1} (f_{W_i}f_{R_i} + \sum_{j=1,j\neq i}^{q-1} f_{U_{ji}}f_{R_i})\\\hfill
=\> \sum_{i=1}^{q-1} (f_{W_i} + \sum_{j=1,j\neq i}^{q-1} f_{U_{ji}})f_{R_i}
= \sum_{i=1}^{q-1} f_{R_i} \text{ over a subspace } 
f_{R_1}^{-1}(Y_{e_1}^\prime) \cap f_{R_2}^{-1}(Y_{e_2}^\prime) \cap \cdots \cap  f_{R_{q-1}}^{-1}(Y_{e_{q-1}}^\prime) \text{ of } A
\end{IEEEeqnarray*}
\begin{IEEEeqnarray*}{l}
%
%
%
\text{Also, } f_{M_3}f_{D_6} + f_{D_3}f_{D_6} = f_{D_6} \text{ over a subspace } f_{D_6}^{-1}(Y_{6,8}^\prime) \text{ of }A.\\
\text{Note that from equation (\ref{a}) we have: }
f_{D_6} + \sum_{i=1}^{q-1} f_{R_i} = I \text{ over a subspace } A^\prime \text{ of } A
\end{IEEEeqnarray*}
\begin{IEEEeqnarray*}{l}
\text{So, } f_{D_1}f_{M_3}f_{D_6} + \sum_{i=1}^{q-1} f_{P_i}f_{M_3}f_{D_6} + \sum_{i=1}^{q-1} f_{D_2}f_{W_i}f_{R_i} + \sum_{i=1}^{q-1} (f_{Q_i}( \sum_{j=1}^{q-1} f_{W_j}f_{R_j}) ) + \sum_{i=1}^{q-1} (\sum_{j=1,j\neq i}^{q-1} f_{U_{ij}}f_{R_j}) + f_{D_3}f_{D_6} = I\\\hfill \text{over a subspace } A^{\prime\prime} \text{ of } A.\\
\text{where, } A^{\prime\prime} = f_{D_6}^{-1}f_{M_3}^{-1}(Y_{1,3}^\prime) \cap (\sum_{i=1}^{q-1} f_{W_i}f_{R_i})^{-1}(Y_{2,4}^{\prime})  \cap f_{R_1}^{-1}(Y_{e_1}^\prime) \cap f_{R_2}^{-1}(Y_{e_2}^\prime)  \cap \cdots \cap  f_{R_{q-1}}^{-1}(Y_{e_{q-1}}^\prime) \cap f_{D_6}^{-1}(Y_{6,8}^\prime) \cap A^\prime\\
\text{So, } codim_{A}A^{\prime\prime} \leq codim_{A}(f_{D_6}^{-1}f_{M_3}^{-1}(Y_{1,3}^\prime)  +  codim_{A}((\sum_{i=1}^{q-1} f_{W_i}f_{R_i})^{-1}(Y_{2,4}^{\prime}))  + codim_{A}(f_{R_1}^{-1}(Y_{e_1}^\prime)) + codim_{A}(f_{R_2}^{-1}(Y_{e_2}^\prime)) \\\hfill +\> \cdots +  codim_{A}(f_{R_{q-1}}^{-1}(Y_{e_{q-1}}^\prime)) + codim_{A}(f_{D_6}^{-1}(Y_{6,8}^\prime)) + codim_{A}(A^\prime)\\
\text{Using Lemma~\ref{Lem1} and Lemma~\ref{Lem2} we have:}\\
codim_{A}(A^{\prime\prime}) \leq codim_{Y_{1,3}}(Y_{1,3}^\prime) + codim_{Y_{2,4}}(Y_{2,4}^\prime) + codim_{Y_{e_1}}(Y_{e_1}^\prime) + codim_{Y_{e_2}}(Y_{e_2}^\prime) + \cdots + codim_{Y_{e_{q-1}}}(Y_{e_{q-1}}^\prime) \\\hfill +\> codim_{Y_{6,8}}(Y_{6,8}^\prime) + codim_{A}(A^\prime)\IEEEeqnarraynumspace\IEEEyesnumber\label{aprime}
\end{IEEEeqnarray*}
Then from Lemma~\ref{Lem4}, over a subspace $\bar{A}$ equations (\ref{aa}), (\ref{abi}) and (\ref{ac}) hold, such that
\begin{IEEEeqnarray*}{l}
codim_{A^{\prime\prime}}(\bar{A}) \leq H(A) + \sum_{i=1}^{q-1} H(B_{i}) + H(C) - H(A,B_1,B_2,\ldots ,B_{q-1},C)\IEEEeqnarraynumspace\IEEEyesnumber\label{abar}\\
\text{Now since }
codim_{A}(\bar{A}) = codim_{A}(A^{\prime\prime}) + codim_{A^{\prime\prime}}(\bar{A}), \text{ from equations (\ref{aprime}) and (\ref{abar}) we have:}\\
codim_{A}(\bar{A}) \leq codim_{Y_{1,3}}(Y_{1,3}^\prime) + codim_{Y_{2,4}}(Y_{2,4}^\prime) + codim_{Y_{e_1}}(Y_{e_1}^\prime)  + \cdots + codim_{Y_{e_{q-1}}}(Y_{e_{q-1}}^\prime) \\\hfill +\> codim_{Y_{6,8}}(Y_{6,8}^\prime) + codim_{A}(A^\prime) + H(A) + \sum_{i=1}^{q-1} H(B_{i}) + H(C) - H(A,B_1,B_2,\ldots ,B_{q-1},C)\IEEEeqnarraynumspace\IEEEyesnumber\label{xa}
\end{IEEEeqnarray*}

For $1\leq i\leq q-1$ we now find an upper-bound on the co-dimension of a subspace $\bar{C_i}$ of $C$ over which the following identities hold:
\begin{IEEEeqnarray}{l}
f_{Q_i}f_{W_i}f_{V_i} + f_{E_i} = 0\label{cbi}\\
\text{for } 1\leq j\leq (q-1), j\neq i:\; (f_{Q_j}f_{W_i} + f_{U_{ji}})f_{V_i} = 0\IEEEeqnarraynumspace\label{cbj}\\
f_{D_2}f_{W_i}f_{V_i} - I = 0\label{cc}
\end{IEEEeqnarray}
Here also notice the similarity between equations (\ref{cbj}) and (\ref{etcix}); and between equations (\ref{cc}) and (\ref{etcc3}).

Using equations (\ref{y24}) and (\ref{yei}) we have:
\begin{IEEEeqnarray*}{l}
f_{D_2}f_{W_i}f_{V_i} + \sum_{j=1}^{q-1} f_{Q_j}f_{W_i}f_{V_i} = f_{W_i}f_{V_i} \text{ over a subspace } f_{V_i}^{-1}f_{W_i}^{-1}(Y_{2,4}^\prime) \text{ of } C\\
f_{W_i}f_{V_i} + \sum_{j=1,j\neq i}^{q-1} f_{U_{ji}}f_{V_i} = f_{V_i} \text{ over a subspace }   f_{V_i}^{-1}(Y_{e_i}^\prime) \text{ of } C\\
\text{And from equation (\ref{ci}) we know: }
f_{V_i} + f_{E_i} = I \text{ over a subspace }  C^\prime_i \text{ of } C
\end{IEEEeqnarray*}
So over a subspace $C^{\prime\prime}_i$ of $C$ we have 
\begin{IEEEeqnarray*}{l}
f_{D_2}f_{W_i}f_{V_i} + \sum_{j=1}^{q-1} f_{Q_j}f_{W_i}f_{V_i} + \sum_{j=1,j\neq i}^{q-1} f_{U_{ji}}f_{V_i} + f_{E_i} = I
\text{ where } C^{\prime\prime}_i = f_{V_i}^{-1}f_{W_i}^{-1}(Y_{2,4}^\prime) \cap f_{V_i}^{-1}(Y_{e_i}^\prime) \cap C^\prime_i\\
\text{So, applying Lemma~\ref{Lem1} and Lemma~\ref{Lem2} we have: }\\
codim_{C}(C^{\prime\prime}_i) \leq codim_{Y_{2,4}}(Y_{2,4}^\prime) + codim_{Y_{e_i}}(Y_{e_i}^\prime) + codim_{C}(C_i^\prime) \IEEEeqnarraynumspace\IEEEyesnumber\label{ciprime}
\end{IEEEeqnarray*}
Now according to Lemma~\ref{Lem4} over a subspace $\bar{C_i}$ equations (\ref{cbi}), (\ref{cbj}), and (\ref{cc}) holds, where 
\begin{IEEEeqnarray*}{l}
codim_{C^{\prime\prime}_i}(\bar{C_i}) \leq \sum_{j=1}^{q-1} H(B_{j}) + H(C) - H(B_1,B_2,\ldots ,B_{q-1},C)\\
\text{Now, }
codim_{C}(\bar{C_i}) = codim_{C}(C^{\prime\prime}_i) + codim_{C^{\prime\prime}_i}(\bar{C_i}). 
\text{ So Using equations (\ref{ciprime}) we have:}\\
codim_{C}(\bar{C_i}) \leq  codim_{Y_{2,4}}(Y_{2,4}^\prime) + codim_{Y_{e_i}}(Y_{e_i}^\prime) + codim_{C}(C^\prime_i) + \sum_{j=1}^{q-1} H(B_{j}) + H(C) - H(B_1,\ldots ,B_{q-1},C)\IEEEeqnarraynumspace\IEEEyesnumber\label{xci}
\end{IEEEeqnarray*}
%
%
We now form some equations analogous to the equations that were pivotal for the proof lemma~\ref{lem3}. Consider the following vector subspaces.
\begin{IEEEeqnarray}{l}
\text{for } 1\leq i\leq q-1:\; S_{B_i} = \{ u \in B_i | f_{M_4}f_{K_i}(u) \in f_{D_5}(\bar{C})\}\label{m8}
\end{IEEEeqnarray}
Hence, equation (\ref{cob}) holds over $S_{B_i}$ when $f_{D_5}$ is replaced by $f_{M_4}f_{K_i}$. So over $S_{B_i}$ we have:
\begin{equation}
\text{for } 1\leq i\leq (q-1): (f_{P_i}f_{M_1} + f_{Q_i}f_{M_2})f_{M_4}f_{K_i} = 0\IEEEeqnarraynumspace\label{cobm1}
\end{equation}
Since equation (\ref{bibi}) holds over $\bar{B_i}$, from equations (\ref{bibi}) and (\ref{cobm1}), over a subspace $\bar{B_i} \cap S_{B_i}$ we have:
\begin{equation}
f_{P_i}f_{M_3}f_{M_5}f_{K_i} = I \label{m2}
\end{equation}
Notice the similarity between equation (\ref{etbb4}) and equation (\ref{m2}).

Now consider the following subspaces.
\begin{IEEEeqnarray*}{l}
\text{for } 1\leq i\leq q-1:\; R_{B_i} = \{ u \in B_i | f_{M_5}f_{K_i}(u) \in f_{D_6}(\bar{A}) \}\\
\text{for } 1\leq i\leq q-1:\; L_{B_i} = \{ u \in B_i | f_{M_1}f_{M_4}f_{K_i}(u) \in f_{M_3}f_{D_6}(\bar{A}) \}
\end{IEEEeqnarray*}
So $f_{M_5}f_{K_i}(R_{B_i})$ is a subspace of $f_{D_6}(\bar{A})$. Then, since from equation (\ref{aa}) $f_{D_1}$ is invertible over $f_{M_3}f_{D_6}(\bar{A})$; $f_{D_1}$ is also invertible over $f_{M_3}f_{M_5}f_{K_i}(R_{B_i})$. Similarly, $f_{M_1}f_{M_4}f_{K_i}(L_{B_i})$ is a subspace of $f_{M_3}f_{D_6}(\bar{A})$. Hence $f_{D_1}$ is also invertible over $f_{M_1}f_{M_4}f_{K_i}(L_{B_i})$. Hence over a subspace $R_{B_i} \cap L_{B_i}$ from equation (\ref{bia}) we have:
\begin{equation}
(f_{M_1}f_{M_4} + f_{M_3}f_{M_5})f_{K_i} = 0 \label{mo2}
\end{equation}
Applying this equation in equation (\ref{bibi}), over a subspace $\bar{B_i} \cap R_{B_i} \cap L_{B_i}$ we have:
\begin{equation}
f_{Q_i}f_{M_2}f_{M_4}f_{K_i} = I \label{m3}
\end{equation}
Notice the similarity between equation (\ref{newmethod1}) and equation (\ref{m3}).
Now consider the following subspace:
\begin{IEEEeqnarray*}{l}
\text{for } 1\leq i\leq q-1:\; S_{A_i} = \{ u \in A | f_{R_i}(u) \in f_{V_i}(\bar{C_i}) \}
\end{IEEEeqnarray*}
Hence for $1\leq i,j\leq (q-1), j\neq i$, $(f_{Q_j}f_{W_i} + f_{U_{ji}})f_{R_i}(S_{A_i})$ is a subspace of $(f_{Q_j}f_{W_i} + f_{U_{ji}})f_{V_i}(\bar{C_i})$. Hence from equation (\ref{cbj}), over $S_{A_i}$ we have:
\begin{equation}
\text{for } 1\leq j\leq (q-1), j\neq i:\; (f_{Q_j}f_{W_i} + f_{U_{ji}})f_{R_i} = 0\label{m4}
\end{equation}
Applying equation (\ref{m4}) on equation (\ref{abi}), over a subspace $\cap_{i=1}^{q-1} S_{A_i} \cap \bar{A}$ we have:
\begin{equation}
\text{for } 1\leq j\leq q-1:\; f_{P_j}f_{M_3}f_{D_6} + f_{Q_j}f_{W_j}f_{R_j} = 0\label{m5}
\end{equation}
Note the similarity between equations (\ref{etab3}) and (\ref{m5}).
Let us now consider the following subspaces:
\begin{IEEEeqnarray}{l}
\text{for } 1\leq i\leq q-1:\; L_{A_i} = \{ u \in A | f_{D_6}(u) \in f_{M_5}f_{K_i}(\bar{B_i} \cap S_{B_i})\}\label{j2}\\
\text{for } 1\leq i\leq q-1:\; R_{A_i} = \{ u \in A | f_{W_i}f_{R_i}(u) \in f_{M_2}f_{M_4}f_{K_i}(\bar{B_i} \cap R_{B_i} \cap L_{B_i}) \}\label{j1}\\
S = \bar{A} \cap (\cap_{i=1}^{q-1} L_{A_i}) \cap (\cap_{i=1}^{q-1} R_{A_i}) \cap (\cap_{i=1}^{q-1} S_{A_i})
\end{IEEEeqnarray}
For any $a \in S$, from equation (\ref{ac}) we have:
\begin{IEEEeqnarray*}{l}
f_{D_3}f_{D_6}(a) + \sum_{i=1}^{q-1} f_{D_2}f_{W_i}f_{R_i}(a) = 0\\
\text{From (\ref{j1}) we know there exists a $b_i \in (\bar{B_i} \cap R_{B_i} \cap L_{B_i})$ such that $f_{W_i}f_{R_i}(a) = f_{M_2}f_{M_4}f_{K_i}(b_i)$. So,}\\
f_{D_3}f_{D_6}(a) + \sum_{i=1}^{q-1} f_{D_2}f_{M_2}f_{M_4}f_{K_i}(b_i) = 0\\
\text{From equation (\ref{m3}) we know that $b_i = f_{Q_i}f_{M_2}f_{M_4}f_{K_i}(b_i)$ for any $b_i \in (\bar{B_i} \cap R_{B_i} \cap L_{B_i})$. So,}\\
f_{D_3}f_{D_6}(a) + \sum_{i=1}^{q-1} f_{D_2}f_{M_2}f_{M_4}f_{K_i}f_{Q_i}f_{M_2}f_{M_4}f_{K_i}(b_i) = 0\\
\text{or, }f_{D_3}f_{D_6}(a) + \sum_{i=1}^{q-1} f_{D_2}f_{M_2}f_{M_4}f_{K_i}f_{Q_i}f_{W_i}f_{R_i}(a) = 0\\
\text{Using equation (\ref{m5}) we have:}\\
f_{D_3}f_{D_6}(a) - \sum_{i=1}^{q-1} f_{D_2}f_{M_2}f_{M_4}f_{K_i}f_{P_i}f_{M_3}f_{D_6}(a) = 0\\
\text{From (\ref{j2}) we know there exists a $b_i^\prime \in (\bar{B_i} \cap S_{B_i})$ such that $f_{D_6}(a) = f_{M_5}f_{K_i}(b_i^\prime)$. So,}\\
f_{D_3}f_{D_6}(a) - \sum_{i=1}^{q-1} f_{D_2}f_{M_2}f_{M_4}f_{K_i}f_{P_i}f_{M_3}f_{M_5}f_{K_i}(b_i^\prime) = 0\\
\text{From equation (\ref{m2}) we know that $b_i^\prime = f_{P_i}f_{M_3}f_{M_5}f_{K_i}(b_i^\prime)$ for any $b_i^\prime \in (\bar{B_i} \cap S_{B_i})$. So,}\\
f_{D_3}f_{D_6}(a) - \sum_{i=1}^{q-1} f_{D_2}f_{M_2}f_{M_4}f_{K_i}(b_i^\prime) = 0\\
\text{Since $b_i^\prime \in \bar{B_i}$, using equation (\ref{bic}) we have:}\\
f_{D_3}f_{D_6}(a) + \sum_{i=1}^{q-1} f_{D_3}f_{M_5}f_{K_i}(b_i^\prime) = 0\\
\text{or, }f_{D_3}f_{D_6}(a) + \sum_{i=1}^{q-1} f_{D_3}f_{D_6}(a) = 0\\
qf_{D_3}f_{D_6}(a) = 0\IEEEyesnumber\label{m7}
\end{IEEEeqnarray*}
We now argue that for equation (\ref{m7}) to hold for any $a \in S$, $S$ must be a zero subspace. From equation (\ref{aa}) we know that $f_{D_6}$ is one-to-one over $\bar{A}$. From equation (\ref{m8}) we know that $f_{M_4}f_{K_i}(S_{B_i})$ for $1\leq i\leq (q-1)$ is a subspace of $f_{D_5}(\bar{C})$. Because of equation (\ref{coc}), $f_{D_2}f_{M_2}$ is one-to-one over $f_{D_5}(\bar{C})$. So $f_{D_2}f_{M_2}$ is also one-to-one over $f_{M_4}f_{K_i}(S_{B_i})$. Then, from equation (\ref{bic}) it can be concluded that $f_{D_3}f_{M_5}f_{K_i}$ is one-to-one over $S_{B_i}$. Now, from (\ref{j2}) we know $f_{D_6}(S)$ is a subspace of $f_{M_5}f_{K_i}(\bar{B_i} \cap S_{B_i})$ for any $1\leq i\leq q-1$. So $f_{D_3}$ is one-to-one over $f_{D_6}(S)$. Moreover, as a pre-condition, since the characteristic of the finite field does not belong to $\{p_1,p_2,\ldots,p_l\}$, $q \neq 0$ over the finite field. Hence for equation (\ref{m7}) to hold, $S$ must be a zero subspace. Now,

\begin{IEEEeqnarray*}{l}
dim(A) = dim(A) -dim(S) = codim_A(S) = codim_A(\bar{A} \cap (\cap_{i=1}^{q-1} L_{A_i}) \cap (\cap_{i=1}^{q-1} R_{A_i}) \cap (\cap_{i=1}^{q-1} S_{A_i}))\\
\text{Applying lemma~\ref{Lem1} we have:}\\
dim(A) \leq codim_A(\bar{A}) + \sum_{i=1}^{q-1} codim_A(L_{A_i}) + \sum_{i=1}^{q-1} codim_A(R_{A_i}) + \sum_{i=1}^{q-1} codim_A(S_{A_i})\IEEEyesnumber\label{dima}
\end{IEEEeqnarray*}
We now calculate some values that would help us in computing a bound over $dim(A)$.
\begin{IEEEeqnarray*}{l}
codim_{B_i}(S_{B_i}) = codim_{B_i}(f_{K_i}^{-1}f_{M_4}^{-1}(f_{D_5}(\bar{C})))\\
\text{Applying lemma~\ref{Lem2}; and noting that from equation (\ref{coc}) $f_{D_5}$ is one-to-one over $\bar{C}$ we have:}\\
codim_{B_i}(S_{B_i}) \leq codim_{Y_{5,7}}(f_{D_5}(\bar{C})) = dim(Y_{5,7}) - dim(f_{D_5}(\bar{C})) = dim(Y_{5,7}) - dim(\bar{C})\\
\text{or, } codim_{B_i}(S_{B_i}) \leq H(Y_{5,7}) + codim_{C}(\bar{C}) - H(C)\IEEEyesnumber\label{sbi}
\end{IEEEeqnarray*}

\begin{IEEEeqnarray*}{l}
codim_{B_i}(R_{B_i}) = codim_{B_i}(f_{K_i}^{-1}f_{M_5}^{-1}(f_{D_6}(\bar{A})))\\
\text{Applying lemma~\ref{Lem2}; and noting that from equation (\ref{aa}) $f_{D_6}$ is one-to-one over $\bar{A}$ we have:}\\
codim_{B_i}(R_{B_i}) \leq codim_{Y_{6,8}}(f_{D_6}(\bar{A})) = dim(Y_{6,8}) - dim(f_{D_6}(\bar{A})) = dim(Y_{6,8}) - dim(\bar{A})\\
\text{or, } codim_{B_i}(R_{B_i}) \leq H(Y_{6,8}) + codim_{A}(\bar{A}) - H(A)\IEEEyesnumber\label{rbi}
\end{IEEEeqnarray*}

\begin{IEEEeqnarray*}{l}
codim_{B_i}(L_{B_i}) = codim_{B_i}(f_{K_i}^{-1}f_{M_4}^{-1}f_{M_1}^{-1}(f_{M_3}f_{D_6}(\bar{A})))\\
\text{Applying lemma~\ref{Lem2}; and noting that from equation (\ref{aa}) $f_{M_3}f_{D_6}$ is one-to-one over $\bar{A}$ we have:}\\
codim_{B_i}(L_{B_i}) \leq codim_{Y_{1,3}}(f_{M_3}f_{D_6}(\bar{A})) = dim(Y_{1,3}) - dim(f_{M_3}f_{D_6}(\bar{A})) = dim(Y_{1,3}) - dim(\bar{A})\\
\text{or, } codim_{B_i}(L_{B_i}) \leq H(Y_{1,3}) + codim_{A}(\bar{A}) - H(A)\IEEEyesnumber\label{lbi}
\end{IEEEeqnarray*}

\begin{IEEEeqnarray*}{l}
codim_{A}(S_{A_i}) = codim_A(f_{R_i}^{-1}(f_{V_i}(\bar{C_i})))\\
\text{Applying lemma~\ref{Lem2}; and noting from equation (\ref{cc}) that $f_{V_i}$ is one-to-one over $\bar{C}$ we have:}\\
codim_{A}(S_{A_i}) \leq codim_{Y_{e_i}}(f_{V_i}(\bar{C_i})) = H(Y_{e_i}) - dim(f_{V_i}(\bar{C_i})) =  H(Y_{e_i}) - dim(\bar{C_i})\\
\text{or, } codim_{A}(S_{A_i}) \leq H(Y_{e_i}) + codim_{C}(\bar{C_i}) - H(C)\IEEEyesnumber\label{sai}
\end{IEEEeqnarray*}

\begin{IEEEeqnarray*}{l}
codim_{A}(R_{A_i}) = codim_{A}(f_{R_i}^{-1}f_{W_i}^{-1}(f_{M_2}f_{M_4}f_{K_i}(\bar{B_i} \cap R_{B_i} \cap L_{B_i})))\\
\text{Applying lemma~\ref{Lem2} we have:}\\
codim_{A}(R_{A_i}) \leq codim_{Y_{2,4}}(f_{M_2}f_{M_4}f_{K_i}(\bar{B_i} \cap R_{B_i} \cap L_{B_i})) = dim(Y_{2,4}) - dim(f_{M_2}f_{M_4}f_{K_i}(\bar{B_i} \cap R_{B_i} \cap L_{B_i}))\\
\text{From equation (\ref{m3}) we know that $f_{M_2}f_{M_4}f_{K_i}$ is one-to-one over $\bar{B_i} \cap R_{B_i} \cap L_{B_i}$. So,}\\
codim_{A}(R_{A_i}) \leq H(Y_{2,4}) - dim(\bar{B_i} \cap R_{B_i} \cap L_{B_i}) = H(Y_{2,4}) + codim_{B_i}(\bar{B_i} \cap R_{B_i} \cap L_{B_i}) - H(B_i)\\
\text{Applying lemma~\ref{Lem1} and then substituting $codim_{B_i}(R_{B_i})$ and $codim_{B_i}(L_{B_i})$ from equations (\ref{rbi}) and (\ref{lbi}) we have:}\\
codim_{A}(R_{A_i}) \leq H(Y_{2,4}) + codim_{B_i}(\bar{B_i}) + codim_{B_i}(R_{B_i}) + codim_{B_i}(L_{B_i}) - H(B_i)\\
\text{or, } codim_{A}(R_{A_i}) \leq H(Y_{2,4}) + codim_{B_i}(\bar{B_i}) + H(Y_{6,8}) + codim_{A}(\bar{A}) - H(A) + H(Y_{1,3}) + codim_{A}(\bar{A}) - H(A) - H(B_i)\\
\text{or, } codim_{A}(R_{A_i}) \leq H(Y_{1,3}) + H(Y_{2,4}) + H(Y_{6,8}) + codim_{B_i}(\bar{B_i}) + 2codim_{A}(\bar{A}) - 2H(A) - H(B_i)\IEEEyesnumber\label{rai}
\end{IEEEeqnarray*}

\begin{IEEEeqnarray*}{l}
codim_{A}(L_{A_i}) = codim_{A}(f_{D_6}^{-1}(f_{M_5}f_{K_i}(\bar{B_i} \cap S_{B_i})))\\
\text{Applying lemma~\ref{Lem2} we have:}\\
codim_{A}(L_{A_i}) \leq codim_{Y_{6,8}}(f_{M_5}f_{K_i}(\bar{B_i} \cap S_{B_i})) = dim(Y_{6,8}) - dim(f_{M_5}f_{K_i}(\bar{B_i} \cap S_{B_i}))\\
\text{From equation (\ref{m2}) we know that $f_{M_5}f_{K_i}$ is one-to-one over $\bar{B_i} \cap S_{B_i}$. So,}\\
codim_{A}(L_{A_i}) \leq H(Y_{6,8}) - dim(\bar{B_i} \cap S_{B_i}) = H(Y_{6,8}) + codim_{B_i}(\bar{B_i} \cap S_{B_i}) - H(B_i)\\
\text{Applying lemma~\ref{Lem1} and then substituting $codim_{B_i}(S_{B_i})$ from equation (\ref{sbi}) we have:}\\
codim_{A}(L_{A_i}) \leq H(Y_{6,8}) + codim_{B_i}(\bar{B_i}) + codim_{B_i}(S_{B_i}) - H(B_i)\\
\text{or, } codim_{A}(L_{A_i})\leq H(Y_{6,8}) + codim_{B_i}(\bar{B_i}) + H(Y_{5,7}) + codim_{C}(\bar{C}) - H(C) - H(B_i)\IEEEyesnumber\label{lai}
\end{IEEEeqnarray*}

Substituting equation (\ref{lai}), (\ref{rai}), and (\ref{sai}) in equation (\ref{dima}) we have:
\begin{IEEEeqnarray*}{l}
H(A) \leq (q-1)(H(Y_{1,3}) + H(Y_{2,4}) + H(Y_{5,7}) + 2H(Y_{6,8})) + \sum_{i=1}^{q-1} H(Y_{e_i}) - 2(q-1)H(A) - 2(q-1)H(C) - \sum_{i=1}^{q-1} 2H(B_i) \\\hfill +\> (2q-1)codim_{A}(\bar{A}) + (q-1)codim_{C}(\bar{C}) + \sum_{i=1}^{q-1}  2codim_{B_i}(\bar{B_i}) + \sum_{i=1}^{q-1} codim_{C}(\bar{C_i})\IEEEeqnarraynumspace\IEEEyesnumber\label{ha}
\end{IEEEeqnarray*}

Now substituting equations (\ref{codimbarc2}), (\ref{barbi}), (\ref{xa}) and (\ref{xci}) in equation (\ref{ha}) we have:

\begin{IEEEeqnarray*}{l}
H(A) \leq (q-1)(H(Y_{1,3}) + H(Y_{2,4}) + H(Y_{5,7}) + 2H(Y_{6,8})) + \sum_{i=1}^{q-1} H(Y_{e_i}) - 2(q-1)H(A) - 2(q-1)H(C) - \sum_{i=1}^{q-1} 2H(B_i) \\ +\> (7q-6)codim_{Y_{1,3}}(Y_{1,3}^\prime) + (6q-5)codim_{Y_{2,4}}(Y_{2,4}^\prime) + \sum_{i=1}^{q-1} (2q)codim_{Y_{e_i}}(Y_{e_i}^\prime) + (3q-3)codim_{Y_{5,7}}(Y_{5,7}^\prime) \\+\> (4q-3)codim_{Y_{6,8}}(Y_{6,8}^\prime) + (2q-2)codim_{Y_{9,10}}(Y_{9,10}^\prime) + (2q-1)codim_{A}(A^\prime) + (q-1)codim_{C}(C^\prime) + \sum_{i=1}^{q-1} 2codim_{B}(B_i^\prime) \\+\> \sum_{i=1}^{q-1} codim_{C}(C^\prime_i) + (5q-4)(H(A) - H(A,B_1,\ldots ,B_{q-1},C)) \\+\> (6q-5)(\sum_{i=1}^{q-1} H(B_{i}) + H(C)) - (q-1)H(B_1,\ldots ,B_{q-1},C)
\end{IEEEeqnarray*}
Substituting values from equations (\ref{y13}), (\ref{y24}), (\ref{yei}), (\ref{y57}), (\ref{y68}), (\ref{a}), (\ref{bi}), (\ref{c}), and (\ref{ci}) we get: 
\begin{IEEEeqnarray*}{l}
H(A) \leq (q-1)(H(Y_{1,3}) + H(Y_{2,4}) + H(Y_{5,7}) + 2H(Y_{6,8})) + \sum_{i=1}^{q-1} H(Y_{e_i}) - 2(q-1)H(A) - 2(q-1)H(C) - \sum_{i=1}^{q-1} 2H(B_i) \\ +\> (7q-6)H(Y_{1,3}|A,B_1,\ldots,B_{q-1}) + (6q-5)H(Y_{2,4}|B_1,\ldots,B_{q-1},C) + \sum_{i=1}^{q-1} (2q)H(Y_{e_i}|Y_{2,4},\cup_{j=1,j\neq i}^{q-1} B_j) \\+\> (3q-3)H(Y_{5,7}|Y_{1,3},Y_{2,4}) + (4q-3)H(Y_{6,8}|Y_{1,3},C) + (2q-2)H(Y_{9,10}|Y_{5,7},Y_{6,8}) + (2q-1)H(A|Y_{6,8},Y_{e_1},\ldots ,Y_{e_{q-1}}) \\+\> (q-1)H(C|A,Y_{5,7}) + \sum_{i=1}^{q-1} 2H(B_i|Y_{9,10},B_1,\ldots,B_{i-1},B_{i+1},\ldots, B_{q-1}) + \sum_{i=1}^{q-1} H(C|Y_{e_i},B_i) \\+\> (5q-4)(H(A) - H(A,B_1,\ldots ,B_{q-1},C)) + (6q-5)(\sum_{i=1}^{q-1} H(B_{i}) + H(C)) - (q-1)H(B_1,\ldots ,B_{q-1},C)
\end{IEEEeqnarray*}

Replacing $Y_{1,3}$ by $U$, $Y_{2,4}$ by $Y$, $Y_{5,7}$ by $W$, $Y_{6,8}$ by $X$, $Y_{e_i}$ by $V_i$, and $Y_{9,10}$ by $Z$ we get the desired inequality (\ref{eq0}) of theorem~\ref{Thm1}.

\begin{IEEEeqnarray*}{l}
H(A) \leq (q-1)(H(U) + H(Y) + H(W) + 2H(X)) + \sum_{i=1}^{q-1} H(V_i) - 2(q-1)H(A) - 2(q-1)H(C) - \sum_{i=1}^{q-1} 2H(B_i) \\ +\> (7q-6)H(U|A,B_1,\ldots,B_{q-1}) + (6q-5)H(Y|B_1,\ldots,B_{q-1},C) + \sum_{i=1}^{q-1} (2q)H(V_i|Y,B_1,\ldots, B_{i-1},B_{i+1},\ldots, B_{q-1}) \\+\> (3q-3)H(W|U,Y) + (4q-3)H(X|U,C) + (2q-2)H(Z|W,X) + (2q-1)H(A|X,V_1,\ldots ,V_{q-1}) \\+\> (q-1)H(C|A,W) + \sum_{i=1}^{q-1} 2H(B_i|Z,B_1,\ldots,B_{i-1},B_{i+1},\ldots, B_{q-1}) + \sum_{i=1}^{q-1} H(C|V_i,B_i) \\+\> (5q-4)(H(A) - H(A,B_1,\ldots ,B_{q-1},C)) + (6q-5)(\sum_{i=1}^{q-1} H(B_{i}) + H(C)) - (q-1)H(B_1,\ldots ,B_{q-1},C)
\end{IEEEeqnarray*}



\section{}\label{appB}
\subsection{Proof of lemma~\ref{lem4}:}
Consider a $(d,dn)$ fractional linear network coding solution of the network in Fig.~\ref{nofano1/n}. The local coding matrices are shown along the edges. The matrices $Q_i$ for $1\leq i\leq n$ and $A_{ij}$ for $1\leq i\leq n, 1\leq j\leq q$ are of size $dn \times d$, and left multiplies the massage vector $a_i$. The matrices $C_{ij}$ for $1\leq i\leq q, 1\leq j\leq n$, $B_{ijk}$ for $1\leq i,k\leq q, i\neq k, 1\leq j\leq n$, and $D_{ij}$ for $1\leq i\leq q, 1\leq j\leq n$ left multiplies $b_{ij}$ and are of size $dn \times d$. The matrices $M_1,M_2,M_3$ and $K_i,R_i$ and $U_i$ for $1\leq i\leq q$ are of sizes $dn \times dn$. And the matrices $E_j,G_{ij}$ and $V_j$ for $1\leq i\leq q$ and $1\leq j\leq n$ are of sizes $d \times dn$. Also let $I_d$ be a $d\times d$ identity matrix. The following comes from the definition of network coding.
\begin{IEEEeqnarray}{l}
Y_{e_a} = \sum_{i=1}^n Q_ia_i + \sum_{i=1}^q \sum_{j=1}^n C_{ij}b_{ij}\\
\text{for } 1\leq i\leq q:\; Y_{e_i} = \sum_{j=1}^n A_{ji}a_j + \sum_{j=1,j\neq i}^q \sum_{k=1}^n B_{jki}b_{jk}\\
Y_{e_b} = \sum_{i=1}^q \sum_{j=1}^n D_{ij}b_{ij}\\
Y_{e_a^\prime} = M_1Y_{e_a} + M_2Y_{e_b} = \sum_{i=1}^n M_1Q_ia_i + \sum_{i=1}^q \sum_{j=1}^n (M_1C_{ij} + M_2D_{ij})b_{ij}\IEEEeqnarraynumspace\label{nt1}\\
\text{for } 1\leq i\leq q:\; Y_{e_i^\prime} = K_iY_{e_a} + R_iY_{e_i} = \sum_{j=1}^n (K_iQ_j + R_iA_{ji})a_j + \sum_{k=1}^n K_iC_{ik}b_{ik} + \sum_{j=1,j\neq i}^q \sum_{k=1}^n (K_iC_{jk} + R_iB_{jki})b_{jk}\IEEEeqnarraynumspace\label{nt2}\\
Y_{e_b^\prime} = \sum_{i=1}^q U_iY_{e_i} + M_3Y_{e_b} =  \sum_{i=1}^q \sum_{j=1}^n U_iA_{ji}a_{j} + \sum_{j=1}^q \sum_{k=1}^n ( \sum_{i=1,i\neq j}^q U_iB_{jki} + M_3D_{jk} )b_{jk} \IEEEeqnarraynumspace\label{nt3}
\end{IEEEeqnarray}

Because of the demands of the terminals the following inequalities must be satisfied. Since any terminal $t_i\in T_{a_1}$ computes $a_i$, using equation (\ref{nt1}) we have, for $1\leq i,j\leq n, j\neq i$:
\begin{IEEEeqnarray}{l}
E_iM_1Q_i = I\label{gq1}\\
E_iM_1Q_j = 0\label{gq2}
\end{IEEEeqnarray}
\begin{IEEEeqnarray}{l}
\text{Let, } E = \begin{bmatrix} E_1 & E_2 & \cdots & E_n \end{bmatrix}^T\label{e}\\
\text{and, }Q = \begin{bmatrix} Q_1 & Q_2 & \cdots & Q_n \end{bmatrix}\\
\text{Then, }M_1Q = \begin{bmatrix} M_1Q_1 & M_1Q_2 & \cdots & M_1Q_n \end{bmatrix}\label{m2q}
\end{IEEEeqnarray}
Applying Lemma~\ref{lem1} on equations (\ref{gq1}) and (\ref{gq2}) and using the matrices in equation (\ref{e}) and (\ref{m2q}) we get:
\begin{equation}
EM_1Q = I \label{em2q}
\end{equation}
At $t_k\in T_{a}$ the component of $b_{ij}$ is zero. So for $1\leq i\leq q, 1\leq j,k\leq n$, using equation (\ref{nt1}) we have:
\begin{equation}
E_k(M_1C_{ij} + M_2D_{ij}) = 0\label{gq3}
\end{equation}
\begin{IEEEeqnarray}{l}
\text{Let } C_i = \begin{bmatrix} C_{i1} & C_{i2} & \cdots & C_{in} \end{bmatrix}\label{l1}\\
\text{and } D_i = \begin{bmatrix} D_{i1} & D_{i2} & \cdots & D_{in} \end{bmatrix}\label{l3}\\
\text{Then }M_1C_{i} + M_2D_{i} = \begin{bmatrix} M_1C_{i1} + M_2D_{i1} & M_1C_{i2} + M_2D_{i2} & \cdots & M_1C_{in} + M_2D_{in} \end{bmatrix}\label{m2ci}
\end{IEEEeqnarray}
From Corollary~\ref{coro1} and equations (\ref{gq3}), (\ref{e}) and (\ref{m2ci}) we get:
\begin{equation}
\text{for }1\leq i\leq q:\; E(M_1C_{i} + M_2D_{i}) = 0\label{em2ci}
\end{equation}
Now consider the terminals in the set $T_{b_i}$ for $1\leq i\leq q$. Since at any terminal $t_j\in T_{b_i}$ for $1\leq j\leq n$ the component of $a_k$ in equation (\ref{nt2}) for $1\leq k\leq n$ is zero, we have:
\begin{equation}
G_{ij}(K_iQ_k + R_iA_{ki}) = 0\label{gq4}
\end{equation}
\begin{IEEEeqnarray}{l}
\text{Let } G_i = \begin{bmatrix} G_{i1} & G_{i2} & \cdots & G_{in} \end{bmatrix}^T\label{gi}\\
\text{and } A_i = \begin{bmatrix} A_{1i} & A_{2i} & \cdots & A_{ni} \end{bmatrix}\label{l2}\\
\text{So, } K_iQ + R_iA_{i} = \begin{bmatrix} K_iQ_1 + R_iA_{1i} & K_iQ_2 + R_iA_{2i} & \cdots & K_iQ_n + R_iA_{ni} \end{bmatrix}\label{kiq}
\end{IEEEeqnarray}
Using Corollary~\ref{coro1} and equations (\ref{gq4}), (\ref{gi}) and (\ref{kiq}) we get:
\begin{equation}
\text{for }1\leq i\leq q:\; G_{i}(K_iQ + R_iA_{i}) = 0\label{gikiq}
\end{equation}
Because $t_j\in T_{b_i}$ computes $b_{ij}$ for $1\leq i\leq q, 1\leq j,k\leq n, k\neq j$, from equation (\ref{nt2}) we have:
\begin{IEEEeqnarray}{l}
G_{ij}(K_iC_{ij}) = I\label{gq5}\\
G_{ij}(K_iC_{ik}) = 0\label{gq6}
\end{IEEEeqnarray}
\begin{IEEEeqnarray}{l}
\text{From the matrix in (\ref{l1}) we already have: }K_iC_{i} = \begin{bmatrix} K_iC_{i1} & K_iC_{i2} & \cdots & K_iC_{in} \end{bmatrix}\label{kici}
\end{IEEEeqnarray}
Using Lemma~\ref{lem1} and equations (\ref{gq5}), (\ref{gq6}), (\ref{gi}) and (\ref{kici}) we get:
\begin{equation}
\text{for }1\leq i\leq q:\; G_{i}(K_iC_{i}) = I\label{gikici}
\end{equation}
As the component of any $b_{kr}$ at $t_j\in T_{b_i}$ is zero if $k\neq i$, for $1\leq i,k\leq q, i\neq k, 1\leq j,r\leq n$ we have:
\begin{equation}
G_{ij}(K_iC_{kr} + R_iB_{kri}) = 0\label{gq7}
\end{equation}
\begin{IEEEeqnarray}{l}
\text{Let } B_{ki} = \begin{bmatrix} B_{k1i} & B_{k2i} & \cdots & B_{kni} \end{bmatrix}.\label{l4}\\
\text{Then }K_iC_{k} + R_iB_{ki} = \begin{bmatrix} K_iC_{k1} + R_iB_{k1i} & K_iC_{k2} + R_iB_{k2i} & \cdots & K_iC_{kn} + R_iB_{kni} \end{bmatrix}\label{kick}
\end{IEEEeqnarray}
From Corollary~\ref{coro1} and equations (\ref{gq7}), (\ref{gi}) and (\ref{kick}) we have:
\begin{equation}
\text{for } 1\leq i,k\leq n, i\neq k:\; G_{i}(K_iC_{k} + R_iB_{ki}) = 0\label{gikick}
\end{equation}
We now consider the set $T_{a_2}$. Since the terminal $t_i\in T_{a_2}$ computes $a_i$ we have, for $1\leq i,j\leq n, j\neq i$, using equation (\ref{nt3}) we have:
\begin{IEEEeqnarray}{l}
V_i(\sum_{k=1}^q U_kA_{ik}) = I\label{gq8}\\
V_i(\sum_{k=1}^q U_kA_{jk}) = 0\label{gq9}
\end{IEEEeqnarray}
\begin{IEEEeqnarray}{l}
\text{Let, } V = \begin{bmatrix} V_1 & V_2 & \cdots & V_n \end{bmatrix}^T\label{v}\\
\text{Using the matrix in (\ref{l2}): } \sum_{k=1}^q U_kA_{k} = \begin{bmatrix} \sum_{k=1}^q U_kA_{1k} & \sum_{k=1}^q U_kA_{2k} & \cdots & \sum_{k=1}^q U_kA_{nk}\end{bmatrix}\label{sumukak}
\end{IEEEeqnarray}
Applying lemma~\ref{lem1} and equations (\ref{gq8}), (\ref{gq9}), (\ref{v}) and (\ref{sumukak}) we have:
\begin{equation}
V(\sum_{k=1}^q U_kA_{k}) = I\label{vsum}
\end{equation}
The component of $b_{jk}$ is zero at $t_i\in T_{a_2}$ for $1\leq j\leq q, 1\leq i,k\leq n$, and hence using equation (\ref{nt3}) we have:
\begin{equation}
V_i( \sum_{r=1,r\neq j}^q U_rB_{jkr} + M_3D_{jk}) = 0\label{gq10}
\end{equation}
\begin{IEEEeqnarray}{l}
\text{Using (\ref{l3}) and (\ref{l4}): }\sum_{r=1,r\neq j}^q U_rB_{jr} + M_3D_{j} = \begin{bmatrix} \sum_{r=1,r\neq j}^q U_rB_{j1r} + M_3D_{j1} & \cdots & \sum_{r=1,r\neq j}^q U_rB_{jnr} + M_3D_{jn} \end{bmatrix}\IEEEeqnarraynumspace\label{urbjr}
\end{IEEEeqnarray}
From Corollary~\ref{coro1} and equations (\ref{gq10}), (\ref{v}) and (\ref{urbjr}) we have, for $1\leq j\leq q$:
\begin{equation}
V(\sum_{r=1,r\neq j}^q U_rB_{jr} + M_3D_{j}) = 0\label{vbigsum}
\end{equation}
%

The matrices $E, M_1$ and $Q$ are invertible from equation (\ref{em2q}). Matrices $G_i,K_i$ and $C_i$ for $1\leq i\leq q$ are invertible from equation (\ref{gikici}). Matrix $V$ is invertible from equation (\ref{vsum}).
Since $E$ is invertible we have from equation (\ref{em2ci}):
\begin{equation}
M_1C_{i} + M_2D_{i} = 0 \label{1}
\end{equation}
As both $M_1$ and $C_i$ are invertible matrices, from equation (\ref{1}) $M_2$ is an invertible matrix. 
Since $G_i$ is invertible for $1\leq i\leq q$, we have from equation (\ref{gikiq}):
\begin{equation}
K_iQ + R_iA_{i} = 0 \label{2}
\end{equation}
Since both $K_i$ and $Q$ are invertible matrices, their product is a full rank matrix, and hence $R_i$ is an invertible matrix for $1\leq i\leq q$. 
Also, from equation (\ref{gikick}) we have, for $1\leq i,k\leq q, i\neq k$:
\begin{equation}
K_iC_{k} + R_iB_{ki} = 0 \label{3}
\end{equation}
And since $V$ is invertible, we have from equation (\ref{vbigsum}), for $1\leq i\leq q$:
\begin{equation}
( \sum_{r=1,r\neq i}^p U_rB_{ir}) + M_3D_{i} = 0 \label{4}
\end{equation}
Substituting $D_i$ from equation (\ref{1}) in equation (\ref{4}) we get, for $1\leq i\leq q$:
\begin{IEEEeqnarray*}{l}
( \sum_{r=1,r\neq i}^q U_rB_{ir}) - M_3M_2^{-1}M_1C_{i} = 0\\
\text{Substituting $B_{ir}$ from equation (\ref{3}) we get}:\\
-( \sum_{r=1,r\neq i}^q\!\! U_rR_r^{-1}K_rC_i) - M_3M_2^{-1}M_1C_{i} = 0\\
\text{Substituting $R_r^{-1}K_r$ from equation (\ref{2}) we get}:\\
( \sum_{r=1,r\neq i}^q U_rA_rQ^{-1}C_i) - M_3M_2^{-1}M_1C_{i} = 0\\
\text{Substituting $Q^{-1}$ from equation (\ref{em2q}) we get:}\\
( \sum_{r=1,r\neq i}^q U_rA_rEM_1C_i) - M_3M_2^{-1}M_1C_{i} = 0\\
\text{or, }( \sum_{r=1,r\neq i}^q U_rA_rE - M_3M_2^{-1})M_1C_{i} = 0\\
\text{Since $M_1$ and $C_i$ both are invertible, we have}:\\
\sum_{r=1,r\neq i}^q U_rA_rE - M_3M_2^{-1} = 0\IEEEyesnumber\label{op1}
\end{IEEEeqnarray*}
\begin{IEEEeqnarray*}{l}
\text{From equation (\ref{vsum}) we have $\sum_{r=1,r\neq i}^q U_rA_r + U_iA_i = V^{-1}$. Substituting this value in equation (\ref{op1}) we get:}\\
(V^{-1} -U_iA_i)E -M_3M_2^{-1} = 0 \\
\text{or, }V^{-1} -U_iA_i = M_3M_2^{-1}E^{-1}\\
\text{or, }U_iA_i = V^{-1} - M_3M_2^{-1}E^{-1}\IEEEyesnumber\label{5}
\end{IEEEeqnarray*}

Now, substituting equation (\ref{5}) in equation (\ref{vsum}) we get:
\begin{IEEEeqnarray*}{l}
V(\sum_{i=1}^q V^{-1} - M_4M_3^{-1}E^{-1}) = I\\
\sum_{i=1}^q V(V^{-1} - M_4M_3^{-1}E^{-1}) = I\\
\sum_{i=1}^q (I - VM_4M_3^{-1}E^{-1}) = I\\
(q-1)I = qIVM_4M_3^{-1}E^{-1}\IEEEyesnumber\label{6}
\end{IEEEeqnarray*}
In equation (\ref{6}), if $q=0$, then the equation becomes $-I = 0$. Hence $q\neq 0$ is a necessary condition for the network $\mathcal{N}_2^\prime$ to have a rate $\frac{1}{n}$ fractional linear network coding solution. Then, from the fact that an element in a finite field is equal to zero if and only if the characteristic of the finite field divides that element (so $q\neq 0$ if and only if the characteristic of the finite field does not divide $q$), the ``only if'' part of the proposition is proved.

We now show that $\mathcal{N}_2$ has a $(1,n)$ fractional linear network coding solution if the $q$ has an inverse in the finite field. 
Let $\bar{a}_i$ be an $n$-length vector whose $i^{\text{th}}$ component is $a_i$ and all other components are zero. Also let $\bar{b}_{ij}$ be an $n$-length vector whose $j^{\text{th}}$ component is $b_{ij}$ and all other components are zero. 
Then let the following information be transmitted by the corresponding edges. 
\begin{IEEEeqnarray*}{l}
Y_{e_a} = \sum_{j=1}^n \bar{a}_j + \sum_{i=1}^q \sum_{j=1}^n \bar{b}_{ij}\\
\text{for } 1\leq i\leq q: \quad Y_{e_i} = \sum_{j=1}^n \bar{a}_j + \sum_{k=1,k\neq i}^q \sum_{j=1}^n \bar{b}_{kj}\\
Y_{e_b} = \sum_{i=1}^q \sum_{j=1}^n \bar{b}_{ij}\\
Y_{e_a^\prime} = Y_{e_a} - Y_{e_b} = \sum_{j=1}^n \bar{a}_j\\
\text{for } 1\leq i\leq q: \quad Y_{e_i^\prime} = Y_{e_a} - Y_{e_i} = \sum_{j=1}^n \bar{b}_{ij}\\
Y_{e_b^\prime} = q^{-1}\{\sum_{i=1}^q Y_{e_i} - (q-1)Y_{e_b}\} = q^{-1}\{q(\sum_{j=1}^n \bar{a}_j) + (q-1)(\sum_{i=1}^q \sum_{j=1}^n \bar{b}_{ij}) - (q-1)(\sum_{i=1}^q \sum_{j=1}^n \bar{b}_{ij})\} = \sum_{j=1}^n \bar{a}_j
\end{IEEEeqnarray*}
Let $\check{u}(j)$ be a unit row vector of length $n$ which has $j^{\text{th}}$ component equal to one and all other components are equal to zero. Then from the dot product of $\check{u}(j)$ and $\sum_{j=1}^n \bar{a}_j$, message $a_j$ can be retrieved. Similarly from the dot product of $\check{u}(j)$ and $\sum_{j=1}^n \bar{b}_{ij}$, $b_{ij}$ can be determined.

\subsection{Proof of Theorem~\ref{Thm2}}\label{appD}
To obtain this inequality, we apply DFZ method to the network shown in Fig.~\ref{nofano1/n} for $n=1$ and $q = p_1\times \cdots \times p_l$. 
%
Corresponding to each of the messages in the network, define vector subspaces $A$, $B_1,B_2,\ldots,B_q$, $Y_{e_a}$, $Y_{e_1},\ldots,Y_{e_q}$, $Y_{e_b}$ of a finite dimensional vector space $V$. Now consider the following linear functions.
\begin{IEEEeqnarray*}{rllll}
&f_{Q}: Y_{e_a} \rightarrow A & f_{M_1}: A \rightarrow Y_{e_a} \qquad & f_{M_2}: A \rightarrow Y_{e_b} \qquad & f_{M_3}: A \rightarrow Y_{e_b}\\
\text{ for } 1\leq i\leq q:\;\;\;& f_{C_i}: Y_{e_a} \rightarrow B_i \qquad & f_{A_i}: Y_{e_i} \rightarrow A & f_{D_i}: Y_{e_b} \rightarrow B_i\\
\text{ for } 1\leq i\leq q:\;\;\;& f_{K_i}: B_i \rightarrow Y_{e_a} & f_{R_i}: B_i \rightarrow Y_{e_i}  & f_{U_i}: A \rightarrow Y_{e_i}\\
\text{ for } 1\leq i,j\leq q, i\neq j:\;\;\; & f_{B_{ji}}: Y_{e_i} \rightarrow B_j
\end{IEEEeqnarray*}
Due to Lemma~\ref{Lem3}, the following holds:
\begin{IEEEeqnarray*}{l}
f_{Q} + \sum_{i=1}^{q} f_{C_i} = I \text{ over a subspace } Y_{e_a}^\prime \text{ of } Y_{e_a} \text{ where } codim_{Y_{e_a}}(Y_{e_a}^\prime) \leq H(Y_a|A,B_1,\ldots ,B_q)\IEEEeqnarraynumspace\IEEEyesnumber\label{y_a}\\
f_{A_i} + \sum_{j=1,j\neq i}^q B_{ji} = I  \text{ over a subspace } Y_{e_i}^\prime \text{ of } Y_{e_i} \text{ where } codim_{Y_{e_i}}(Y_{e_i}^\prime) \leq H(Y_{e_i}|A,\cup_{j=1,j\neq i}^q B_j)\IEEEeqnarraynumspace\IEEEyesnumber\label{y_i}\\
\sum_{i=1}^q f_{D_i} = I \text{ over a subspace } Y_{e_b}^\prime \text{ of } Y_{e_b}  \text{ where } codim_{Y_{e_b}}(Y_{e_b}^\prime) \leq H(Y_{e_b}|B_1,\ldots ,B_q)\IEEEeqnarraynumspace\IEEEyesnumber\label{y_b}\\
f_{M_1} + f_{M_2} = I \text{ over a subspace } A^\prime \text{ of } A  \text{ where } codim_{A}(A^\prime) \leq H(A|Y_{e_a},Y_{e_b})\IEEEeqnarraynumspace\IEEEyesnumber\label{apime}\\
\text{for } 1\leq i\leq q:\quad f_{K_i} + f_{R_i} = I \text{ over a subspace } B_i^\prime \text{ of } B_i \text{ where } codim_{B_i}(B_i^\prime) \leq H(B_i|Y_{e_a},Y_{e_i})\IEEEeqnarraynumspace\IEEEyesnumber\label{b_iprime}\\
\sum_{i=1}^q f_{U_i} + f_{M_3} = I \text{ over a subspace } A^{\prime\prime} \text{ of } A \text{ where } codim_{A}(A^{\prime\prime}) \leq H(A|Y_{e_1},\ldots ,Y_{e_q},Y_{e_b})\IEEEeqnarraynumspace\IEEEyesnumber\label{a_primeprime}
\end{IEEEeqnarray*}
Now we have:
\begin{IEEEeqnarray*}{l}
f_{Q}f_{M_1} + \sum_{i=1}^q f_{C_i}f_{M_1} = f_{M_1} \text{ over a subspace } f_{M_1}^{-1}(Y_{e_a}^\prime) \text{ of } A\\
\sum_{i=1}^q f_{D_i}f_{M_2} = f_{M_2} \text{ over a subspace } f_{M_2}^{-1}(Y_{e_b}^\prime) \text{ of } A\\
\text{So, } f_{Q}f_{M_1} + \sum_{i=1}^q f_{C_i}f_{M_1} + \sum_{i=1}^q f_{D_i}f_{M_2} = f_{M_1} + f_{M_2} = I \text{ over a subspace } A^{\prime\prime\prime} = f_{M_1}^{-1}(Y_{e_a}^\prime) \cap f_{M_2}^{-1}(Y_{e_b}^\prime) \cap A^\prime.\\
\text{Hence from lemma~\ref{Lem1}: } codim_A(A^{\prime\prime\prime}) \leq codim_A(f_{M_1}^{-1}(Y_{e_a}^\prime)) + codim_A(f_{M_2}^{-1}(Y_{e_b}^\prime)) + codim_A(A^\prime)\\
\text{or, using lemma~\ref{Lem2}: } codim_A(A^{\prime\prime\prime}) \leq codim_{Y_{e_a}}(Y_{e_a}^\prime) + codim_{Y_{e_b}}(Y_{e_b}^\prime) + codim_A(A^\prime)\IEEEeqnarraynumspace\IEEEyesnumber\label{aprimeprime}
\end{IEEEeqnarray*}
So, due to Lemma~\ref{Lem4} there exists a subspace $\bar{A}$ of $A^{\prime\prime\prime}$ over which:
\begin{IEEEeqnarray}{l}
f_{Q}f_{M_1} - I = 0\label{naa}\\
\text{for } 1\leq i \leq q: f_{C_i}f_{M_1} + f_{D_i}f_{M_2} = 0  \label{nabi}
\end{IEEEeqnarray}
where,
\begin{IEEEeqnarray*}{l}
codim_{A^{\prime\prime\prime}}(\bar{A}) \leq H(A) + \sum_{i=1}^q H(B_i) - H(A,B_1,\ldots ,B_q)\\
\text{So, }codim_{A}(\bar{A}) = codim_A(A^{\prime\prime\prime}) + codim_{A^{\prime\prime\prime}}(\bar{A}) \leq codim_{Y_{e_a}}(Y_{e_a}^\prime) + codim_{Y_{e_b}}(Y_{e_b}^\prime) + codim_A(A^\prime) \\\hfill +\> H(A) + \sum_{i=1}^q H(B_i) - H(A,B_1,\ldots ,B_q)\IEEEeqnarraynumspace\IEEEyesnumber\label{codima}
%
\end{IEEEeqnarray*}
Similarly, note the following.
\begin{IEEEeqnarray*}{l}
f_{Q}f_{K_i} + \sum_{j=1}^q f_{C_j}f_{K_i} = f_{K_i} \text{ over a subspace }  f_{K_i}^{-1}(Y_{e_a}^\prime) \text{ of } B_i\\
f_{A_i}f_{R_i} + \sum_{j=1,j\neq i}^q f_{B_{ji}}f_{R_i} = f_{R_i} \text{ over a subspace } f_{R_i}^{-1}(Y_{e_i}^\prime) \text{ of } B_i\\
\text{So, } f_{Q}f_{K_i} + f_{A_i}f_{R_i} + f_{C_i}f_{K_i} + \sum_{j=1,j\neq i}^q (f_{C_j}f_{K_i} + f_{B_{ji}}f_{R_i}) = I
\text{ over a subspace } B_i^{\prime\prime} =  f_{K_i}^{-1}(Y_{e_a}^\prime) \cap f_{R_i}^{-1}(Y_{e_i}^\prime) \cap B_i^\prime \text{ of } B_i.\\\text{So, from Lemma~\ref{Lem1} and Lemma~\ref{Lem2} we have: } codim_{B_i}(B_i^{\prime\prime}) \leq codim_{Y_{e_a}}(Y_{e_a}^\prime) + codim_{Y_{e_i}}(Y_{e_i}^\prime) + codim_{B_i}(B_i^\prime)
\end{IEEEeqnarray*}
So according to Lemma~\ref{Lem4} there exists a subspace $\bar{B}_i$ over which the following identities hold:
\begin{IEEEeqnarray}{l}
f_{Q}f_{K_i} + f_{A_i}f_{R_i} = 0  \label{nbia}\\
f_{C_i}f_{K_i} = I\label{nbi}\\
\text{for } 1\leq j\leq q, j\neq i:\;
f_{C_j}f_{K_i} + f_{B_{ji}}f_{R_i} = 0 \IEEEeqnarraynumspace \label{nbibj}
\end{IEEEeqnarray}
where,
\begin{IEEEeqnarray*}{l}
codim_{B_i^{\prime\prime}}(\bar{B}_i) \leq H(A) + \sum_{i=1}^q H(B_i) - H(A,B_1,\ldots ,B_q)\\
\text{So, } codim_{B_i}(\bar{B}_i) = codim_{B_i}(B_i^{\prime\prime}) + codim_{B_i^{\prime\prime}}(\bar{B}_i)\\
\text{or, } codim_{B_i}(\bar{B}_i) \leq codim_{Y_{e_a}}(Y_{e_a}^\prime) + codim_{Y_{e_i}}(Y_{e_i}^\prime) + codim_{B_i}(B_i^\prime)  + H(A) + \sum_{i=1}^q H(B_i) - H(A,B_1,\ldots ,B_q)\IEEEeqnarraynumspace\IEEEyesnumber\label{codimbi}
\end{IEEEeqnarray*}

Similarly we have,
\begin{IEEEeqnarray*}{l}
\text{for } 1\leq i\leq q:\; f_{A_i}f_{U_i} + \sum_{j=1,j\neq i}^q f_{B_{ji}}f_{U_i} = f_{U_i} \text{ over a subspace } f_{U_i}^{-1}(Y_{e_i}) \text{ of } A\\
\sum_{i=1}^q f_{D_i}f_{M_3} = f_{M_3} \text{ over a subspace } f_{M_3}^{-1}(Y_{e_b}) \text{ of } A\\
\text{Hence, } \sum_{i=1}^q (f_{A_i}f_{U_i} + \sum_{j=1,j\neq i}^q f_{B_{ji}}f_{U_i}) + \sum_{i=1}^q f_{D_i}f_{M_3} = I \text{ over  a subspace } A^{\prime\prime\prime\prime} = \cap_{i=1}^q f_{U_i}^{-1}(Y_{e_i}) \cap f_{M_3}^{-1}(Y_{e_b}) \cap A^{\prime\prime} \text{ of } A \\\text{Using Lemma~\ref{Lem1} and Lemma~\ref{Lem2} we have: }\\
codim_{A}(A^{\prime\prime\prime\prime}) \leq \sum_{i=1}^{q} codim_{Y_{e_i}}(Y_{e_i}) + codim_{Y_{e_b}}(Y_{e_b}) + codim_{A}(A^{\prime\prime})\\
\end{IEEEeqnarray*}
From Lemma~\ref{Lem4}, over a subspace $\hat{A}$ of $A$ we have:
\begin{IEEEeqnarray}{l}
\sum_{i=1}^q f_{A_i}f_{U_i} = I\label{ncc}\\
\text{for } 1\leq i\leq q:\; \sum_{j=1,j\neq i}^q f_{B_{ij}}f_{U_j} + f_{D_i}f_{M_3} = 0\label{ncb}
\end{IEEEeqnarray}
where,
\begin{IEEEeqnarray*}{l}
codim_{A^{\prime\prime\prime\prime}}(\hat{A}) \leq H(A) + \sum_{i=1}^q H(B_i) - H(A,B_1,\ldots ,B_q)\\
\text{So, } codim_{A}(\hat{A}) = codim_{A}(A^{\prime\prime\prime\prime}) + codim_{A^{\prime\prime\prime\prime}}(\hat{A})\\ 
\text{or, } codim_{A}(\hat{A}) \leq \sum_{i=1}^{q} codim_{Y_{e_i}}(Y_{e_i}) + codim_{Y_{e_b}}(Y_{e_b}) + codim_{A}(A^{\prime\prime}) + H(A) + \sum_{i=1}^q H(B_i) - H(A,B_1,\ldots ,B_q)\IEEEeqnarraynumspace\IEEEyesnumber\label{codimhata}
\end{IEEEeqnarray*}
One way to find the respective set $S$ for this proof is to use the proof of lemma~\ref{lem4} in the above subsection --- like lemma~\ref{lem3} was used to find the set $S$ for the proof of theorem~\ref{Thm1}. In Section~\ref{use4} we show the the inequality that would result if we indeed follow this method. However, we have found that using a different technique, which is a generalization of the proof of Theorem~15 in p. 2502 of \cite{rateregion}, a tighter upper-bound on the linear coding capacity of $\mathcal{N}_2$ can be obtained. Towards this end, let us define the following subspaces:

\begin{IEEEeqnarray*}{l}
A^{*} = f_{M_1}(\bar{A})\qquad
\text{for } 1\leq i\leq q:\; B_{i}^{*} = f_{K_i}(\bar{B_i})\\
A^{**} = A^* \cap B_1^* \qquad B_1^{**} = B_1^* \cap B_2^* \cap \cdots \cap B_q^* \qquad \text{for } 2\leq i\leq q:\; B_i^{**} = B_1^* \cap B_i^*\\
A^{***} = f_{Q}(A^{**}) \qquad \text{for } 1\leq i\leq q:\; B_i^{***} = f_{C_i}(B_i^{**})
\end{IEEEeqnarray*}
From equation~(\ref{naa}) we know that $f_{Q}$ is one-to-one over $A^*$. Then, as $f_{Q}f_{M_1}(A^{***}) = A^{***} = f_{Q}(A^{**})$, we must have $A^{**} = f_{M_1}(A^{***})$. With similar reasoning we have: $B_i^{**} = f_{K_i}(B_i^{***})$ for $1\leq i\leq q$.

\noindent Let us define the following subspaces:
\begin{IEEEeqnarray}{l}
S_a = \{ a \in A | f_{M_3}(a) \in f_{M_2}(A^{***}) \}\label{r1}\\
\text{for } 1\leq i\leq q:\; S_i = \{ a \in A | f_{U_i}(a) \in f_{R_i}({B_i}^{***}) \}\label{r2}\\
S = \hat{A} \cap S_a \cap S_1 \cap S_2 \cap \cdots \cap S_q
\end{IEEEeqnarray}

Let $\hat{a}\in S$. Then $f_{U_i}(\hat{a}) = f_{R_i}(b_i)$ for some 
$b_i \in B_i^{***}$ where $1\leq i\leq q$. Also $f_{M_3}(\hat{a}) = f_{M_2}(a)$ for some $a \in A^{***}$. So from equations~(\ref{ncc}) and~(\ref{ncb}) respectively we have:
\begin{IEEEeqnarray}{l}
\sum_{i=1}^q f_{A_i}f_{R_i}(b_i) = \hat{a} \label{ncc1}\\
\text{and for } 1\leq i\leq q: \sum_{j=1,j\neq i}^q f_{B_{ij}}f_{R_j}(b_j) + f_{D_i}f_{M_2}(a) = 0 \label{ncb1}
\end{IEEEeqnarray}
Summing equation~(\ref{nbia}) for $1\leq i\leq q$ we have:
\begin{equation}
\sum_{i=1}^q (f_{Q}f_{K_i} + f_{A_i}f_{R_i})(b_i) = 0 \label{temp1}
\end{equation}
Substituting $\sum_{i=1}^q f_{A_i}f_{R_i}(b_i)$ from equation~(\ref{ncc1}) in equation~(\ref{temp1}) we have:
\begin{equation}
\sum_{i=1}^q f_{Q}f_{K_i}(b_i) = -\hat{a} \label{temp2}
\end{equation}
interchanging $i$ and $j$ in equation~(\ref{nbibj}), and then summing for $1\leq j\leq q, j\neq i$ we have:
\begin{equation}
\text{for } 1\leq i\leq q: \sum_{j=1,j\neq i}^q (f_{C_i}f_{K_j} + f_{B_{ij}}f_{R_j})(b_j) = 0 \label{temp3}
\end{equation}
Substituting $\sum_{j=1,j\neq i}^q f_{B_{ij}}f_{R_j}(b_j)$ from equation~(\ref{ncb1}) in equation~(\ref{temp3}) we have:
\begin{equation}
\text{for } 1\leq i\leq q: \sum_{j=1,j\neq i}^q f_{C_i}f_{K_j}(b_j) - f_{D_i}f_{M_2}(a) = 0  \label{temp4}
\end{equation}
Substituting $f_{D_i}f_{M_2}(a)$ from equation~(\ref{nabi}) we have:
\begin{equation}
\text{for } 1\leq i\leq q: \sum_{j=1,j\neq i}^q f_{C_i}f_{K_j}(b_j) + f_{C_i}f_{M_1}(a) = 0 \label{temp5}
\end{equation}
For $i=1$, from equation (\ref{temp5}) we get:
\begin{IEEEeqnarray*}{l}
\sum_{j=2}^q f_{C_1}f_{K_j}(b_j) + f_{C_1}f_{M_1}(a) = 0\\
\text{or, } f_{C_1} (\sum_{j=2}^q f_{K_j}(b_j) + f_{M_1}(a)) = 0\\
\text{Since } f_{K_j}(b_j) \in (B_1^{*}\cap B_j^{*}) \text{ for } 2\leq j\leq q; \text{ and } f_{M_1}(a) \in (A^* \cap B_1^*); \text{ and as } f_{C_1} \text{ is invertible over } B_1^{*}, \text{ we have:}\\ 
\sum_{j=2}^q f_{K_j}(b_j) + f_{M_1}(a) = 0 \IEEEyesnumber\label{temp6}
\end{IEEEeqnarray*}
For $2\leq i\leq q$, from equation (\ref{temp5}) we get:
\begin{IEEEeqnarray*}{l}
f_{C_i}(\sum_{j=1,j\neq i}^q f_{K_j}(b_j) + f_{M_1}(a)) = 0\\
\text{or, } f_{C_i}(f_{K_1}(b_1) - f_{K_i}(b_i) + \sum_{j=2}^q f_{K_j}(b_j) + f_{M_1}(a)) = 0\\
\text{or, } f_{C_i}(f_{K_1}(b_1) - f_{K_i}(b_i))  = 0\hfill \text{ [using equation~(\ref{temp6})]}\\
\text{or, } (f_{K_1}(b_1) - f_{K_i}(b_i))  = 0 \qquad\qquad\qquad\qquad\qquad\qquad\text{[Since } f_{K_1}(b_1) \in B_i^{*} \text{ and } f_{C_i} \text{ is invertible over } B_i^* \text{]}\\
\text{or, } f_{K_1}(b_1) = f_{K_i}(b_i) \IEEEyesnumber\label{temp7}
\end{IEEEeqnarray*}

Substituting equation~(\ref{temp7}) in equation~(\ref{temp2}) we get:
\begin{IEEEeqnarray*}{l}
\sum_{i=1}^q f_{Q}f_{K_1}(b_1) = -\hat{a}\\
\text{or, } qf_{Q}f_{K_1}(b_1) = -\hat{a}\\
\text{Since the characteristic of the finite field divides $q$, we must have $q=0$. So, }
-\hat{a} = 0
\end{IEEEeqnarray*} 
Since this is true for any arbitrary $\hat{a} \in S$, we must have $S = \{ 0 \}$, which implies $dim(S) = 0$.

\begin{IEEEeqnarray*}{l}
\text{Now, }dim(A) = dim(A) - dim(S) = codim_{A}(S) = codim_{A}(\hat{A} \cap S_a \cap S_1 \cap S_2 \cap \cdots \cap S_q)\\
\leq codim_A(\hat{A}) + codim_A(S_a) + \sum_{i=1}^q codim_A(S_i)\hfill \text{ [applying lemma~\ref{Lem1}]}\IEEEeqnarraynumspace\IEEEyesnumber\label{mq}
\end{IEEEeqnarray*}
From (\ref{r1}) we have:
\begin{IEEEeqnarray*}{l}
codim_{A}(S_a) = codim_{A}(f_{M_3}^{-1}(f_{M_2}(A^{***}))) \leq codim_{Y_{e_b}}(f_{M_2}(A^{***})) \hfill\text{ [from lemma~\ref{Lem2}]}\\
\text{or, } codim_{A}(S_a) \leq dim(Y_{e_b}) - dim(f_{M_2}(A^{***}))\IEEEyesnumber\label{r3}\\
\text{Since } f_{M_1}(A^{***}) \text{ is a subspace of } B_1^*,  f_{C_1}f_{M_1} \text{ is invertible over } A^{***}; \text{ from equation~(\ref{nabi}) } f_{D_1}f_{M_2}, \text{ and hence } f_{M_2} \text{ is}\\\text{invertible over } A^{***}.\text{ Then, } dim(f_{M_2}(A^{***})) = dim(A^{***}). \text{ From eqn. (\ref{r3}) we have:}\\
codim_{A}(S_a) \leq dim(Y_{e_b}) - dim(A^{***}) = dim(Y_{e_b}) - dim(f_{Q}(A^{**})) \\\text{Now, $A^{**}$ is a subspace of $f_{M_1}(\bar{A})$, and over $f_{M_1}(\bar{A})$ $f_Q$ is invertible because of eqn. (\ref{naa}). So,}\\
codim_{A}(S_a) \leq dim(Y_{e_b}) - dim(A^{**}) = dim(Y_{e_b}) - dim(A^* \cap B_1^*) = dim(Y_{e_b}) + codim_{Y_{e_a}}(A^* \cap B_1^*) - dim(Y_{e_a})\\
\text{or, } codim_{A}(S_a) \leq dim(Y_{e_b}) + codim_{Y_{e_a}}(A^*) + codim_{Y_{e_a}}(B_1^*) - dim(Y_{e_a})\hfill \text{ [from lemma~\ref{Lem1}]}\\
\text{or, } codim_{A}(S_a) \leq dim(Y_{e_b}) + dim(Y_{e_a}) - dim(A^*) + dim(Y_{e_a}) - dim(B_1^*) - dim(Y_{e_a})\\
\text{or, } codim_{A}(S_a) \leq dim(Y_{e_b}) + dim(Y_{e_a}) - dim(f_{M_1}(\bar{A})) - dim(f_{K_1}(\bar{B_1}))\\
\text{or, } codim_{A}(S_a) \leq dim(Y_{e_b}) + dim(Y_{e_a}) - dim(\bar{A}) - dim(\bar{B_1})\\
\text{or, } codim_{A}(S_a) \leq dim(Y_{e_b}) + dim(Y_{e_a}) + codim_A(\bar{A}) + codim_{B_1}(\bar{B_1}) -dim(A) -dim(B_1)\IEEEeqnarraynumspace\IEEEyesnumber\label{tm0}
\end{IEEEeqnarray*}
for $1\leq i\leq q$, from (\ref{r2}) we have:
\begin{IEEEeqnarray*}{l}
codim_A(S_i) = codim_A(f_{U_i}^{-1}(f_{R_i}({B_i}^{***}))) \leq codim_{Y_{e_i}}(f_{R_i}({B_i}^{***})) \hfill \text{[from Lemma~\ref{Lem2}]}\\
\text{or, } codim_A(S_i) \leq dim(Y_{e_i}) - dim(f_{R_i}({B_i}^{***}))\\
\text{Since } f_{K_i}(B_i^{***}) \text{ is a subspace of } B_1^*, f_{C_1}f_{K_i} \text{ is invertible over } B_i^{***}; \text{ from equation~(\ref{nbibj}) } f_{B_{1i}}f_{R_i}, \text{ and hence } f_{R_i} \\\text{must be invertible over } B_i^{***}. \text{ So,}\\
codim_A(S_i) \leq dim(Y_{e_i}) - dim({B_i}^{***}) = dim(Y_{e_i}) - dim(f_{C_i}({B_i}^{**})) \IEEEyesnumber\label{tm1}\\
\text{Now, $B_i^{**}$ is a subspace of $f_{K_i}(\bar{B_i})$, and over $f_{K_i}(\bar{B_i})$ $f_{C_i}$ is invertible from equation (\ref{nbi}). So, for } 2\leq i\leq q \text{ we have:}\\
codim_A(S_i) \leq dim(Y_{e_i}) - dim({B_i}^{**}) = dim(Y_{e_i}) - dim({B_i}^{*} \cap B_1^*) = dim(Y_{e_i}) + codim_{Y_{e_a}}({B_i}^{*} \cap B_1^*) - dim(Y_{e_a})\\
\leq dim(Y_{e_i}) + codim_{Y_{e_a}}({B_i}^{*}) + codim_{Y_{e_a}}(B_1^*) - dim(Y_{e_a})\hfill \text{ [using lemma~\ref{Lem1}]}\\
=\> dim(Y_{e_i}) + dim(Y_{e_a}) - dim(B_i^*) + dim(Y_{e_a}) - dim(B_1^*) - dim(Y_{e_a})\\
=\> dim(Y_{e_i}) + dim(Y_{e_a}) - dim(f_{K_i}(\bar{B_i})) - dim(f_{K_1}(\bar{B_1}))\\
=\> dim(Y_{e_i}) + dim(Y_{e_a}) - dim(\bar{B_i}) - dim(\bar{B_1})\hfill \text{ [using equation (\ref{nbi})]}\\
=\> dim(Y_{e_i}) + dim(Y_{e_a}) + codim_{B_i}(\bar{B_i}) + codim_{B_1}(\bar{B_1}) - dim(B_i) - dim(B_1)\IEEEyesnumber\label{tm2}\\
\text{for } i=1 \text{ form equation~(\ref{tm1}) we have:}\\
codim_A(S_1) \leq dim(Y_{e_1}) - dim({B_1}^{**}) = dim(Y_{e_1}) - dim({B_1}^{*} \cap B_2^* \cap \cdots \cap B_q^*)\\
=\> dim(Y_{e_1}) + codim_{Y_{e_a}}({B_1}^{*} \cap B_2^* \cap \cdots \cap B_q^*) - dim(Y_{e_a})\\
\leq dim(Y_{e_1}) + \sum_{i=1}^q codim_{Y_{e_a}}({B_i}^{*}) - dim(Y_{e_a})\hfill \text{ [from lemma~\ref{Lem1}]}\\
=\> dim(Y_{e_1}) + (q)dim(Y_{e_a}) - \sum_{i=1}^q dim({B_i}^{*}) - dim(Y_{e_a})\\
=\> dim(Y_{e_1}) + (q-1)dim(Y_{e_a}) - \sum_{i=1}^q dim(f_{K_i}(\bar{B_i})) \\
=\> dim(Y_{e_1}) + (q-1)dim(Y_{e_a}) - \sum_{i=1}^q dim(\bar{B_i})\hfill \text{ [using equation (\ref{nbi})]}\\
=\> dim(Y_{e_1}) + (q-1)dim(Y_{e_a}) + \sum_{i=1}^q codim_{B_i}(\bar{B_i}) - \sum_{i=1}^q dim(B_i)\IEEEyesnumber\label{tm3}\\
\end{IEEEeqnarray*}
So, substituting equations (\ref{tm0}), (\ref{tm2}), and (\ref{tm3}) in equation (\ref{mq}) we have:
\begin{IEEEeqnarray*}{l}
H(A) \leq codim_A(\hat{A}) + codim_A(S_a) + \sum_{i=1}^q codim_A(S_i)\\
\text{or, }H(A) \leq codim_A(\hat{A}) + dim(Y_{e_b}) + dim(Y_{e_a}) + codim_A(\bar{A}) + codim_{B_1}(\bar{B_1}) -dim(A) -dim(B_1) + dim(Y_{e_1}) \\+\> (q-1)dim(Y_{e_a}) + \sum_{i=1}^q codim_{B_i}(\bar{B_i}) - \sum_{i=1}^q dim(B_i) + \sum_{i=2}^q dim(Y_{e_i}) + (q-1)dim(Y_{e_a}) + \sum_{i=2}^q codim_{B_i}(\bar{B_i}) \\\hfill+\> (q-1)codim_{B_1}(\bar{B_1}) - \sum_{i=2}^q dim(B_i) - (q-1)dim(B_1)\\
\text{or, }H(A) \leq codim_A(\hat{A}) + H(Y_{e_b}) + (2q-1)H(Y_{e_a}) + codim_A(\bar{A}) - H(A) -(q+1)H(B_1)  + \sum_{i=1}^q H(Y_{e_i})  \\\hfill +\> \sum_{i=2}^q 2codim_{B_i}(\bar{B_i}) + (q+1)codim_{B_1}(\bar{B_1}) - \sum_{i=2}^q 2H(B_i)\\
%
\text{Substituting values from equations (\ref{codima}), (\ref{codimbi}), and (\ref{codimhata}) we get:}\\
H(A) \leq \sum_{i=1}^{q} codim_{Y_{e_i}}(Y_{e_i}) + codim_{Y_{e_b}}(Y_{e_b}) + codim_{A}(A^{\prime\prime}) + H(Y_{e_b}) + (2q-1)H(Y_{e_a}) + codim_{Y_{e_a}}(Y_{e_a}^\prime) \\+\> codim_{Y_{e_b}}(Y_{e_b}^\prime) + codim_A(A^\prime) - H(A) -(q+1)H(B_1)  + \sum_{i=1}^q H(Y_{e_i})  + \sum_{i=2}^q 2(codim_{Y_{e_a}}(Y_{e_a}^\prime) + codim_{Y_{e_i}}(Y_{e_i}^\prime) \\+\> codim_{B_i}(B_i^\prime)) + (q+1)(codim_{Y_{e_a}}(Y_{e_a}^\prime) + codim_{Y_{e_1}}(Y_{e_1}^\prime) + codim_{B_1}(B_1^\prime)) - \sum_{i=2}^q 2H(B_i) \\+\> (3q+1)(H(A) + \sum_{i=1}^q H(B_i) - H(A,B_1,\ldots ,B_q))\\
\text{or, }H(A) \leq (q+2)codim_{Y_{e_1}}(Y_{e_1}^\prime) + \sum_{i=2}^{q} 3codim_{Y_{e_i}}(Y_{e_i}^\prime) + 2codim_{Y_{e_b}}(Y_{e_b}^\prime) + codim_{A}(A^{\prime\prime}) + H(Y_{e_b}) + (2q-1)H(Y_{e_a}) \\+\> (3q)codim_{Y_{e_a}}(Y_{e_a}^\prime)  + codim_A(A^\prime) - H(A) -(q+1)H(B_1)  + \sum_{i=1}^q H(Y_{e_i})  + \sum_{i=2}^q 2codim_{B_i}(B_i^\prime) \\+\> (q+1)codim_{B_1}(B_1^\prime) - \sum_{i=2}^q 2H(B_i) + (3q+1)(H(A) + \sum_{i=1}^q H(B_i) - H(A,B_1,\ldots ,B_q))\\
\text{Substituting values from equations (\ref{y_a}), (\ref{y_b}), (\ref{y_i}), (\ref{apime}), (\ref{b_iprime}) and (\ref{a_primeprime}) we get:}\\
\text{or, }H(A) \leq (q+2)H(Y_{e_1}|A,\cup_{j=2}^q B_j) + \sum_{i=2}^{q} 3H(Y_{e_i}|A,\cup_{j=1,j\neq i}^q B_j) + 2H(Y_{e_b}|B_1,\ldots ,B_q) + H(A|Y_{e_1},\ldots ,Y_{e_q},Y_{e_b}) \\+\> H(Y_{e_b}) + (2q-1)H(Y_{e_a}) + (3q)H(Y_a|A,B_1,\ldots ,B_q)  + H(A|Y_{e_a},Y_{e_b}) - H(A) -(q+1)H(B_1)  + \sum_{i=1}^q H(Y_{e_i})  \\+\> \sum_{i=2}^q 2 H(B_i|Y_{e_a},Y_{e_i}) + (q+1)H(B_1|Y_{e_a},Y_{e_1}) - \sum_{i=2}^q 2H(B_i) + (3q+1)(H(A) + \sum_{i=1}^q H(B_i) - H(A,B_1,\ldots ,B_q))\\
\end{IEEEeqnarray*}
Replacing $Y_{e_a}$ by $X$, $Y_{e_i}$ by $Y_i$ and $Y_{e_b}$ by $Z$ we get the desired inequality.
\begin{IEEEeqnarray*}{l}
H(A) \leq (q+2)H(Y_1|A,\cup_{j=2}^q B_j) + \sum_{i=2}^{q} 3H(Y_i|A,\cup_{j=1,j\neq i}^q B_j) + 2H(Z|B_1,\ldots ,B_q) + H(A|Y_1,\ldots ,Y_q,Z) + H(Z) \\+\> (2q-1)H(X) + (3q)H(X|A,B_1,\ldots ,B_q)  + H(A|X,Z) - H(A) -(q+1)H(B_1)  + \sum_{i=1}^q H(Y_i)  + \sum_{i=2}^q 2 H(B_i|X,Y_i) \\+\> (q+1)H(B_1|X,Y_1) - \sum_{i=2}^q 2H(B_i) + (3q+1)(H(A) + \sum_{i=1}^q H(B_i) - H(A,B_1,\ldots ,B_q))
%
\end{IEEEeqnarray*}
Rearranging terms we get:
\begin{IEEEeqnarray*}{l}
2H(A) + (q+1)H(B_1) + \sum_{i=2}^q 2H(B_i) \leq (2q-1)H(X) + \sum_{i=1}^q H(Y_i) + H(Z) + (3q)H(X|A,B_1,\ldots ,B_q) \\
+\> (q+2)H(Y_1|A,\cup_{j=2}^q B_j)  + \sum_{i=2}^{q} 3H(Y_i|A,\cup_{j=1,j\neq i}^q B_j)  + 2H(Z|B_1,\ldots ,B_q) + H(A|Y_1,\ldots ,Y_q,Z) + H(A|X,Z) \\+\> (q+1)H(B_1|X,Y_1) + \sum_{i=2}^q 2H(B_i|X,Y_i) + (3q+1)(H(A) + \sum_{i=1}^q H(B_i) - H(A,B_1,\ldots ,B_q))
\end{IEEEeqnarray*}

\subsection{Using the proof of lemma~\ref{lem4} to find the set $S$}\label{use4}
We now show that if we had used the proof of lemma~\ref{lem4} for finding the set $S$, in order to compute a characteristic-dependent rank inequality from the network in Fig.~\ref{nofano1/n} for $n=1$, then the upper-bound on the linear coding capacity of $\mathcal{N}_2$ produced by the resultant inequality would have been greater than the respective upper-bound produced by the inequality in (\ref{thmeq1}).
Towards this end, we first define some subspaces which will be required to obtain equations analogous to equations (\ref{5}) and (\ref{6}).
\begin{equation}
\text{for } 1\leq i\leq q:\; S_{A_i} = \{ u \in A | f_{M_1}(u) \in f_{K_i}(\bar{B_i}) \}
\end{equation}
Now note that over $f_{K_i}(\bar{B_i})$, $f_{C_i}$ is one-to-one from equation (\ref{nbi}); and $f_{M_1}$ is one-to-one over $\bar{A}$ from equation (\ref{naa}). Then, over a subspace $\bar{A} \cap S_{A_i}$, $f_{C_i}f_{M_1}$ is one-to-one. Hence from equation (\ref{nabi}), both $f_{D_i}$ and $f_{M_2}$ are one-to-one over $\bar{A} \cap S_{A_i}$.

Now note that from equation (\ref{naa}), we have, over $f_{M_1}(\bar{A})$:
\begin{equation}
f_{M_1}f_{Q} = I \label{la1}
\end{equation}
Multiplying both sides of equation (\ref{nbia}) by $f_{M_1}$ we have:
\begin{equation}
f_{M_1}f_{Q}f_{K_i} + f_{M_1}f_{A_i}f_{R_i} = 0 \label{uu0} 
\end{equation}
Consider the below subspaces:
\begin{equation}
\text{for } 1\leq i\leq q:\; S_{B_i} = \{ u \in B_i | f_{K_i}(u) \in f_{M_1}(\bar{A}) \}
\end{equation}
Then over $\bar{B_i} \cap S_{B_i})$, from equation (\ref{uu0}) we have:
\begin{equation}
f_{K_i} + f_{M_1}f_{A_i}f_{R_i} = 0\label{la4}
\end{equation}
Multiplying both sides of equation (\ref{la4}) by $f_{C_i}$ we get:
\begin{IEEEeqnarray}{l}
f_{C_i}f_{K_i} + f_{C_i}f_{M_1}f_{A_i}f_{R_i} = 0\IEEEnonumber\\
\text{or, from equation (\ref{nbi}): } f_{C_i}f_{M_1}f_{A_i}f_{R_i} = -I \label{la6}
\end{IEEEeqnarray}

We define the set $S$ as following:
\begin{IEEEeqnarray*}{l}
S_{\hat{A}} = \{ u \in \hat{A} | f_{M_3}(u) \in f_{M_2}(\bar{A} \cap \cap_{i=1}^q S_{A_i})\}\label{la2}\\
R_{\hat{A_i}} = \{ u \in \hat{A} | f_{U_i}(u) \in f_{R_i}(\bar{B_i} \cap S_{B_i}) \}\label{la3}\\
S = \hat{A} \cap \bar{A} \cap \cap_{i=1}^q S_{A_i} \cap S_{\hat{A}} \cap \cap_{i=1}^q R_{\hat{A_i}}\label{la5}
\end{IEEEeqnarray*}
Let $\hat{a} \in S$. Then from equation (\ref{ncb}), for $1\leq i\leq q$ we have:
\begin{IEEEeqnarray*}{l}
\sum_{j=1,j\neq i}^q f_{B_{ij}}f_{U_j}(\hat{a}) + f_{D_i}f_{M_3}(\hat{a}) = 0\\
\text{From (\ref{la2}) we know there exists a $a \in \bar{A} \cap \cap_{i=1}^q S_{A_i}$ such that $f_{M_3}(\hat{a}) = f_{M_2}(a)$. So,}\\
\sum_{j=1,j\neq i}^q f_{B_{ij}}f_{U_j}(\hat{a}) + f_{D_i}f_{M_2}(a) = 0\\
\text{Substituting $f_{D_i}f_{M_2}(a)$ from equation (\ref{nabi}) we have:}\\
\sum_{j=1,j\neq i}^q f_{B_{ij}}f_{U_j}(\hat{a}) - f_{C_i}f_{M_1}(a) = 0\\
\text{Since $f_{M_2}$ is invertible over $\bar{A} \cap \cup_{i=1}^q S_{A_i}$, we can write:}\\
\sum_{j=1,j\neq i}^q f_{B_{ij}}f_{U_j}(\hat{a}) - f_{C_i}f_{M_1}f_{M_2}^{-1}f_{M_2}(a) = 0\\
\text{or, }\sum_{j=1,j\neq i}^q f_{B_{ij}}f_{U_j}(\hat{a}) - f_{C_i}f_{M_1}f_{M_2}^{-1}f_{M_3}(\hat{a}) = 0\\
\text{From (\ref{la3}) we know there exists a $b_j \in \bar{B_j} \cap S_{B_j}$ such that $f_{U_j}(\hat{a}) = f_{R_j}(b_j)$. So,}\\
\sum_{j=1,j\neq i}^q f_{B_{ij}}f_{R_j}(b_j) - f_{C_i}f_{M_1}f_{M_2}^{-1}f_{M_3}(\hat{a}) = 0\\
\text{Substituting $f_{B_{ij}}f_{R_j}(b_j)$ from equation (\ref{nbibj}) we have:}\\
\sum_{j=1,j\neq i}^q -f_{C_i}f_{K_j}(b_j) - f_{C_i}f_{M_1}f_{M_2}^{-1}f_{M_3}(\hat{a})= 0\\
\text{Substituting $f_{K_j}(b_j)$ from equation (\ref{la4}) we have:}\\
\sum_{j=1,j\neq i}^q f_{C_i}f_{M_1}f_{A_j}f_{R_j}(b_j) - f_{C_i}f_{M_1}f_{M_2}^{-1}f_{M_3}(\hat{a}) = 0\\
\text{or, }\sum_{j=1,j\neq i}^q f_{C_i}f_{M_1}f_{A_j}f_{U_j}(\hat{a}) - f_{C_i}f_{M_1}f_{M_2}^{-1}f_{M_3}(\hat{a}) = 0\\
\text{or, } f_{C_i}f_{M_1}(\sum_{j=1,j\neq i}^q f_{A_j}f_{U_j} - f_{M_2}^{-1}f_{M_3})(\hat{a}) = 0\\
\text{Now from equation (\ref{ncc}) we have $(f_{A_i}f_{U_i} + \sum_{j=1,j\neq i}^q f_{A_j}f_{U_j})(\hat{a}) = \hat{a}$. So,}\\
f_{C_i}f_{M_1}(\hat{a} - f_{A_i}f_{U_i}(\hat{a}) - f_{M_2}^{-1}f_{M_3}(\hat{a})) = 0\\
\text{Using (\ref{la5}), (\ref{la6}) and (\ref{la2}) we have:}\\
\hat{a} - f_{A_i}f_{U_i}(\hat{a}) - f_{M_2}^{-1}f_{M_3}(\hat{a}) = 0\\
\text{or, } f_{A_i}f_{U_i}(\hat{a}) = \hat{a} - f_{M_2}^{-1}f_{M_3}(\hat{a})\IEEEyesnumber\label{la7}
\end{IEEEeqnarray*}
As equation (\ref{la7}) holds for $1\leq i\leq q$ we have:
\begin{IEEEeqnarray*}{l}
\sum_{i=1}^q f_{A_i}f_{U_i}(\hat{a}) = q\hat{a} - qf_{M_2}^{-1}f_{M_3}(\hat{a})\\
\text{Using equation (\ref{ncc}): } \hat{a} = q\hat{a} - qf_{M_2}^{-1}f_{M_3}(\hat{a})\\
\text{As $q = 0$ over the finite field: } \hat{a} = 0
\end{IEEEeqnarray*}
As this holds for any $\hat{a} \in S$, we must have $S = \{ 0 \}$.
Now we calculate some values that help us compute an upper-bound over $dim(A)$.%

\begin{IEEEeqnarray*}{l}
codim_A(S_{A_i}) = codim_{A}(f_{M_1}^{-1}(f_{K_i}(\bar{B_i}))) \leq codim_{Y_{e_a}}(f_{K_i}(\bar{B_i})) = dim(Y_{e_a}) + dim(f_{K_i}(\bar{B_i}))\\
\text{or, } codim_A(S_{A_i}) \leq H(Y_{e_a}) + dim(\bar{B_i}) = H(Y_{e_a}) + codim_{B_i}(\bar{B_i}) - H(B_i)\IEEEyesnumber\label{la10}
\end{IEEEeqnarray*}

\begin{IEEEeqnarray*}{l}
codim_{B_i}(S_{B_i}) = codim_{B_i}(f_{K_i}^{-1}(f_{M_1}(\bar{A}))) \leq codim_{Y_{e_a}}(f_{M_1}(\bar{A})) = H(Y_{e_a}) - dim(f_{M_1}(\bar{A}))\\
codim_{B_i}(S_{B_i}) \leq H(Y_{e_a}) - dim(\bar{A}) = H(Y_{e_a}) + codim_{A}(\bar{A}) - H(A)\IEEEyesnumber\label{la9}
\end{IEEEeqnarray*}

\begin{IEEEeqnarray*}{l}
H(A) =  dim(A) - dim(S) =  codim_{A}(S)\\
\text{or, } H(A) \leq codim_{A}(\hat{A}) + codim_{A}(\bar{A}) + \sum_{i=1}^q codim_{A}(S_{A_i}) + codim_{A}(f_{M_3}^{-1}(f_{M_2}(\bar{A} \cap \cap_{i=1}^q S_{A_i}))) \\\hfill +\> \sum_{i=1}^q codim_{A}(f_{U_i}^{-1}(f_{R_i}(\bar{B_i} \cap S_{B_i})))\\
\text{or, } H(A) \leq codim_{A}(\hat{A}) + codim_{A}(\bar{A}) + \sum_{i=1}^q codim_{A}(S_{A_i}) + codim_{Y_{e_b}}(f_{M_2}(\bar{A} \cap \cap_{i=1}^q S_{A_i})) + \sum_{i=1}^q codim_{Y_{e_i}}(f_{R_i}(\bar{B_i} \cap S_{B_i}))\\
\text{As over $\bar{A} \cap \cap_{i=1}^q S_{A_i}$, $f_{M_2}$ is one-to-one, and as from equation (\ref{la4}) over $\bar{B_i} \cap S_{B_i})$, $f_{R_i}$ is one-to-one: }\\
\text{or, } H(A) \leq codim_{A}(\hat{A}) + codim_{A}(\bar{A}) + \sum_{i=1}^q codim_{A}(S_{A_i}) + codim_{Y_{e_b}}(\bar{A} \cap \cap_{i=1}^q S_{A_i}) + \sum_{i=1}^q codim_{Y_{e_i}}(\bar{B_i} \cap S_{B_i})\\
\text{or, } H(A) \leq codim_{A}(\hat{A}) + codim_{A}(\bar{A}) + \sum_{i=1}^q codim_{A}(S_{A_i}) + H(Y_{e_b}) + codim_{A}(\bar{A} \cap \cap_{i=1}^q S_{A_i}) - H(A) + \sum_{i=1}^q H(Y_{e_i}) \\\hfill +\> \sum_{i=1}^q codim_{B_i}(\bar{B_i} \cap S_{B_i}) -  \sum_{i=1}^q H(B_i)\\
\text{or, } H(A) \leq codim_{A}(\hat{A}) + codim_{A}(\bar{A}) + \sum_{i=1}^q codim_{A}(S_{A_i}) + H(Y_{e_b}) + codim_{A}(\bar{A}) +  \sum_{i=1}^q codim_{A}(S_{A_i}) - H(A) \\\hfill +\> \sum_{i=1}^q H(Y_{e_i}) + \sum_{i=1}^q codim_{B_i}(\bar{B_i}) +  \sum_{i=1}^q codim_{B_i}(S_{B_i}) -  \sum_{i=1}^q H(B_i)\\
\text{or, } H(A) \leq codim_{A}(\hat{A}) + 2codim_{A}(\bar{A}) + \sum_{i=1}^q 2codim_{A}(S_{A_i}) + H(Y_{e_b})  - H(A) + \sum_{i=1}^q H(Y_{e_i}) + \sum_{i=1}^q codim_{B_i}(\bar{B_i}) \\\hfill +\>  \sum_{i=1}^q codim_{B_i}(S_{B_i}) -  \sum_{i=1}^q H(B_i)\\
\text{Substituting values from equations (\ref{la9}) and (\ref{la10}) we have:}\\
\text{or, } H(A) \leq codim_{A}(\hat{A}) + 2codim_{A}(\bar{A}) + \sum_{i=1}^q 2(H(Y_{e_a}) + codim_{B_i}(\bar{B_i}) - H(B_i)) + H(Y_{e_b})  - H(A) + \sum_{i=1}^q H(Y_{e_i}) \\\hfill +\> \sum_{i=1}^q codim_{B_i}(\bar{B_i}) +  \sum_{i=1}^q (H(Y_{e_a}) + codim_{A}(\bar{A}) - H(A)) -  \sum_{i=1}^q H(B_i)\\
\text{or, } H(A) \leq codim_{A}(\hat{A}) + (q+2)codim_{A}(\bar{A}) + 3qH(Y_{e_a}) + \sum_{i=1}^q 3codim_{B_i}(\bar{B_i}) - \sum_{i=1}^q  3H(B_i) + H(Y_{e_b})  - (q+1)H(A) \\\hfill +\> \sum_{i=1}^q H(Y_{e_i})\IEEEeqnarraynumspace\IEEEyesnumber\label{la8}
%
%
\end{IEEEeqnarray*}
Now substituting values from equations (\ref{codima}), (\ref{codimbi}), and (\ref{codimhata}) it can be seen that when equation (\ref{la8}) is applied to the network $\mathcal{N}_2$, it results an upper-bound equal to $\frac{(4q+1)k}{(4q+2)n}$.

\end{document}